%% file: P02_LFI_Processing_2013.tex
\documentclass[traditabstract,longauth]{aa}
\usepackage{epsfig}
\usepackage{graphicx}
\usepackage{url}
\usepackage[dvips]{color}
\usepackage{natbib}
\usepackage[breaklinks,colorlinks,citecolor=blue]{hyperref}
\usepackage{amssymb,,amsmath}
\usepackage{longtable}
\usepackage{multirow}
\usepackage{fixltx2e}
\usepackage{breakurl}
\usepackage{txfonts}
\bibpunct{(}{)}{;}{a}{}{,}

\def\lsim{\mathrel{\raise .4ex\hbox{\rlap{$<$}\lower 1.2ex\hbox{$\sim$}}}}
\def\gsim{\mathrel{\raise .4ex\hbox{\rlap{$>$}\lower 1.2ex\hbox{$\sim$}}}}
\def\arcm{\ifmmode {^{\scriptscriptstyle\prime}}
          \else $^{\scriptscriptstyle\prime}$\fi}

\input Planck.tex

\input Planck_updated_lfi_values.tex

\begin{document}
\title{\textit{Planck} 2013 results. II. Low Frequency Instrument data processing}

\input{AuthorList_P02_LFI_Processing_authors_and_institutes.tex}

\titlerunning{LFI data processing}
\authorrunning{Planck Collaboration}


\input{00_abstract}

\maketitle
\allearlypapers

\section{Introduction}
\label{sec_introduction}
    \input{01_introduction}

\section{In-flight behaviour and operations}
\label{sec_flightbahavior}
    \input{02-00_inflight_behav_intro}

    \subsection{Operations}
    \label{sec_operations}
        \input{02-01_operations}

    \subsection{Instrument performance update} \label{sec_instrupdate}
    \label{sec_instrupdate}
        \input{02-02_instrupdate_summary}

\section{Data processing overview}
\label{sec_dataproc_overview}
    \input{03-00_data_proc_over}

\section{Time ordered information (TOI) processing}
\label{sec_toiprocessing}
    \input{04-00_toi_processing}

    \subsection{Input flags}
    \label{sec_flagging}
        \input{04-01_flagging}

    \subsection{ADC linearity correction}
    \label{sec_adc_nonlinearity}
        \input{04-02_adc}

    \subsection{Corrections for electronic spikes}
    \label{sec_electronic_spikes}
        \input{04-03_spikes}

    \subsection{Demodulation: Gain modulation factor estimation and application}
    \label{sec_gain_modulation}
        \input{04-04_gain_modulation}

    \subsection{Combining diodes}
    \label{sec_comb_diodes}
        \input{04-05_comb_diodes}

\section{Pointing}
\label{sec_pointing}
    \input{05-00_pointing}

\section{Main beams and the geometrical calibration of the focal plane}
\label{sec_beamrecovery}
    \input{06-00_beamrecovery}

    \subsection{Scanning beams}
    \label{sec_scanningbeam}
        \input{06-01_scanningbeam}

    \subsection{Effective beams}
    \label{sec_effectivebeam}
        \input{06-02_effectivebeam}

\section{Photometric calibration}
\label{sec_calibration}
    \input{07-00_calibration}

    \subsection{Iterative calibration}
    \label{sec_iterative_calib}
        \input{07-01_iterative_calib}

    \subsection{Calibration using 4K reference load signal}
    \label{sec_4k_calib}
        \input{07-02_4k_calib}

    \subsection{Color correction}
    \label{sec_color_correction}
        \input{07-03_colcorr}

\section{Noise estimation}
\label{sec_noise}
    \input{08-00_noise}

    \subsection{Updated noise properties}
    \label{sec_update_noise}
        \input{08-01_instrupdate_noise}

\section{Mapmaking}
\label{sec_mmaking_intro}
    \input{09-00_mmaking_intro}

    \subsection{{\tt MADAM} pipeline for frequency maps}
    \label{sec_madam}
        \input{09-01_madam}

    \subsection{Low-resolution data set}
    \label{sec_low_cov}
        \input{09-02_low_cov}

    \subsection{Half-ring noise maps}
    \label{sec_halfring}
        \input{09-03_halfring}

    \subsection{Noise Monte Carlo simulations}
    \label{sec_noise_simulations}
        \input{09-04_noise_simulations}

    \subsection{Overview of LFI map properties}
    \label{sec_OverviewMaps}
        \input{09-05_Overview_maps}

\section{Polarization}
\label{sec_polarization}
\input{10-00_polarization}

\section{Power spectra}
\label{sec_power_spectra}
\input{11-00_power_spectra}

\section{Data validation}
\label{sec_dataval_intro}
\input{12-00_dataval_intro}

    \subsection{Null test results}
    \label{sec_null_test}
        \input{12-01_null_test}

    \subsection{Half-ring test}
    \label{sec_half_ring_test}
        \input{12-02_half_ring_test}

    \subsection{Intra-frequency consistency check}
    \label{sec_consistency_scatter}
        \input{12-03_consistency_scatter}

    \subsection{70\,GHz internal consistency check}
    \label{sec_70_int_consis}
        \input{12-04_int_consis}

    \subsection{Updated systematic effects assessment}
    \label{sec_systematics}
        \input{12-05_systematics}

\section{Infrastructure overview}
\label{infrast}
\input{13_infrast}
\section{Discussion and conclusions}
\label{conclusion}
\input{14_conclusion}

\input{15_acknow}

\allearlypapers

\bibliographystyle{aa}
\bibliography{Planck_bib,LFI_DPC_bib}

\raggedright
\end{document}

%% file: Planck.tex
\def\setsymbol#1#2{\expandafter\def\csname #1\endcsname{#2}}
\def\getsymbol#1{\csname #1\endcsname}

\def\Planck{\textit{Planck}}

\def\allearlypapers{\nocite{planck2011-1.1, planck2011-1.3, planck2011-1.4, planck2011-1.5, planck2011-1.6, planck2011-1.7, planck2011-1.10, planck2011-1.10sup, planck2011-5.1a, planck2011-5.1b, planck2011-5.2a, planck2011-5.2b, planck2011-5.2c, planck2011-6.1, planck2011-6.2, planck2011-6.3a, planck2011-6.4a, planck2011-6.4b, planck2011-6.6, planck2011-7.0, planck2011-7.2, planck2011-7.3, planck2011-7.7a, planck2011-7.7b, planck2011-7.12, planck2011-7.13}}

\def\all2013resultspapers{\nocite{planck2013-p01, planck2013-p02, planck2013-p02a, planck2013-p02d, planck2013-p02b, planck2013-p03, planck2013-p03c, planck2013-p03f, planck2013-p03d, planck2013-p03e, planck2013-p06b, planck2013-p06, planck2013-p03a, planck2013-pip88, planck2013-p08, planck2013-p11, planck2013-p12, planck2013-p13, planck2013-p14, planck2013-p15, planck2013-p05b, planck2013-p17, planck2013-p09, planck2013-p09a, planck2013-p20, planck2013-p19, planck2013-pipaberration, planck2013-p05, planck2013-p05a, planck2013-pip56, planck2013-p01a}}

\newbox\tablebox    \newdimen\tablewidth
\def\leaderfil{\leaders\hbox to 5pt{\hss.\hss}\hfil}
\def\endPlancktable{\tablewidth=\columnwidth 
    $$\hss\copy\tablebox\hss$$
    \vskip-\lastskip\vskip -2pt}
\def\endPlancktablewide{\tablewidth=\textwidth 
    $$\hss\copy\tablebox\hss$$
    \vskip-\lastskip\vskip -2pt}
\def\tablenote#1 #2\par{\begingroup \parindent=0.8em
    \abovedisplayshortskip=0pt\belowdisplayshortskip=0pt
    \noindent
    $$\hss\vbox{\hsize\tablewidth \hangindent=\parindent \hangafter=1 \noindent
    \hbox to \parindent{$^#1$\hss}\strut#2\strut\par}\hss$$
    \endgroup}
\def\doubleline{\vskip 3pt\hrule \vskip 1.5pt \hrule \vskip 5pt}

\def\L2{\ifmmode L_2\else $L_2$\fi}

\def\DeltaT{\ifmmode \Delta T\else $\Delta T$\fi}
\def\deltat{\ifmmode \Delta t\else $\Delta t$\fi}
\def\fknee{\ifmmode f_{\rm knee}\else $f_{\rm knee}$\fi}
\def\Fmax{\ifmmode F_{\rm max}\else $F_{\rm max}$\fi}
\def\solar{\ifmmode{\rm M}_{\mathord\odot}\else${\rm M}_{\mathord\odot}$\fi}
\def\Msolar{\ifmmode{\rm M}_{\mathord\odot}\else${\rm M}_{\mathord\odot}$\fi}
\def\Lsolar{\ifmmode{\rm L}_{\mathord\odot}\else${\rm L}_{\mathord\odot}$\fi}

\def\inv{\ifmmode^{-1}\else$^{-1}$\fi}
\def\mo{\ifmmode^{-1}\else$^{-1}$\fi}
\def\sup#1{\ifmmode ^{\rm #1}\else $^{\rm #1}$\fi}
\def\expo#1{\ifmmode \times 10^{#1}\else $\times 10^{#1}$\fi}
\def\,{\thinspace}
\def\lsim{\mathrel{\raise .4ex\hbox{\rlap{$<$}\lower 1.2ex\hbox{$\sim$}}}}
\def\gsim{\mathrel{\raise .4ex\hbox{\rlap{$>$}\lower 1.2ex\hbox{$\sim$}}}}

\def\simprop{\mathrel{\raise .4ex\hbox{\rlap{$\propto$}\lower 1.2ex\hbox{$\sim$}}}}
\def\deg{\ifmmode^\circ\else$^\circ$\fi}
\def\pdeg{\ifmmode $\setbox0=\hbox{$^{\circ}$}\rlap{\hskip.11\wd0 .}$^{\circ}
          \else \setbox0=\hbox{$^{\circ}$}\rlap{\hskip.11\wd0 .}$^{\circ}$\fi}
\def\arcs{\ifmmode {^{\scriptstyle\prime\prime}}
          \else $^{\scriptstyle\prime\prime}$\fi}
\def\arcm{\ifmmode {^{\scriptstyle\prime}}
          \else $^{\scriptstyle\prime}$\fi}
\newdimen\sa  \newdimen\sb
\def\parcs{\sa=.07em \sb=.03em
     \ifmmode \hbox{\rlap{.}}^{\scriptstyle\prime\kern -\sb\prime}\hbox{\kern -\sa}
     \else \rlap{.}$^{\scriptstyle\prime\kern -\sb\prime}$\kern -\sa\fi}
\def\parcm{\sa=.08em \sb=.03em
     \ifmmode \hbox{\rlap{.}\kern\sa}^{\scriptstyle\prime}\hbox{\kern-\sb}
     \else \rlap{.}\kern\sa$^{\scriptstyle\prime}$\kern-\sb\fi}
\def\ra[#1 #2 #3.#4]{#1\sup{h}#2\sup{m}#3\sup{s}\llap.#4}
\def\dec[#1 #2 #3.#4]{#1\deg#2\arcm#3\arcs\llap.#4}
\def\deco[#1 #2 #3]{#1\deg#2\arcm#3\arcs}
\def\rra[#1 #2]{#1\sup{h}#2\sup{m}}

\def\dots{\relax\ifmmode \ldots\else $\ldots$\fi}
\def\WHzsr{\ifmmode $W\,Hz\mo\,sr\mo$\else W\,Hz\mo\,sr\mo\fi}
\def\mHz{\ifmmode $\,mHz$\else \,mHz\fi}
\def\GHz{\ifmmode $\,GHz$\else \,GHz\fi}
\def\mKs{\ifmmode $\,mK\,s$^{1/2}\else \,mK\,s$^{1/2}$\fi}
\def\muKs{\ifmmode \,\mu$K\,s$^{1/2}\else \,$\mu$K\,s$^{1/2}$\fi}
\def\muKRJs{\ifmmode \,\mu$K$_{\rm RJ}$\,s$^{1/2}\else \,$\mu$K$_{\rm RJ}$\,s$^{1/2}$\fi}
\def\muKHz{\ifmmode \,\mu$K\,Hz$^{-1/2}\else \,$\mu$K\,Hz$^{-1/2}$\fi}
\def\MJysr{\ifmmode \,$MJy\,sr\mo$\else \,MJy\,sr\mo\fi}
\def\MJysrmK{\ifmmode \,$MJy\,sr\mo$\,mK$_{\rm CMB}\mo\else \,MJy\,sr\mo\,mK$_{\rm CMB}\mo$\fi}
\def\microns{\ifmmode \,\mu$m$\else \,$\mu$m\fi}

\def\muK{\ifmmode \,\mu$K$\else \,$\mu$\hbox{K}\fi}
\def\microK{\ifmmode \,\mu$K$\else \,$\mu$\hbox{K}\fi}
\def\muW{\ifmmode \,\mu$W$\else \,$\mu$\hbox{W}\fi}
\def\kms{\ifmmode $\,km\,s$^{-1}\else \,km\,s$^{-1}$\fi}
\def\kmsMpc{\ifmmode $\,\kms\,Mpc\mo$\else \,\kms\,Mpc\mo\fi}
\setsymbol{LFI:center:frequency:70GHz:units}{70.3\,GHz}
\setsymbol{LFI:center:frequency:44GHz:units}{44.1\,GHz}
\setsymbol{LFI:center:frequency:30GHz:units}{28.5\,GHz}

\setsymbol{LFI:center:frequency:70GHz}{70.3}
\setsymbol{LFI:center:frequency:44GHz}{44.1}
\setsymbol{LFI:center:frequency:30GHz}{28.5}

\setsymbol{LFI:center:frequency:LFI18:Rad:M:units}{71.7\GHz}
\setsymbol{LFI:center:frequency:LFI19:Rad:M:units}{67.5\GHz}
\setsymbol{LFI:center:frequency:LFI20:Rad:M:units}{69.2\GHz}
\setsymbol{LFI:center:frequency:LFI21:Rad:M:units}{70.4\GHz}
\setsymbol{LFI:center:frequency:LFI22:Rad:M:units}{71.5\GHz}
\setsymbol{LFI:center:frequency:LFI23:Rad:M:units}{70.8\GHz}
\setsymbol{LFI:center:frequency:LFI24:Rad:M:units}{44.4\GHz}
\setsymbol{LFI:center:frequency:LFI25:Rad:M:units}{44.0\GHz}
\setsymbol{LFI:center:frequency:LFI26:Rad:M:units}{43.9\GHz}
\setsymbol{LFI:center:frequency:LFI27:Rad:M:units}{28.3\GHz}
\setsymbol{LFI:center:frequency:LFI28:Rad:M:units}{28.8\GHz}
\setsymbol{LFI:center:frequency:LFI18:Rad:S:units}{70.1\GHz}
\setsymbol{LFI:center:frequency:LFI19:Rad:S:units}{69.6\GHz}
\setsymbol{LFI:center:frequency:LFI20:Rad:S:units}{69.5\GHz}
\setsymbol{LFI:center:frequency:LFI21:Rad:S:units}{69.5\GHz}
\setsymbol{LFI:center:frequency:LFI22:Rad:S:units}{72.8\GHz}
\setsymbol{LFI:center:frequency:LFI23:Rad:S:units}{71.3\GHz}
\setsymbol{LFI:center:frequency:LFI24:Rad:S:units}{44.1\GHz}
\setsymbol{LFI:center:frequency:LFI25:Rad:S:units}{44.1\GHz}
\setsymbol{LFI:center:frequency:LFI26:Rad:S:units}{44.1\GHz}
\setsymbol{LFI:center:frequency:LFI27:Rad:S:units}{28.5\GHz}
\setsymbol{LFI:center:frequency:LFI28:Rad:S:units}{28.2\GHz}

\setsymbol{LFI:center:frequency:LFI18:Rad:M}{71.7}
\setsymbol{LFI:center:frequency:LFI19:Rad:M}{67.5}
\setsymbol{LFI:center:frequency:LFI20:Rad:M}{69.2}
\setsymbol{LFI:center:frequency:LFI21:Rad:M}{70.4}
\setsymbol{LFI:center:frequency:LFI22:Rad:M}{71.5}
\setsymbol{LFI:center:frequency:LFI23:Rad:M}{70.8}
\setsymbol{LFI:center:frequency:LFI24:Rad:M}{44.4}
\setsymbol{LFI:center:frequency:LFI25:Rad:M}{44.0}
\setsymbol{LFI:center:frequency:LFI26:Rad:M}{43.9}
\setsymbol{LFI:center:frequency:LFI27:Rad:M}{28.3}
\setsymbol{LFI:center:frequency:LFI28:Rad:M}{28.8}
\setsymbol{LFI:center:frequency:LFI18:Rad:S}{70.1}
\setsymbol{LFI:center:frequency:LFI19:Rad:S}{69.6}
\setsymbol{LFI:center:frequency:LFI20:Rad:S}{69.5}
\setsymbol{LFI:center:frequency:LFI21:Rad:S}{69.5}
\setsymbol{LFI:center:frequency:LFI22:Rad:S}{72.8}
\setsymbol{LFI:center:frequency:LFI23:Rad:S}{71.3}
\setsymbol{LFI:center:frequency:LFI24:Rad:S}{44.1}
\setsymbol{LFI:center:frequency:LFI25:Rad:S}{44.1}
\setsymbol{LFI:center:frequency:LFI26:Rad:S}{44.1}
\setsymbol{LFI:center:frequency:LFI27:Rad:S}{28.5}
\setsymbol{LFI:center:frequency:LFI28:Rad:S}{28.2}

\setsymbol{LFI:white:noise:sensitivity:70GHz:units}{134.7\muKs}
\setsymbol{LFI:white:noise:sensitivity:44GHz:units}{164.7\muKs}
\setsymbol{LFI:white:noise:sensitivity:30GHz:units}{143.4\muKs}

\setsymbol{LFI:white:noise:sensitivity:70GHz}{134.7}
\setsymbol{LFI:white:noise:sensitivity:44GHz}{164.7}
\setsymbol{LFI:white:noise:sensitivity:30GHz}{143.4}

\setsymbol{LFI:white:noise:sensitivity:LFI18:Rad:M:units}{512.0\muKs}
\setsymbol{LFI:white:noise:sensitivity:LFI19:Rad:M:units}{581.4\muKs}
\setsymbol{LFI:white:noise:sensitivity:LFI20:Rad:M:units}{590.8\muKs}
\setsymbol{LFI:white:noise:sensitivity:LFI21:Rad:M:units}{455.2\muKs}
\setsymbol{LFI:white:noise:sensitivity:LFI22:Rad:M:units}{492.0\muKs}
\setsymbol{LFI:white:noise:sensitivity:LFI23:Rad:M:units}{507.7\muKs}
\setsymbol{LFI:white:noise:sensitivity:LFI24:Rad:M:units}{462.2\muKs}
\setsymbol{LFI:white:noise:sensitivity:LFI25:Rad:M:units}{413.6\muKs}
\setsymbol{LFI:white:noise:sensitivity:LFI26:Rad:M:units}{478.6\muKs}
\setsymbol{LFI:white:noise:sensitivity:LFI27:Rad:M:units}{277.7\muKs}
\setsymbol{LFI:white:noise:sensitivity:LFI28:Rad:M:units}{312.3\muKs}
\setsymbol{LFI:white:noise:sensitivity:LFI18:Rad:S:units}{465.7\muKs}
\setsymbol{LFI:white:noise:sensitivity:LFI19:Rad:S:units}{555.6\muKs}
\setsymbol{LFI:white:noise:sensitivity:LFI20:Rad:S:units}{623.2\muKs}
\setsymbol{LFI:white:noise:sensitivity:LFI21:Rad:S:units}{564.1\muKs}
\setsymbol{LFI:white:noise:sensitivity:LFI22:Rad:S:units}{534.4\muKs}
\setsymbol{LFI:white:noise:sensitivity:LFI23:Rad:S:units}{542.4\muKs}
\setsymbol{LFI:white:noise:sensitivity:LFI24:Rad:S:units}{399.2\muKs}
\setsymbol{LFI:white:noise:sensitivity:LFI25:Rad:S:units}{392.6\muKs}
\setsymbol{LFI:white:noise:sensitivity:LFI26:Rad:S:units}{418.6\muKs}
\setsymbol{LFI:white:noise:sensitivity:LFI27:Rad:S:units}{302.9\muKs}
\setsymbol{LFI:white:noise:sensitivity:LFI28:Rad:S:units}{285.3\muKs}

\setsymbol{LFI:white:noise:sensitivity:LFI18:Rad:M}{512.0}
\setsymbol{LFI:white:noise:sensitivity:LFI19:Rad:M}{581.4}
\setsymbol{LFI:white:noise:sensitivity:LFI20:Rad:M}{590.8}
\setsymbol{LFI:white:noise:sensitivity:LFI21:Rad:M}{455.2}
\setsymbol{LFI:white:noise:sensitivity:LFI22:Rad:M}{492.0}
\setsymbol{LFI:white:noise:sensitivity:LFI23:Rad:M}{507.7}
\setsymbol{LFI:white:noise:sensitivity:LFI24:Rad:M}{462.2}
\setsymbol{LFI:white:noise:sensitivity:LFI25:Rad:M}{413.6}
\setsymbol{LFI:white:noise:sensitivity:LFI26:Rad:M}{478.6}
\setsymbol{LFI:white:noise:sensitivity:LFI27:Rad:M}{277.7}
\setsymbol{LFI:white:noise:sensitivity:LFI28:Rad:M}{312.3}
\setsymbol{LFI:white:noise:sensitivity:LFI18:Rad:S}{465.7}
\setsymbol{LFI:white:noise:sensitivity:LFI19:Rad:S}{555.6}
\setsymbol{LFI:white:noise:sensitivity:LFI20:Rad:S}{623.2}
\setsymbol{LFI:white:noise:sensitivity:LFI21:Rad:S}{564.1}
\setsymbol{LFI:white:noise:sensitivity:LFI22:Rad:S}{534.4}
\setsymbol{LFI:white:noise:sensitivity:LFI23:Rad:S}{542.4}
\setsymbol{LFI:white:noise:sensitivity:LFI24:Rad:S}{399.2}
\setsymbol{LFI:white:noise:sensitivity:LFI25:Rad:S}{392.6}
\setsymbol{LFI:white:noise:sensitivity:LFI26:Rad:S}{418.6}
\setsymbol{LFI:white:noise:sensitivity:LFI27:Rad:S}{302.9}
\setsymbol{LFI:white:noise:sensitivity:LFI28:Rad:S}{285.3}

\setsymbol{LFI:knee:frequency:70GHz:units}{29.5\mHz}
\setsymbol{LFI:knee:frequency:44GHz:units}{56.2\mHz}
\setsymbol{LFI:knee:frequency:30GHz:units}{113.7\mHz}

\setsymbol{LFI:knee:frequency:70GHz}{29.5}
\setsymbol{LFI:knee:frequency:44GHz}{56.2}
\setsymbol{LFI:knee:frequency:30GHz}{113.7}

\setsymbol{LFI:knee:frequency:LFI18:Rad:M:units}{16.3\mHz}
\setsymbol{LFI:knee:frequency:LFI19:Rad:M:units}{15.1\mHz}
\setsymbol{LFI:knee:frequency:LFI20:Rad:M:units}{18.7\mHz}
\setsymbol{LFI:knee:frequency:LFI21:Rad:M:units}{37.2\mHz}
\setsymbol{LFI:knee:frequency:LFI22:Rad:M:units}{12.7\mHz}
\setsymbol{LFI:knee:frequency:LFI23:Rad:M:units}{34.6\mHz}
\setsymbol{LFI:knee:frequency:LFI24:Rad:M:units}{46.2\mHz}
\setsymbol{LFI:knee:frequency:LFI25:Rad:M:units}{24.9\mHz}
\setsymbol{LFI:knee:frequency:LFI26:Rad:M:units}{67.6\mHz}
\setsymbol{LFI:knee:frequency:LFI27:Rad:M:units}{187.4\mHz}
\setsymbol{LFI:knee:frequency:LFI28:Rad:M:units}{122.2\mHz}
\setsymbol{LFI:knee:frequency:LFI18:Rad:S:units}{17.7\mHz}
\setsymbol{LFI:knee:frequency:LFI19:Rad:S:units}{22.0\mHz}
\setsymbol{LFI:knee:frequency:LFI20:Rad:S:units}{8.7\mHz}
\setsymbol{LFI:knee:frequency:LFI21:Rad:S:units}{25.9\mHz}
\setsymbol{LFI:knee:frequency:LFI22:Rad:S:units}{15.8\mHz}
\setsymbol{LFI:knee:frequency:LFI23:Rad:S:units}{129.8\mHz}
\setsymbol{LFI:knee:frequency:LFI24:Rad:S:units}{100.9\mHz}
\setsymbol{LFI:knee:frequency:LFI25:Rad:S:units}{38.9\mHz}
\setsymbol{LFI:knee:frequency:LFI26:Rad:S:units}{58.9\mHz}
\setsymbol{LFI:knee:frequency:LFI27:Rad:S:units}{104.4\mHz}
\setsymbol{LFI:knee:frequency:LFI28:Rad:S:units}{40.7\mHz}

\setsymbol{LFI:knee:frequency:LFI18:Rad:M}{16.3}
\setsymbol{LFI:knee:frequency:LFI19:Rad:M}{15.1}
\setsymbol{LFI:knee:frequency:LFI20:Rad:M}{18.7}
\setsymbol{LFI:knee:frequency:LFI21:Rad:M}{37.2}
\setsymbol{LFI:knee:frequency:LFI22:Rad:M}{12.7}
\setsymbol{LFI:knee:frequency:LFI23:Rad:M}{34.6}
\setsymbol{LFI:knee:frequency:LFI24:Rad:M}{46.2}
\setsymbol{LFI:knee:frequency:LFI25:Rad:M}{24.9}
\setsymbol{LFI:knee:frequency:LFI26:Rad:M}{67.6}
\setsymbol{LFI:knee:frequency:LFI27:Rad:M}{187.4}
\setsymbol{LFI:knee:frequency:LFI28:Rad:M}{122.2}
\setsymbol{LFI:knee:frequency:LFI18:Rad:S}{17.7}
\setsymbol{LFI:knee:frequency:LFI19:Rad:S}{22.0}
\setsymbol{LFI:knee:frequency:LFI20:Rad:S}{8.7}
\setsymbol{LFI:knee:frequency:LFI21:Rad:S}{25.9}
\setsymbol{LFI:knee:frequency:LFI22:Rad:S}{15.8}
\setsymbol{LFI:knee:frequency:LFI23:Rad:S}{129.8}
\setsymbol{LFI:knee:frequency:LFI24:Rad:S}{100.9}
\setsymbol{LFI:knee:frequency:LFI25:Rad:S}{38.9}
\setsymbol{LFI:knee:frequency:LFI26:Rad:S}{58.9}
\setsymbol{LFI:knee:frequency:LFI27:Rad:S}{104.4}
\setsymbol{LFI:knee:frequency:LFI28:Rad:S}{40.7}

\setsymbol{LFI:slope:70GHz:units}{$-1.03$\mHz}
\setsymbol{LFI:slope:44GHz:units}{$-0.89$\mHz}
\setsymbol{LFI:slope:30GHz:units}{$-0.87$\mHz}

\setsymbol{LFI:slope:70GHz}{$-1.03$}
\setsymbol{LFI:slope:44GHz}{$-0.89$}
\setsymbol{LFI:slope:30GHz}{$-0.87$}

\setsymbol{LFI:slope:LFI18:Rad:M:units}{$-1.04$\mHz}
\setsymbol{LFI:slope:LFI19:Rad:M:units}{$-1.09$\mHz}
\setsymbol{LFI:slope:LFI20:Rad:M:units}{$-0.69$\mHz}
\setsymbol{LFI:slope:LFI21:Rad:M:units}{$-1.56$\mHz}
\setsymbol{LFI:slope:LFI22:Rad:M:units}{$-1.01$\mHz}
\setsymbol{LFI:slope:LFI23:Rad:M:units}{$-0.96$\mHz}
\setsymbol{LFI:slope:LFI24:Rad:M:units}{$-0.83$\mHz}
\setsymbol{LFI:slope:LFI25:Rad:M:units}{$-0.91$\mHz}
\setsymbol{LFI:slope:LFI26:Rad:M:units}{$-0.95$\mHz}
\setsymbol{LFI:slope:LFI27:Rad:M:units}{$-0.87$\mHz}
\setsymbol{LFI:slope:LFI28:Rad:M:units}{$-0.88$\mHz}
\setsymbol{LFI:slope:LFI18:Rad:S:units}{$-1.15$\mHz}
\setsymbol{LFI:slope:LFI19:Rad:S:units}{$-1.00$\mHz}
\setsymbol{LFI:slope:LFI20:Rad:S:units}{$-0.95$\mHz}
\setsymbol{LFI:slope:LFI21:Rad:S:units}{$-0.92$\mHz}
\setsymbol{LFI:slope:LFI22:Rad:S:units}{$-1.01$\mHz}
\setsymbol{LFI:slope:LFI23:Rad:S:units}{$-0.95$\mHz}
\setsymbol{LFI:slope:LFI24:Rad:S:units}{$-0.73$\mHz}
\setsymbol{LFI:slope:LFI25:Rad:S:units}{$-1.16$\mHz}
\setsymbol{LFI:slope:LFI26:Rad:S:units}{$-0.79$\mHz}
\setsymbol{LFI:slope:LFI27:Rad:S:units}{$-0.82$\mHz}
\setsymbol{LFI:slope:LFI28:Rad:S:units}{$-0.91$\mHz}

\setsymbol{LFI:slope:LFI18:Rad:M}{$-1.04$}
\setsymbol{LFI:slope:LFI19:Rad:M}{$-1.09$}
\setsymbol{LFI:slope:LFI20:Rad:M}{$-0.69$}
\setsymbol{LFI:slope:LFI21:Rad:M}{$-1.56$}
\setsymbol{LFI:slope:LFI22:Rad:M}{$-1.01$}
\setsymbol{LFI:slope:LFI23:Rad:M}{$-0.96$}
\setsymbol{LFI:slope:LFI24:Rad:M}{$-0.83$}
\setsymbol{LFI:slope:LFI25:Rad:M}{$-0.91$}
\setsymbol{LFI:slope:LFI26:Rad:M}{$-0.95$}
\setsymbol{LFI:slope:LFI27:Rad:M}{$-0.87$}
\setsymbol{LFI:slope:LFI28:Rad:M}{$-0.88$}
\setsymbol{LFI:slope:LFI18:Rad:S}{$-1.15$}
\setsymbol{LFI:slope:LFI19:Rad:S}{$-1.00$}
\setsymbol{LFI:slope:LFI20:Rad:S}{$-0.95$}
\setsymbol{LFI:slope:LFI21:Rad:S}{$-0.92$}
\setsymbol{LFI:slope:LFI22:Rad:S}{$-1.01$}
\setsymbol{LFI:slope:LFI23:Rad:S}{$-0.95$}
\setsymbol{LFI:slope:LFI24:Rad:S}{$-0.73$}
\setsymbol{LFI:slope:LFI25:Rad:S}{$-1.16$}
\setsymbol{LFI:slope:LFI26:Rad:S}{$-0.79$}
\setsymbol{LFI:slope:LFI27:Rad:S}{$-0.82$}
\setsymbol{LFI:slope:LFI28:Rad:S}{$-0.91$}

\setsymbol{LFI:FWHM:70GHz:units}{13\parcm01}
\setsymbol{LFI:FWHM:44GHz:units}{27\parcm92}
\setsymbol{LFI:FWHM:30GHz:units}{32\parcm65}

\setsymbol{LFI:FWHM:70GHz}{13.01}
\setsymbol{LFI:FWHM:44GHz}{27.92}
\setsymbol{LFI:FWHM:30GHz}{32.65}

\setsymbol{LFI:FWHM:LFI18:units}{13\parcm39}
\setsymbol{LFI:FWHM:LFI19:units}{13\parcm01}
\setsymbol{LFI:FWHM:LFI20:units}{12\parcm75}
\setsymbol{LFI:FWHM:LFI21:units}{12\parcm74}
\setsymbol{LFI:FWHM:LFI22:units}{12\parcm87}
\setsymbol{LFI:FWHM:LFI23:units}{13\parcm27}
\setsymbol{LFI:FWHM:LFI24:units}{22\parcm98}
\setsymbol{LFI:FWHM:LFI25:units}{30\parcm46}
\setsymbol{LFI:FWHM:LFI26:units}{30\parcm31}
\setsymbol{LFI:FWHM:LFI27:units}{32\parcm65}
\setsymbol{LFI:FWHM:LFI28:units}{32\parcm66}

\setsymbol{LFI:FWHM:LFI18}{13.39}
\setsymbol{LFI:FWHM:LFI19}{13.01}
\setsymbol{LFI:FWHM:LFI20}{12.75}
\setsymbol{LFI:FWHM:LFI21}{12.74}
\setsymbol{LFI:FWHM:LFI22}{12.87}
\setsymbol{LFI:FWHM:LFI23}{13.27}
\setsymbol{LFI:FWHM:LFI24}{22.98}
\setsymbol{LFI:FWHM:LFI25}{30.46}
\setsymbol{LFI:FWHM:LFI26}{30.31}
\setsymbol{LFI:FWHM:LFI27}{32.65}
\setsymbol{LFI:FWHM:LFI28}{32.66}

\setsymbol{LFI:FWHM:uncertainty:LFI18:units}{0.170\arcm}
\setsymbol{LFI:FWHM:uncertainty:LFI19:units}{0.174\arcm}
\setsymbol{LFI:FWHM:uncertainty:LFI20:units}{0.170\arcm}
\setsymbol{LFI:FWHM:uncertainty:LFI21:units}{0.156\arcm}
\setsymbol{LFI:FWHM:uncertainty:LFI22:units}{0.164\arcm}
\setsymbol{LFI:FWHM:uncertainty:LFI23:units}{0.171\arcm}
\setsymbol{LFI:FWHM:uncertainty:LFI24:units}{0.652\arcm}
\setsymbol{LFI:FWHM:uncertainty:LFI25:units}{1.075\arcm}
\setsymbol{LFI:FWHM:uncertainty:LFI26:units}{1.131\arcm}
\setsymbol{LFI:FWHM:uncertainty:LFI27:units}{1.266\arcm}
\setsymbol{LFI:FWHM:uncertainty:LFI28:units}{1.287\arcm}

\setsymbol{LFI:FWHM:uncertainty:LFI18}{0.170}
\setsymbol{LFI:FWHM:uncertainty:LFI19}{0.174}
\setsymbol{LFI:FWHM:uncertainty:LFI20}{0.170}
\setsymbol{LFI:FWHM:uncertainty:LFI21}{0.156}
\setsymbol{LFI:FWHM:uncertainty:LFI22}{0.164}
\setsymbol{LFI:FWHM:uncertainty:LFI23}{0.171}
\setsymbol{LFI:FWHM:uncertainty:LFI24}{0.652}
\setsymbol{LFI:FWHM:uncertainty:LFI25}{1.075}
\setsymbol{LFI:FWHM:uncertainty:LFI26}{1.131}
\setsymbol{LFI:FWHM:uncertainty:LFI27}{1.266}
\setsymbol{LFI:FWHM:uncertainty:LFI28}{1.287}

\setsymbol{HFI:center:frequency:100GHz:units}{100\,GHz}
\setsymbol{HFI:center:frequency:143GHz:units}{143\,GHz}
\setsymbol{HFI:center:frequency:217GHz:units}{217\,GHz}
\setsymbol{HFI:center:frequency:353GHz:units}{353\,GHz}
\setsymbol{HFI:center:frequency:545GHz:units}{545\,GHz}
\setsymbol{HFI:center:frequency:857GHz:units}{857\,GHz}

\setsymbol{HFI:center:frequency:100GHz}{100}
\setsymbol{HFI:center:frequency:143GHz}{143}
\setsymbol{HFI:center:frequency:217GHz}{217}
\setsymbol{HFI:center:frequency:353GHz}{353}
\setsymbol{HFI:center:frequency:545GHz}{545}
\setsymbol{HFI:center:frequency:857GHz}{857}

\setsymbol{HFI:Ndetectors:100GHz}{8}
\setsymbol{HFI:Ndetectors:143GHz}{11}
\setsymbol{HFI:Ndetectors:217GHz}{12}
\setsymbol{HFI:Ndetectors:353GHz}{12}
\setsymbol{HFI:Ndetectors:545GHz}{3}
\setsymbol{HFI:Ndetectors:857GHz}{4}

\setsymbol{HFI:FWHM:Maps:100GHz:units}{9\parcm88}
\setsymbol{HFI:FWHM:Maps:143GHz:units}{7\parcm18}
\setsymbol{HFI:FWHM:Maps:217GHz:units}{4\parcm87}
\setsymbol{HFI:FWHM:Maps:353GHz:units}{4\parcm65}
\setsymbol{HFI:FWHM:Maps:545GHz:units}{4\parcm72}
\setsymbol{HFI:FWHM:Maps:857GHz:units}{4\parcm39}
\setsymbol{HFI:FWHM:Maps:100GHz}{9.88}
\setsymbol{HFI:FWHM:Maps:143GHz}{7.18}
\setsymbol{HFI:FWHM:Maps:217GHz}{4.87}
\setsymbol{HFI:FWHM:Maps:353GHz}{4.65}
\setsymbol{HFI:FWHM:Maps:545GHz}{4.72}
\setsymbol{HFI:FWHM:Maps:857GHz}{4.39}

\setsymbol{HFI:beam:ellipticity:Maps:100GHz}{1.15}
\setsymbol{HFI:beam:ellipticity:Maps:143GHz}{1.01}
\setsymbol{HFI:beam:ellipticity:Maps:217GHz}{1.06}
\setsymbol{HFI:beam:ellipticity:Maps:353GHz}{1.05}
\setsymbol{HFI:beam:ellipticity:Maps:545GHz}{1.14}
\setsymbol{HFI:beam:ellipticity:Maps:857GHz}{1.19}

\setsymbol{HFI:FWHM:Mars:100GHz:units}{9\parcm37}
\setsymbol{HFI:FWHM:Mars:143GHz:units}{7\parcm04}
\setsymbol{HFI:FWHM:Mars:217GHz:units}{4\parcm68}
\setsymbol{HFI:FWHM:Mars:353GHz:units}{4\parcm43}
\setsymbol{HFI:FWHM:Mars:545GHz:units}{3\parcm80}
\setsymbol{HFI:FWHM:Mars:857GHz:units}{3\parcm67}

\setsymbol{HFI:FWHM:Mars:100GHz}{9.37}
\setsymbol{HFI:FWHM:Mars:143GHz}{7.04}
\setsymbol{HFI:FWHM:Mars:217GHz}{4.68}
\setsymbol{HFI:FWHM:Mars:353GHz}{4.43}
\setsymbol{HFI:FWHM:Mars:545GHz}{3.80}
\setsymbol{HFI:FWHM:Mars:857GHz}{3.67}

\setsymbol{HFI:beam:ellipticity:Mars:100GHz}{1.18}
\setsymbol{HFI:beam:ellipticity:Mars:143GHz}{1.03}
\setsymbol{HFI:beam:ellipticity:Mars:217GHz}{1.14}
\setsymbol{HFI:beam:ellipticity:Mars:353GHz}{1.09}
\setsymbol{HFI:beam:ellipticity:Mars:545GHz}{1.25}
\setsymbol{HFI:beam:ellipticity:Mars:857GHz}{1.03}

\setsymbol{HFI:CMB:relative:calibration:100GHz}{$\lsim 1\%$}
\setsymbol{HFI:CMB:relative:calibration:143GHz}{$\lsim 1\%$}
\setsymbol{HFI:CMB:relative:calibration:217GHz}{$\lsim 1\%$}
\setsymbol{HFI:CMB:relative:calibration:353GHz}{$\lsim 1\%$}
\setsymbol{HFI:CMB:relative:calibration:545GHz}{}
\setsymbol{HFI:CMB:relative:calibration:857GHz}{}

\setsymbol{HFI:CMB:absolute:calibration:100GHz}{$\lsim 2\%$}
\setsymbol{HFI:CMB:absolute:calibration:143GHz}{$\lsim 2\%$}
\setsymbol{HFI:CMB:absolute:calibration:217GHz}{$\lsim 2\%$}
\setsymbol{HFI:CMB:absolute:calibration:353GHz}{$\lsim 2\%$}
\setsymbol{HFI:CMB:absolute:calibration:545GHz}{}
\setsymbol{HFI:CMB:absolute:calibration:857GHz}{}

\setsymbol{HFI:FIRAS:gain:calibration:accuracy:statistical:100GHz}{}
\setsymbol{HFI:FIRAS:gain:calibration:accuracy:statistical:143GHz}{}
\setsymbol{HFI:FIRAS:gain:calibration:accuracy:statistical:217GHz}{}
\setsymbol{HFI:FIRAS:gain:calibration:accuracy:statistical:353GHz}{2.5\%}
\setsymbol{HFI:FIRAS:gain:calibration:accuracy:statistical:545GHz}{1\%}
\setsymbol{HFI:FIRAS:gain:calibration:accuracy:statistical:857GHz}{0.5\%}

\setsymbol{HFI:FIRAS:gain:calibration:accuracy:systematic:100GHz}{}
\setsymbol{HFI:FIRAS:gain:calibration:accuracy:systematic:143GHz}{}
\setsymbol{HFI:FIRAS:gain:calibration:accuracy:systematic:217GHz}{}
\setsymbol{HFI:FIRAS:gain:calibration:accuracy:systematic:353GHz}{}
\setsymbol{HFI:FIRAS:gain:calibration:accuracy:systematic:545GHz}{7\%}
\setsymbol{HFI:FIRAS:gain:calibration:accuracy:systematic:857GHz}{7\%}

\setsymbol{HFI:FIRAS:zero:point:accuracy:100GHz:units}{0.8\MJysr}
\setsymbol{HFI:FIRAS:zero:point:accuracy:143GHz:units}{}
\setsymbol{HFI:FIRAS:zero:point:accuracy:217GHz:units}{}
\setsymbol{HFI:FIRAS:zero:point:accuracy:353GHz:units}{1.4\MJysr}
\setsymbol{HFI:FIRAS:zero:point:accuracy:545GHz:units}{2.2\MJysr}
\setsymbol{HFI:FIRAS:zero:point:accuracy:857GHz:units}{1.7\MJysr}

\setsymbol{HFI:FIRAS:zero:point:accuracy:100GHz}{0.8}
\setsymbol{HFI:FIRAS:zero:point:accuracy:143GHz}{}
\setsymbol{HFI:FIRAS:zero:point:accuracy:217GHz}{}
\setsymbol{HFI:FIRAS:zero:point:accuracy:353GHz}{1.4}
\setsymbol{HFI:FIRAS:zero:point:accuracy:545GHz}{2.2}
\setsymbol{HFI:FIRAS:zero:point:accuracy:857GHz}{1.7}

\setsymbol{HFI:unit:conversion:100GHz:units}{0.2415\MJysrmK}
\setsymbol{HFI:unit:conversion:143GHz:units}{0.3694\MJysrmK}
\setsymbol{HFI:unit:conversion:217GHz:units}{0.4811\MJysrmK}
\setsymbol{HFI:unit:conversion:353GHz:units}{0.2883\MJysrmK}
\setsymbol{HFI:unit:conversion:545GHz:units}{0.05826\MJysrmK}
\setsymbol{HFI:unit:conversion:857GHz:units}{0.002238\MJysrmK}

\setsymbol{HFI:unit:conversion:100GHz}{0.2415}
\setsymbol{HFI:unit:conversion:143GHz}{0.3694}
\setsymbol{HFI:unit:conversion:217GHz}{0.4811}
\setsymbol{HFI:unit:conversion:353GHz}{0.2883}
\setsymbol{HFI:unit:conversion:545GHz}{0.05826}
\setsymbol{HFI:unit:conversion:857GHz}{0.002238}

\setsymbol{HFI:colour:correction:alpha=-2:V1.01:100GHz}{0.9893}
\setsymbol{HFI:colour:correction:alpha=-2:V1.01:143GHz}{0.9759}
\setsymbol{HFI:colour:correction:alpha=-2:V1.01:217GHz}{1.0007}
\setsymbol{HFI:colour:correction:alpha=-2:V1.01:353GHz}{1.0028}
\setsymbol{HFI:colour:correction:alpha=-2:V1.01:545GHz}{1.0019}
\setsymbol{HFI:colour:correction:alpha=-2:V1.01:857GHz}{0.9889}

\setsymbol{HFI:colour:correction:alpha=0:V1.01:100GHz}{1.0008}
\setsymbol{HFI:colour:correction:alpha=0:V1.01:143GHz}{1.0148}
\setsymbol{HFI:colour:correction:alpha=0:V1.01:217GHz}{0.9909}
\setsymbol{HFI:colour:correction:alpha=0:V1.01:353GHz}{0.9888}
\setsymbol{HFI:colour:correction:alpha=0:V1.01:545GHz}{0.9878}
\setsymbol{HFI:colour:correction:alpha=0:V1.01:857GHz}{1.0014}

\providecommand{\sorthelp}[1]{}

%% file: Planck_updated_lfi_values.tex
\def\setsymbol#1#2{\expandafter\def\csname #1\endcsname{#2}}
\def\getsymbol#1{\csname #1\endcsname}

\def\Planck{{\it Planck\/}}

\def\allearlypapers{\nocite{planck2011-1.1, planck2011-1.3, planck2011-1.4, planck2011-1.5, planck2011-1.6, planck2011-1.7, planck2011-1.10, planck2011-1.10sup, planck2011-5.1a, planck2011-5.1b, planck2011-5.2a, planck2011-5.2b, planck2011-5.2c, planck2011-6.1, planck2011-6.2, planck2011-6.3a, planck2011-6.4a, planck2011-6.4b, planck2011-6.6, planck2011-7.0, planck2011-7.2, planck2011-7.3, planck2011-7.7a, planck2011-7.7b, planck2011-7.12, planck2011-7.13}}

\newbox\tablebox    \newdimen\tablewidth
\def\leaderfil{\leaders\hbox to 5pt{\hss.\hss}\hfil}
\def\endPlancktable{\tablewidth=\columnwidth 
    $$\hss\copy\tablebox\hss$$
    \vskip-\lastskip\vskip -2pt}
\def\endPlancktablewide{\tablewidth=\textwidth 
    $$\hss\copy\tablebox\hss$$
    \vskip-\lastskip\vskip -2pt}
\def\tablenote#1 #2\par{\begingroup \parindent=0.8em
    \abovedisplayshortskip=0pt\belowdisplayshortskip=0pt
    \noindent
    $$\hss\vbox{\hsize\tablewidth \hangindent=\parindent \hangafter=1 \noindent
    \hbox to \parindent{\sup{\rm #1}\hss}\strut#2\strut\par}\hss$$
    \endgroup}
\def\doubleline{\vskip 3pt\hrule \vskip 1.5pt \hrule \vskip 5pt}

\def\L2{\ifmmode L_2\else $L_2$\fi}

\def\DeltaT{\ifmmode \Delta T\else $\Delta T$\fi}
\def\deltat{\ifmmode \Delta t\else $\Delta t$\fi}
\def\fknee{\ifmmode f_{\rm knee}\else $f_{\rm knee}$\fi}
\def\Fmax{\ifmmode F_{\rm max}\else $F_{\rm max}$\fi}
\def\solar{\ifmmode{\rm M}_{\mathord\odot}\else${\rm M}_{\mathord\odot}$\fi}

\def\inv{\ifmmode^{-1}\else$^{-1}$\fi}
\def\mo{\ifmmode^{-1}\else$^{-1}$\fi}
\def\sup#1{\ifmmode ^{\rm #1}\else $^{\rm #1}$\fi}
\def\expo#1{\ifmmode \times 10^{#1}\else $\times 10^{#1}$\fi}
\def\,{\thinspace}
\def\lsim{\mathrel{\raise .4ex\hbox{\rlap{$<$}\lower 1.2ex\hbox{$\sim$}}}}
\def\gsim{\mathrel{\raise .4ex\hbox{\rlap{$>$}\lower 1.2ex\hbox{$\sim$}}}}

\def\simprop{\mathrel{\raise .4ex\hbox{\rlap{$\propto$}\lower 1.2ex\hbox{$\sim$}}}}
\def\deg{\ifmmode^\circ\else$^\circ$\fi}
\def\pdeg{\ifmmode $\setbox0=\hbox{$^{\circ}$}\rlap{\hskip.11\wd0 .}$^{\circ}
          \else \setbox0=\hbox{$^{\circ}$}\rlap{\hskip.11\wd0 .}$^{\circ}$\fi}
\def\arcs{\ifmmode {^{\scriptstyle\prime\prime}}
          \else $^{\scriptstyle\prime\prime}$\fi}
\def\arcm{\ifmmode {^{\scriptstyle\prime}}
          \else $^{\scriptstyle\prime}$\fi}
\newdimen\sa  \newdimen\sb
\def\parcs{\sa=.07em \sb=.03em
     \ifmmode \hbox{\rlap{.}}^{\scriptstyle\prime\kern -\sb\prime}\hbox{\kern -\sa}
     \else \rlap{.}$^{\scriptstyle\prime\kern -\sb\prime}$\kern -\sa\fi}
\def\parcm{\sa=.08em \sb=.03em
     \ifmmode \hbox{\rlap{.}\kern\sa}^{\scriptstyle\prime}\hbox{\kern-\sb}
     \else \rlap{.}\kern\sa$^{\scriptstyle\prime}$\kern-\sb\fi}
\def\ra[#1 #2 #3.#4]{#1\sup{h}#2\sup{m}#3\sup{s}\llap.#4}
\def\dec[#1 #2 #3.#4]{#1\deg#2\arcm#3\arcs\llap.#4}
\def\deco[#1 #2 #3]{#1\deg#2\arcm#3\arcs}
\def\rra[#1 #2]{#1\sup{h}#2\sup{m}}

\def\dots{\relax\ifmmode \ldots\else $\ldots$\fi}
\def\WHzsr{\ifmmode $W\,Hz\mo\,sr\mo$\else W\,Hz\mo\,sr\mo\fi}
\def\mHz{\ifmmode $\,mHz$\else \,mHz\fi}
\def\GHz{\ifmmode $\,GHz$\else \,GHz\fi}
\def\mKs{\ifmmode $\,mK\,s$^{1/2}\else \,mK\,s$^{1/2}$\fi}
\def\muKs{\ifmmode \,\mu$K\,s$^{1/2}\else \,$\mu$K\,s$^{1/2}$\fi}
\def\muKRJs{\ifmmode \,\mu$K$_{\rm RJ}$\,s$^{1/2}\else \,$\mu$K$_{\rm RJ}$\,s$^{1/2}$\fi}
\def\muKCMBs{\ifmmode \,\mu$K$_{\rm CMB}$\,s$^{1/2}\else \,$\mu$K$_{\rm CMB}$\,s$^{1/2}$\fi} 
\def\muKHz{\ifmmode \,\mu$K\,Hz$^{-1/2}\else \,$\mu$K\,Hz$^{-1/2}$\fi}
\def\MJysr{\ifmmode \,$MJy\,sr\mo$\else \,MJy\,sr\mo\fi}
\def\MJysrmK{\ifmmode \,$MJy\,sr\mo$\,mK$_{\rm CMB}\mo\else \,MJy\,sr\mo\,mK$_{\rm CMB}\mo$\fi}
\def\microns{\ifmmode \,\mu$m$\else \,$\mu$m\fi}

\def\muK{\ifmmode \,\mu$K$\else \,$\mu$\hbox{K}\fi}
\def\muKRJ{\ifmmode \,\mu$K$_{\rm RJ}$\else \,$\mu$\hbox{K}$_{\rm RJ}$\fi}    
\def\muKCMB{\ifmmode \,\mu$K$_{\rm CMB}$\else \,$\mu$\hbox{K}$_{\rm CMB}$\fi} 
\def\microK{\ifmmode \,\mu$K$\else \,$\mu$\hbox{K}\fi}
\def\muW{\ifmmode \,\mu$W$\else \,$\mu$\hbox{W}\fi}
\def\kms{\ifmmode $\,km\,s$^{-1}\else \,km\,s$^{-1}$\fi}
\def\kmsMpc{\ifmmode $\,\kms\,Mpc\mo$\else \,\kms\,Mpc\mo\fi}
\setsymbol{LFI:center:frequency:70GHz:units}{$70.4$\,GHz}
\setsymbol{LFI:center:frequency:44GHz:units}{$44.1$\,GHz}
\setsymbol{LFI:center:frequency:30GHz:units}{$28.4$\,GHz}

\setsymbol{LFI:center:frequency:70GHz}{$70.4$}
\setsymbol{LFI:center:frequency:44GHz}{$44.1$}
\setsymbol{LFI:center:frequency:30GHz}{$28.4$}

\setsymbol{LFI:center:frequency:LFI18:Rad:M:units}{$71.7$\GHz}
\setsymbol{LFI:center:frequency:LFI19:Rad:M:units}{$67.5$\GHz}
\setsymbol{LFI:center:frequency:LFI20:Rad:M:units}{$69.2$\GHz}
\setsymbol{LFI:center:frequency:LFI21:Rad:M:units}{$70.4$\GHz}
\setsymbol{LFI:center:frequency:LFI22:Rad:M:units}{$71.5$\GHz}
\setsymbol{LFI:center:frequency:LFI23:Rad:M:units}{$70.8$\GHz}
\setsymbol{LFI:center:frequency:LFI24:Rad:M:units}{$44.4$\GHz}
\setsymbol{LFI:center:frequency:LFI25:Rad:M:units}{$44.0$\GHz}
\setsymbol{LFI:center:frequency:LFI26:Rad:M:units}{$43.9$\GHz}
\setsymbol{LFI:center:frequency:LFI27:Rad:M:units}{$28.3$\GHz}
\setsymbol{LFI:center:frequency:LFI28:Rad:M:units}{$28.8$\GHz}
\setsymbol{LFI:center:frequency:LFI18:Rad:S:units}{$70.1$\GHz}
\setsymbol{LFI:center:frequency:LFI19:Rad:S:units}{$69.6$\GHz}
\setsymbol{LFI:center:frequency:LFI20:Rad:S:units}{$69.5$\GHz}
\setsymbol{LFI:center:frequency:LFI21:Rad:S:units}{$69.5$\GHz}
\setsymbol{LFI:center:frequency:LFI22:Rad:S:units}{$72.8$\GHz}
\setsymbol{LFI:center:frequency:LFI23:Rad:S:units}{$71.3$\GHz}
\setsymbol{LFI:center:frequency:LFI24:Rad:S:units}{$44.1$\GHz}
\setsymbol{LFI:center:frequency:LFI25:Rad:S:units}{$44.1$\GHz}
\setsymbol{LFI:center:frequency:LFI26:Rad:S:units}{$44.1$\GHz}
\setsymbol{LFI:center:frequency:LFI27:Rad:S:units}{$28.5$\GHz}
\setsymbol{LFI:center:frequency:LFI28:Rad:S:units}{$28.2$\GHz}

\setsymbol{LFI:center:frequency:LFI18:Rad:M}{$71.7$}
\setsymbol{LFI:center:frequency:LFI19:Rad:M}{$67.5$}
\setsymbol{LFI:center:frequency:LFI20:Rad:M}{$69.2$}
\setsymbol{LFI:center:frequency:LFI21:Rad:M}{$70.4$}
\setsymbol{LFI:center:frequency:LFI22:Rad:M}{$71.5$}
\setsymbol{LFI:center:frequency:LFI23:Rad:M}{$70.8$}
\setsymbol{LFI:center:frequency:LFI24:Rad:M}{$44.4$}
\setsymbol{LFI:center:frequency:LFI25:Rad:M}{$44.0$}
\setsymbol{LFI:center:frequency:LFI26:Rad:M}{$43.9$}
\setsymbol{LFI:center:frequency:LFI27:Rad:M}{$28.3$}
\setsymbol{LFI:center:frequency:LFI28:Rad:M}{$28.8$}
\setsymbol{LFI:center:frequency:LFI18:Rad:S}{$70.1$}
\setsymbol{LFI:center:frequency:LFI19:Rad:S}{$69.6$}
\setsymbol{LFI:center:frequency:LFI20:Rad:S}{$69.5$}
\setsymbol{LFI:center:frequency:LFI21:Rad:S}{$69.5$}
\setsymbol{LFI:center:frequency:LFI22:Rad:S}{$72.8$}
\setsymbol{LFI:center:frequency:LFI23:Rad:S}{$71.3$}
\setsymbol{LFI:center:frequency:LFI24:Rad:S}{$44.1$}
\setsymbol{LFI:center:frequency:LFI25:Rad:S}{$44.1$}
\setsymbol{LFI:center:frequency:LFI26:Rad:S}{$44.1$}
\setsymbol{LFI:center:frequency:LFI27:Rad:S}{$28.5$}
\setsymbol{LFI:center:frequency:LFI28:Rad:S}{$28.2$}

\setsymbol{LFI:white:noise:sensitivity:70GHz:units}{$151.9$\muKs}
\setsymbol{LFI:white:noise:sensitivity:44GHz:units}{$173.2$\muKs}
\setsymbol{LFI:white:noise:sensitivity:30GHz:units}{$148.5$\muKs}

\setsymbol{LFI:white:noise:sensitivity:70GHz}{$151.9$}
\setsymbol{LFI:white:noise:sensitivity:44GHz}{$173.2$}
\setsymbol{LFI:white:noise:sensitivity:30GHz}{$148.5$}

\setsymbol{LFI:white:noise:sensitivity:LFI18:Rad:M:units}{511.7\muKs}
\setsymbol{LFI:white:noise:sensitivity:LFI19:Rad:M:units}{579.8\muKs}
\setsymbol{LFI:white:noise:sensitivity:LFI20:Rad:M:units}{587.5\muKs}
\setsymbol{LFI:white:noise:sensitivity:LFI21:Rad:M:units}{451.6\muKs}
\setsymbol{LFI:white:noise:sensitivity:LFI22:Rad:M:units}{489.9\muKs}
\setsymbol{LFI:white:noise:sensitivity:LFI23:Rad:M:units}{503.4\muKs}
\setsymbol{LFI:white:noise:sensitivity:LFI24:Rad:M:units}{461.0\muKs}
\setsymbol{LFI:white:noise:sensitivity:LFI25:Rad:M:units}{413.5\muKs}
\setsymbol{LFI:white:noise:sensitivity:LFI26:Rad:M:units}{480.8\muKs}
\setsymbol{LFI:white:noise:sensitivity:LFI27:Rad:M:units}{282.2\muKs}
\setsymbol{LFI:white:noise:sensitivity:LFI28:Rad:M:units}{318.2\muKs}
\setsymbol{LFI:white:noise:sensitivity:LFI18:Rad:S:units}{466.3\muKs}
\setsymbol{LFI:white:noise:sensitivity:LFI19:Rad:S:units}{554.1\muKs}
\setsymbol{LFI:white:noise:sensitivity:LFI20:Rad:S:units}{619.7\muKs}
\setsymbol{LFI:white:noise:sensitivity:LFI21:Rad:S:units}{560.9\muKs}
\setsymbol{LFI:white:noise:sensitivity:LFI22:Rad:S:units}{531.0\muKs}
\setsymbol{LFI:white:noise:sensitivity:LFI23:Rad:S:units}{538.8\muKs}
\setsymbol{LFI:white:noise:sensitivity:LFI24:Rad:S:units}{398.2\muKs}
\setsymbol{LFI:white:noise:sensitivity:LFI25:Rad:S:units}{393.3\muKs}
\setsymbol{LFI:white:noise:sensitivity:LFI26:Rad:S:units}{419.1\muKs}
\setsymbol{LFI:white:noise:sensitivity:LFI27:Rad:S:units}{304.7\muKs}
\setsymbol{LFI:white:noise:sensitivity:LFI28:Rad:S:units}{286.8\muKs}

\setsymbol{LFI:white:noise:sensitivity:uncertainty:LFI18:Rad:M:units}{1.7\muKs} 
\setsymbol{LFI:white:noise:sensitivity:uncertainty:LFI19:Rad:M:units}{1.6\muKs} 
\setsymbol{LFI:white:noise:sensitivity:uncertainty:LFI20:Rad:M:units}{1.4\muKs} 
\setsymbol{LFI:white:noise:sensitivity:uncertainty:LFI21:Rad:M:units}{1.7\muKs} 
\setsymbol{LFI:white:noise:sensitivity:uncertainty:LFI22:Rad:M:units}{1.5\muKs} 
\setsymbol{LFI:white:noise:sensitivity:uncertainty:LFI23:Rad:M:units}{1.8\muKs} 
\setsymbol{LFI:white:noise:sensitivity:uncertainty:LFI24:Rad:M:units}{1.3\muKs} 
\setsymbol{LFI:white:noise:sensitivity:uncertainty:LFI25:Rad:M:units}{1.5\muKs} 
\setsymbol{LFI:white:noise:sensitivity:uncertainty:LFI26:Rad:M:units}{1.5\muKs} 
\setsymbol{LFI:white:noise:sensitivity:uncertainty:LFI27:Rad:M:units}{2.1\muKs} 
\setsymbol{LFI:white:noise:sensitivity:uncertainty:LFI28:Rad:M:units}{1.9\muKs} 
\setsymbol{LFI:white:noise:sensitivity:uncertainty:LFI18:Rad:S:units}{1.7\muKs} 
\setsymbol{LFI:white:noise:sensitivity:uncertainty:LFI19:Rad:S:units}{1.6\muKs} 
\setsymbol{LFI:white:noise:sensitivity:uncertainty:LFI20:Rad:S:units}{2.1\muKs} 
\setsymbol{LFI:white:noise:sensitivity:uncertainty:LFI21:Rad:S:units}{1.7\muKs} 
\setsymbol{LFI:white:noise:sensitivity:uncertainty:LFI22:Rad:S:units}{2.1\muKs} 
\setsymbol{LFI:white:noise:sensitivity:uncertainty:LFI23:Rad:S:units}{1.9\muKs} 
\setsymbol{LFI:white:noise:sensitivity:uncertainty:LFI24:Rad:S:units}{1.3\muKs} 
\setsymbol{LFI:white:noise:sensitivity:uncertainty:LFI25:Rad:S:units}{3.0\muKs} 
\setsymbol{LFI:white:noise:sensitivity:uncertainty:LFI26:Rad:S:units}{1.9\muKs} 
\setsymbol{LFI:white:noise:sensitivity:uncertainty:LFI27:Rad:S:units}{2.0\muKs} 
\setsymbol{LFI:white:noise:sensitivity:uncertainty:LFI28:Rad:S:units}{2.1\muKs} 

\setsymbol{LFI:white:noise:sensitivity:LFI18:Rad:M}{511.7}
\setsymbol{LFI:white:noise:sensitivity:LFI19:Rad:M}{579.8}
\setsymbol{LFI:white:noise:sensitivity:LFI20:Rad:M}{587.5}
\setsymbol{LFI:white:noise:sensitivity:LFI21:Rad:M}{451.6}
\setsymbol{LFI:white:noise:sensitivity:LFI22:Rad:M}{489.9}
\setsymbol{LFI:white:noise:sensitivity:LFI23:Rad:M}{503.4}
\setsymbol{LFI:white:noise:sensitivity:LFI24:Rad:M}{461.0}
\setsymbol{LFI:white:noise:sensitivity:LFI25:Rad:M}{413.5}
\setsymbol{LFI:white:noise:sensitivity:LFI26:Rad:M}{480.8}
\setsymbol{LFI:white:noise:sensitivity:LFI27:Rad:M}{282.2}
\setsymbol{LFI:white:noise:sensitivity:LFI28:Rad:M}{318.2}
\setsymbol{LFI:white:noise:sensitivity:LFI18:Rad:S}{466.3}
\setsymbol{LFI:white:noise:sensitivity:LFI19:Rad:S}{554.1}
\setsymbol{LFI:white:noise:sensitivity:LFI20:Rad:S}{619.7}
\setsymbol{LFI:white:noise:sensitivity:LFI21:Rad:S}{560.9}
\setsymbol{LFI:white:noise:sensitivity:LFI22:Rad:S}{531.0}
\setsymbol{LFI:white:noise:sensitivity:LFI23:Rad:S}{538.8}
\setsymbol{LFI:white:noise:sensitivity:LFI24:Rad:S}{398.2}
\setsymbol{LFI:white:noise:sensitivity:LFI25:Rad:S}{393.3}
\setsymbol{LFI:white:noise:sensitivity:LFI26:Rad:S}{419.1}
\setsymbol{LFI:white:noise:sensitivity:LFI27:Rad:S}{304.7}
\setsymbol{LFI:white:noise:sensitivity:LFI28:Rad:S}{286.8}

\setsymbol{LFI:white:noise:sensitivity:uncertainty:LFI18:Rad:M}{1.7} 
\setsymbol{LFI:white:noise:sensitivity:uncertainty:LFI19:Rad:M}{1.6} 
\setsymbol{LFI:white:noise:sensitivity:uncertainty:LFI20:Rad:M}{1.4} 
\setsymbol{LFI:white:noise:sensitivity:uncertainty:LFI21:Rad:M}{1.7} 
\setsymbol{LFI:white:noise:sensitivity:uncertainty:LFI22:Rad:M}{1.5} 
\setsymbol{LFI:white:noise:sensitivity:uncertainty:LFI23:Rad:M}{1.8} 
\setsymbol{LFI:white:noise:sensitivity:uncertainty:LFI24:Rad:M}{1.3} 
\setsymbol{LFI:white:noise:sensitivity:uncertainty:LFI25:Rad:M}{1.5} 
\setsymbol{LFI:white:noise:sensitivity:uncertainty:LFI26:Rad:M}{1.5} 
\setsymbol{LFI:white:noise:sensitivity:uncertainty:LFI27:Rad:M}{2.1} 
\setsymbol{LFI:white:noise:sensitivity:uncertainty:LFI28:Rad:M}{1.9} 
\setsymbol{LFI:white:noise:sensitivity:uncertainty:LFI18:Rad:S}{1.7} 
\setsymbol{LFI:white:noise:sensitivity:uncertainty:LFI19:Rad:S}{1.6} 
\setsymbol{LFI:white:noise:sensitivity:uncertainty:LFI20:Rad:S}{2.1} 
\setsymbol{LFI:white:noise:sensitivity:uncertainty:LFI21:Rad:S}{1.7} 
\setsymbol{LFI:white:noise:sensitivity:uncertainty:LFI22:Rad:S}{2.1} 
\setsymbol{LFI:white:noise:sensitivity:uncertainty:LFI23:Rad:S}{1.9} 
\setsymbol{LFI:white:noise:sensitivity:uncertainty:LFI24:Rad:S}{1.3} 
\setsymbol{LFI:white:noise:sensitivity:uncertainty:LFI25:Rad:S}{3.0} 
\setsymbol{LFI:white:noise:sensitivity:uncertainty:LFI26:Rad:S}{1.9} 
\setsymbol{LFI:white:noise:sensitivity:uncertainty:LFI27:Rad:S}{2.0} 
\setsymbol{LFI:white:noise:sensitivity:uncertainty:LFI28:Rad:S}{2.1}

\setsymbol{LFI:knee:frequency:70GHz:units}{$20.2$\mHz}
\setsymbol{LFI:knee:frequency:44GHz:units}{$45.7$\mHz}
\setsymbol{LFI:knee:frequency:30GHz:units}{$114.1$\mHz}

\setsymbol{LFI:knee:frequency:70GHz}{$20.2$}
\setsymbol{LFI:knee:frequency:44GHz}{$45.7$}
\setsymbol{LFI:knee:frequency:30GHz}{$114.1$}

\setsymbol{LFI:knee:frequency:LFI18:Rad:M:units}{15.3\mHz}
\setsymbol{LFI:knee:frequency:LFI19:Rad:M:units}{11.9\mHz}
\setsymbol{LFI:knee:frequency:LFI20:Rad:M:units}{8.4\mHz}
\setsymbol{LFI:knee:frequency:LFI21:Rad:M:units}{39.3\mHz}
\setsymbol{LFI:knee:frequency:LFI22:Rad:M:units}{10.1\mHz}
\setsymbol{LFI:knee:frequency:LFI23:Rad:M:units}{30.2\mHz}
\setsymbol{LFI:knee:frequency:LFI24:Rad:M:units}{26.9\mHz}
\setsymbol{LFI:knee:frequency:LFI25:Rad:M:units}{20.1\mHz}
\setsymbol{LFI:knee:frequency:LFI26:Rad:M:units}{64.4\mHz}
\setsymbol{LFI:knee:frequency:LFI27:Rad:M:units}{175.1\mHz}
\setsymbol{LFI:knee:frequency:LFI28:Rad:M:units}{127.9\mHz}
\setsymbol{LFI:knee:frequency:LFI18:Rad:S:units}{18.3\mHz}
\setsymbol{LFI:knee:frequency:LFI19:Rad:S:units}{14.6\mHz}
\setsymbol{LFI:knee:frequency:LFI20:Rad:S:units}{6.0\mHz}
\setsymbol{LFI:knee:frequency:LFI21:Rad:S:units}{14.0\mHz}
\setsymbol{LFI:knee:frequency:LFI22:Rad:S:units}{15.9\mHz}
\setsymbol{LFI:knee:frequency:LFI23:Rad:S:units}{58.8\mHz}
\setsymbol{LFI:knee:frequency:LFI24:Rad:S:units}{73.0\mHz}
\setsymbol{LFI:knee:frequency:LFI25:Rad:S:units}{46.1\mHz}
\setsymbol{LFI:knee:frequency:LFI26:Rad:S:units}{43.8\mHz}
\setsymbol{LFI:knee:frequency:LFI27:Rad:S:units}{109.6\mHz}
\setsymbol{LFI:knee:frequency:LFI28:Rad:S:units}{43.9\mHz}

\setsymbol{LFI:knee:frequency:uncertainty:LFI18:Rad:M:units}{2.8\mHz}
\setsymbol{LFI:knee:frequency:uncertainty:LFI19:Rad:M:units}{1.3\mHz}
\setsymbol{LFI:knee:frequency:uncertainty:LFI20:Rad:M:units}{1.9\mHz}
\setsymbol{LFI:knee:frequency:uncertainty:LFI21:Rad:M:units}{4.0\mHz}
\setsymbol{LFI:knee:frequency:uncertainty:LFI22:Rad:M:units}{2.1\mHz}
\setsymbol{LFI:knee:frequency:uncertainty:LFI23:Rad:M:units}{1.4\mHz}
\setsymbol{LFI:knee:frequency:uncertainty:LFI24:Rad:M:units}{1.2\mHz}
\setsymbol{LFI:knee:frequency:uncertainty:LFI25:Rad:M:units}{0.6\mHz}
\setsymbol{LFI:knee:frequency:uncertainty:LFI26:Rad:M:units}{2.0\mHz}
\setsymbol{LFI:knee:frequency:uncertainty:LFI27:Rad:M:units}{2.2\mHz}
\setsymbol{LFI:knee:frequency:uncertainty:LFI28:Rad:M:units}{3.8\mHz}
\setsymbol{LFI:knee:frequency:uncertainty:LFI18:Rad:S:units}{1.6\mHz}
\setsymbol{LFI:knee:frequency:uncertainty:LFI19:Rad:S:units}{1.1\mHz}
\setsymbol{LFI:knee:frequency:uncertainty:LFI20:Rad:S:units}{1.7\mHz}
\setsymbol{LFI:knee:frequency:uncertainty:LFI21:Rad:S:units}{2.4\mHz}
\setsymbol{LFI:knee:frequency:uncertainty:LFI22:Rad:S:units}{7.1\mHz}
\setsymbol{LFI:knee:frequency:uncertainty:LFI23:Rad:S:units}{9.0\mHz}
\setsymbol{LFI:knee:frequency:uncertainty:LFI24:Rad:S:units}{7.9\mHz}
\setsymbol{LFI:knee:frequency:uncertainty:LFI25:Rad:S:units}{1.8\mHz}
\setsymbol{LFI:knee:frequency:uncertainty:LFI26:Rad:S:units}{8.9\mHz}
\setsymbol{LFI:knee:frequency:uncertainty:LFI27:Rad:S:units}{2.3\mHz}
\setsymbol{LFI:knee:frequency:uncertainty:LFI28:Rad:S:units}{2.2\mHz}

\setsymbol{LFI:knee:frequency:LFI18:Rad:M}{15.3}
\setsymbol{LFI:knee:frequency:LFI19:Rad:M}{11.9}
\setsymbol{LFI:knee:frequency:LFI20:Rad:M}{8.4}
\setsymbol{LFI:knee:frequency:LFI21:Rad:M}{39.3}
\setsymbol{LFI:knee:frequency:LFI22:Rad:M}{10.1}
\setsymbol{LFI:knee:frequency:LFI23:Rad:M}{30.2}
\setsymbol{LFI:knee:frequency:LFI24:Rad:M}{26.9}
\setsymbol{LFI:knee:frequency:LFI25:Rad:M}{20.1}
\setsymbol{LFI:knee:frequency:LFI26:Rad:M}{64.4}
\setsymbol{LFI:knee:frequency:LFI27:Rad:M}{175.1}
\setsymbol{LFI:knee:frequency:LFI28:Rad:M}{127.9}
\setsymbol{LFI:knee:frequency:LFI18:Rad:S}{18.3}
\setsymbol{LFI:knee:frequency:LFI19:Rad:S}{14.6}
\setsymbol{LFI:knee:frequency:LFI20:Rad:S}{6.0}
\setsymbol{LFI:knee:frequency:LFI21:Rad:S}{14.0}
\setsymbol{LFI:knee:frequency:LFI22:Rad:S}{15.9}
\setsymbol{LFI:knee:frequency:LFI23:Rad:S}{58.8}
\setsymbol{LFI:knee:frequency:LFI24:Rad:S}{73.0}
\setsymbol{LFI:knee:frequency:LFI25:Rad:S}{46.1}
\setsymbol{LFI:knee:frequency:LFI26:Rad:S}{43.8}
\setsymbol{LFI:knee:frequency:LFI27:Rad:S}{109.6}
\setsymbol{LFI:knee:frequency:LFI28:Rad:S}{43.9}

\setsymbol{LFI:knee:frequency:uncertainty:LFI18:Rad:M}{2.8}
\setsymbol{LFI:knee:frequency:uncertainty:LFI19:Rad:M}{1.3}
\setsymbol{LFI:knee:frequency:uncertainty:LFI20:Rad:M}{1.9}
\setsymbol{LFI:knee:frequency:uncertainty:LFI21:Rad:M}{4.0}
\setsymbol{LFI:knee:frequency:uncertainty:LFI22:Rad:M}{2.1}
\setsymbol{LFI:knee:frequency:uncertainty:LFI23:Rad:M}{1.4}
\setsymbol{LFI:knee:frequency:uncertainty:LFI24:Rad:M}{1.2}
\setsymbol{LFI:knee:frequency:uncertainty:LFI25:Rad:M}{0.6}
\setsymbol{LFI:knee:frequency:uncertainty:LFI26:Rad:M}{2.0}
\setsymbol{LFI:knee:frequency:uncertainty:LFI27:Rad:M}{2.2}
\setsymbol{LFI:knee:frequency:uncertainty:LFI28:Rad:M}{3.8}
\setsymbol{LFI:knee:frequency:uncertainty:LFI18:Rad:S}{1.6}
\setsymbol{LFI:knee:frequency:uncertainty:LFI19:Rad:S}{1.1}
\setsymbol{LFI:knee:frequency:uncertainty:LFI20:Rad:S}{1.7}
\setsymbol{LFI:knee:frequency:uncertainty:LFI21:Rad:S}{2.4}
\setsymbol{LFI:knee:frequency:uncertainty:LFI22:Rad:S}{7.1}
\setsymbol{LFI:knee:frequency:uncertainty:LFI23:Rad:S}{9.0}
\setsymbol{LFI:knee:frequency:uncertainty:LFI24:Rad:S}{7.9}
\setsymbol{LFI:knee:frequency:uncertainty:LFI25:Rad:S}{1.8}
\setsymbol{LFI:knee:frequency:uncertainty:LFI26:Rad:S}{8.9}
\setsymbol{LFI:knee:frequency:uncertainty:LFI27:Rad:S}{2.3}
\setsymbol{LFI:knee:frequency:uncertainty:LFI28:Rad:S}{2.2}

\setsymbol{LFI:slope:70GHz:units}{$-1.23$\mHz}
\setsymbol{LFI:slope:44GHz:units}{$-0.90$\mHz}
\setsymbol{LFI:slope:30GHz:units}{$-0.92$\mHz}

\setsymbol{LFI:slope:70GHz}{$-1.23$}
\setsymbol{LFI:slope:44GHz}{$-0.90$}
\setsymbol{LFI:slope:30GHz}{$-0.92$}

\setsymbol{LFI:slope:LFI18:Rad:M}{$-1.07$}
\setsymbol{LFI:slope:LFI19:Rad:M}{$-1.22$}
\setsymbol{LFI:slope:LFI20:Rad:M}{$-1.31$}
\setsymbol{LFI:slope:LFI21:Rad:M}{$-1.26$}
\setsymbol{LFI:slope:LFI22:Rad:M}{$-1.53$}
\setsymbol{LFI:slope:LFI23:Rad:M}{$-1.07$}
\setsymbol{LFI:slope:LFI24:Rad:M}{$-0.94$}
\setsymbol{LFI:slope:LFI25:Rad:M}{$-0.85$}
\setsymbol{LFI:slope:LFI26:Rad:M}{$-0.92$}
\setsymbol{LFI:slope:LFI27:Rad:M}{$-0.93$}
\setsymbol{LFI:slope:LFI28:Rad:M}{$-0.93$}
\setsymbol{LFI:slope:LFI18:Rad:S}{$-1.20$}
\setsymbol{LFI:slope:LFI19:Rad:S}{$-1.12$}
\setsymbol{LFI:slope:LFI20:Rad:S}{$-1.34$}
\setsymbol{LFI:slope:LFI21:Rad:S}{$-1.24$}
\setsymbol{LFI:slope:LFI22:Rad:S}{$-1.20$}
\setsymbol{LFI:slope:LFI23:Rad:S}{$-1.21$}
\setsymbol{LFI:slope:LFI24:Rad:S}{$-0.91$}
\setsymbol{LFI:slope:LFI25:Rad:S}{$-0.90$}
\setsymbol{LFI:slope:LFI26:Rad:S}{$-0.88$}
\setsymbol{LFI:slope:LFI27:Rad:S}{$-0.91$}
\setsymbol{LFI:slope:LFI28:Rad:S}{$-0.91$}

\setsymbol{LFI:slope:uncertainty:LFI18:Rad:M}{$0.11$} 
\setsymbol{LFI:slope:uncertainty:LFI19:Rad:M}{$0.30$} 
\setsymbol{LFI:slope:uncertainty:LFI20:Rad:M}{$0.40$} 
\setsymbol{LFI:slope:uncertainty:LFI21:Rad:M}{$0.09$} 
\setsymbol{LFI:slope:uncertainty:LFI22:Rad:M}{$0.34$} 
\setsymbol{LFI:slope:uncertainty:LFI23:Rad:M}{$0.03$} 
\setsymbol{LFI:slope:uncertainty:LFI24:Rad:M}{$0.01$} 
\setsymbol{LFI:slope:uncertainty:LFI25:Rad:M}{$0.01$} 
\setsymbol{LFI:slope:uncertainty:LFI26:Rad:M}{$0.01$} 
\setsymbol{LFI:slope:uncertainty:LFI27:Rad:M}{$0.01$} 
\setsymbol{LFI:slope:uncertainty:LFI28:Rad:M}{$0.01$} 
\setsymbol{LFI:slope:uncertainty:LFI18:Rad:S}{$0.15$} 
\setsymbol{LFI:slope:uncertainty:LFI19:Rad:S}{$0.16$} 
\setsymbol{LFI:slope:uncertainty:LFI20:Rad:S}{$0.47$} 
\setsymbol{LFI:slope:uncertainty:LFI21:Rad:S}{$0.11$} 
\setsymbol{LFI:slope:uncertainty:LFI22:Rad:S}{$0.36$} 
\setsymbol{LFI:slope:uncertainty:LFI23:Rad:S}{$0.05$} 
\setsymbol{LFI:slope:uncertainty:LFI24:Rad:S}{$0.01$} 
\setsymbol{LFI:slope:uncertainty:LFI25:Rad:S}{$0.01$} 
\setsymbol{LFI:slope:uncertainty:LFI26:Rad:S}{$0.06$} 
\setsymbol{LFI:slope:uncertainty:LFI27:Rad:S}{$0.01$} 
\setsymbol{LFI:slope:uncertainty:LFI28:Rad:S}{$0.02$} 

\setsymbol{LFI:FWHM:70GHz:units}{13\parcm08}
\setsymbol{LFI:FWHM:44GHz:units}{28\parcm09}
\setsymbol{LFI:FWHM:30GHz:units}{33\parcm16}

\setsymbol{LFI:FWHM:70GHz}{$13.08$}
\setsymbol{LFI:FWHM:44GHz}{$28.09$}
\setsymbol{LFI:FWHM:30GHz}{$33.16$}

\setsymbol{LFI:FWHM:LFI18:units}{13\parcm44}
\setsymbol{LFI:FWHM:LFI19:units}{13\parcm11}
\setsymbol{LFI:FWHM:LFI20:units}{12\parcm84}
\setsymbol{LFI:FWHM:LFI21:units}{12\parcm81}
\setsymbol{LFI:FWHM:LFI22:units}{12\parcm95}
\setsymbol{LFI:FWHM:LFI23:units}{13\parcm33}
\setsymbol{LFI:FWHM:LFI24:units}{23\parcm17}
\setsymbol{LFI:FWHM:LFI25:units}{30\parcm60}
\setsymbol{LFI:FWHM:LFI26:units}{30\parcm49}
\setsymbol{LFI:FWHM:LFI27:units}{33\parcm09}
\setsymbol{LFI:FWHM:LFI28:units}{33\parcm23}

\setsymbol{LFI:FWHM:LFI18}{13.44}
\setsymbol{LFI:FWHM:LFI19}{13.11}
\setsymbol{LFI:FWHM:LFI20}{12.84}
\setsymbol{LFI:FWHM:LFI21}{12.81}
\setsymbol{LFI:FWHM:LFI22}{12.95}
\setsymbol{LFI:FWHM:LFI23}{13.33}
\setsymbol{LFI:FWHM:LFI24}{23.17}
\setsymbol{LFI:FWHM:LFI25}{30.60}
\setsymbol{LFI:FWHM:LFI26}{30.49}
\setsymbol{LFI:FWHM:LFI27}{33.09}
\setsymbol{LFI:FWHM:LFI28}{33.23}

\setsymbol{LFI:effFWHM:70GHz:units}{13\parcm25}
\setsymbol{LFI:effFWHM:44GHz:units}{27\parcm01}
\setsymbol{LFI:effFWHM:30GHz:units}{32\parcm24}

\setsymbol{LFI:effFWHM:70GHz}{$13.25$}
\setsymbol{LFI:effFWHM:44GHz}{$27.01$}
\setsymbol{LFI:effFWHM:30GHz}{$32.24$}

\setsymbol{LFI:FWHM:uncertainty:70GHz}{0.040\arcm}
\setsymbol{LFI:FWHM:uncertainty:44GHz}{0.097\arcm}
\setsymbol{LFI:FWHM:uncertainty:30GHz}{0.110\arcm}

\setsymbol{LFI:FWHM:uncertainty:LFI18:units}{0.03\arcm}
\setsymbol{LFI:FWHM:uncertainty:LFI19:units}{0.04\arcm}
\setsymbol{LFI:FWHM:uncertainty:LFI20:units}{0.04\arcm}
\setsymbol{LFI:FWHM:uncertainty:LFI21:units}{0.03\arcm}
\setsymbol{LFI:FWHM:uncertainty:LFI22:units}{0.03\arcm}
\setsymbol{LFI:FWHM:uncertainty:LFI23:units}{0.04\arcm}
\setsymbol{LFI:FWHM:uncertainty:LFI24:units}{0.07\arcm}
\setsymbol{LFI:FWHM:uncertainty:LFI25:units}{0.10\arcm}
\setsymbol{LFI:FWHM:uncertainty:LFI26:units}{0.12\arcm}
\setsymbol{LFI:FWHM:uncertainty:LFI27:units}{0.11\arcm}
\setsymbol{LFI:FWHM:uncertainty:LFI28:units}{0.11\arcm}

\setsymbol{LFI:FWHM:uncertainty:LFI18}{0.03} 
\setsymbol{LFI:FWHM:uncertainty:LFI19}{0.04}
\setsymbol{LFI:FWHM:uncertainty:LFI20}{0.04}
\setsymbol{LFI:FWHM:uncertainty:LFI21}{0.03}
\setsymbol{LFI:FWHM:uncertainty:LFI22}{0.03}
\setsymbol{LFI:FWHM:uncertainty:LFI23}{0.04}
\setsymbol{LFI:FWHM:uncertainty:LFI24}{0.07}
\setsymbol{LFI:FWHM:uncertainty:LFI25}{0.10}
\setsymbol{LFI:FWHM:uncertainty:LFI26}{0.12}
\setsymbol{LFI:FWHM:uncertainty:LFI27}{0.11}
\setsymbol{LFI:FWHM:uncertainty:LFI28}{0.11}

\setsymbol{LFI:ellipticity:70GHz}{$1.27$}
\setsymbol{LFI:ellipticity:44GHz}{$1.25$}
\setsymbol{LFI:ellipticity:30GHz}{$1.37$}

\setsymbol{LFI:ellipticity:LFI18}{1.26}
\setsymbol{LFI:ellipticity:LFI19}{1.27}
\setsymbol{LFI:ellipticity:LFI20}{1.28}
\setsymbol{LFI:ellipticity:LFI21}{1.29}
\setsymbol{LFI:ellipticity:LFI22}{1.28}
\setsymbol{LFI:ellipticity:LFI23}{1.26}
\setsymbol{LFI:ellipticity:LFI24}{1.37}
\setsymbol{LFI:ellipticity:LFI25}{1.19}
\setsymbol{LFI:ellipticity:LFI26}{1.20}
\setsymbol{LFI:ellipticity:LFI27}{1.38}
\setsymbol{LFI:ellipticity:LFI28}{1.37}

\setsymbol{LFI:effellipticity:70GHz}{1.22}
\setsymbol{LFI:effellipticity:44GHz}{1.03}
\setsymbol{LFI:effellipticity:30GHz}{1.32}

\setsymbol{LFI:ellipticity:uncertainty:70GHz}{0.01}
\setsymbol{LFI:ellipticity:uncertainty:44GHz}{0.01}
\setsymbol{LFI:ellipticity:uncertainty:30GHz}{0.01}

\setsymbol{LFI:ellipticity:uncertainty:LFI18}{0.01}
\setsymbol{LFI:ellipticity:uncertainty:LFI19}{0.01}
\setsymbol{LFI:ellipticity:uncertainty:LFI20}{0.01}
\setsymbol{LFI:ellipticity:uncertainty:LFI21}{0.01}
\setsymbol{LFI:ellipticity:uncertainty:LFI22}{0.01}
\setsymbol{LFI:ellipticity:uncertainty:LFI23}{0.01}
\setsymbol{LFI:ellipticity:uncertainty:LFI24}{0.01}
\setsymbol{LFI:ellipticity:uncertainty:LFI25}{0.01}
\setsymbol{LFI:ellipticity:uncertainty:LFI26}{0.01}
\setsymbol{LFI:ellipticity:uncertainty:LFI27}{0.01}
\setsymbol{LFI:ellipticity:uncertainty:LFI28}{0.01}

\setsymbol{LFI:absolute:calibration:uncertainty:70GHz:units}{0.25\%}
\setsymbol{LFI:absolute:calibration:uncertainty:44GHz:units}{0.25\%}
\setsymbol{LFI:absolute:calibration:uncertainty:30GHz:units}{0.25\%}

\setsymbol{LFI:absolute:calibration:uncertainty:70GHz}{0.25}
\setsymbol{LFI:absolute:calibration:uncertainty:44GHz}{0.25}
\setsymbol{LFI:absolute:calibration:uncertainty:30GHz}{0.25}

\setsymbol{LFI:relative:calibration:uncertainty:70GHz:units}{0.1\%}
\setsymbol{LFI:relative:calibration:uncertainty:44GHz:units}{0.2\%}
\setsymbol{LFI:relative:calibration:uncertainty:30GHz:units}{0.2\%}

\setsymbol{LFI:relative:calibration:uncertainty:70GHz}{0.1}
\setsymbol{LFI:relative:calibration:uncertainty:44GHz}{0.2}
\setsymbol{LFI:relative:calibration:uncertainty:30GHz}{0.2}

\setsymbol{LFI:systematic:effects:rms:uncertainty:70GHz:units}{2.00\muKCMB}
\setsymbol{LFI:systematic:effects:rms:uncertainty:44GHz:units}{1.13\muKCMB}
\setsymbol{LFI:systematic:effects:rms:uncertainty:30GHz:units}{4.83\muKCMB}

\setsymbol{LFI:systematic:effects:rms:uncertainty:70GHz}{2.00}
\setsymbol{LFI:systematic:effects:rms:uncertainty:44GHz}{1.13}
\setsymbol{LFI:systematic:effects:rms:uncertainty:30GHz}{4.83}

\setsymbol{LFI:systematic:effects:p2p:uncertainty:70GHz:units}{7.87\muKCMB}
\setsymbol{LFI:systematic:effects:p2p:uncertainty:44GHz:units}{5.61\muKCMB}
\setsymbol{LFI:systematic:effects:p2p:uncertainty:30GHz:units}{21.02\muKCMB}

\setsymbol{LFI:systematic:effects:p2p:uncertainty:70GHz}{7.87}
\setsymbol{LFI:systematic:effects:p2p:uncertainty:44GHz}{5.61}
\setsymbol{LFI:systematic:effects:p2p:uncertainty:30GHz}{21.02}

\setsymbol{LFI:systematic:effects:spikes:rms:uncertainty:70GHz:units}{0.30\muKCMB} 
\setsymbol{LFI:systematic:effects:spikes:rms:uncertainty:44GHz:units}{0.15\muKCMB} 
\setsymbol{LFI:systematic:effects:spikes:rms:uncertainty:30GHz:units}{0.45\muKCMB} 

\setsymbol{LFI:systematic:effects:spikes:rms:uncertainty:70GHz}{0.30} 
\setsymbol{LFI:systematic:effects:spikes:rms:uncertainty:44GHz}{0.15} 
\setsymbol{LFI:systematic:effects:spikes:rms:uncertainty:30GHz}{0.45} 

\setsymbol{LFI:systematic:effects:spikes:p2p:uncertainty:70GHz:units}{2.56\muKCMB} 
\setsymbol{LFI:systematic:effects:spikes:p2p:uncertainty:44GHz:units}{1.51\muKCMB} 
\setsymbol{LFI:systematic:effects:spikes:p2p:uncertainty:30GHz:units}{4.00\muKCMB} 

\setsymbol{LFI:systematic:effects:spikes:p2p:uncertainty:70GHz}{2.56} 
\setsymbol{LFI:systematic:effects:spikes:p2p:uncertainty:44GHz}{1.51} 
\setsymbol{LFI:systematic:effects:spikes:p2p:uncertainty:30GHz}{4.00} 

\setsymbol{LFI:systematic:effects:feu:rms:uncertainty:70GHz:units}{0.21\muKCMB} 
\setsymbol{LFI:systematic:effects:feu:rms:uncertainty:44GHz:units}{0.22\muKCMB} 
\setsymbol{LFI:systematic:effects:feu:rms:uncertainty:30GHz:units}{0.23\muKCMB} 

\setsymbol{LFI:systematic:effects:feu:rms:uncertainty:70GHz}{0.21} 
\setsymbol{LFI:systematic:effects:feu:rms:uncertainty:44GHz}{0.22} 
\setsymbol{LFI:systematic:effects:feu:rms:uncertainty:30GHz}{0.23} 

\setsymbol{LFI:systematic:effects:feu:p2p:uncertainty:70GHz:units}{1.12\muKCMB} 
\setsymbol{LFI:systematic:effects:feu:p2p:uncertainty:44GHz:units}{1.15\muKCMB} 
\setsymbol{LFI:systematic:effects:feu:p2p:uncertainty:30GHz:units}{1.05\muKCMB} 

\setsymbol{LFI:systematic:effects:feu:p2p:uncertainty:70GHz}{1.12} 
\setsymbol{LFI:systematic:effects:feu:p2p:uncertainty:44GHz}{1.15} 
\setsymbol{LFI:systematic:effects:feu:p2p:uncertainty:30GHz}{1.05} 

\setsymbol{LFI:systematic:effects:beu:rms:uncertainty:70GHz:units}{0.24\muKCMB} 
\setsymbol{LFI:systematic:effects:beu:rms:uncertainty:44GHz:units}{0.05\muKCMB} 
\setsymbol{LFI:systematic:effects:beu:rms:uncertainty:30GHz:units}{0.11\muKCMB} 

\setsymbol{LFI:systematic:effects:beu:rms:uncertainty:70GHz}{0.24} 
\setsymbol{LFI:systematic:effects:beu:rms:uncertainty:44GHz}{0.05} 
\setsymbol{LFI:systematic:effects:beu:rms:uncertainty:30GHz}{0.11} 

\setsymbol{LFI:systematic:effects:beu:p2p:uncertainty:70GHz:units}{2.70\muKCMB} 
\setsymbol{LFI:systematic:effects:beu:p2p:uncertainty:44GHz:units}{0.63\muKCMB} 
\setsymbol{LFI:systematic:effects:beu:p2p:uncertainty:30GHz:units}{1.27\muKCMB} 

\setsymbol{LFI:systematic:effects:beu:p2p:uncertainty:70GHz}{2.70} 
\setsymbol{LFI:systematic:effects:beu:p2p:uncertainty:44GHz}{0.63} 
\setsymbol{LFI:systematic:effects:beu:p2p:uncertainty:30GHz}{1.27} 

\setsymbol{LFI:systematic:effects:4k:rms:uncertainty:70GHz:units}{0.16\muKCMB} 
\setsymbol{LFI:systematic:effects:4k:rms:uncertainty:44GHz:units}{0.98\muKCMB} 
\setsymbol{LFI:systematic:effects:4k:rms:uncertainty:30GHz:units}{0.98\muKCMB} 

\setsymbol{LFI:systematic:effects:4k:rms:uncertainty:70GHz}{0.16} 
\setsymbol{LFI:systematic:effects:4k:rms:uncertainty:44GHz}{0.98} 
\setsymbol{LFI:systematic:effects:4k:rms:uncertainty:30GHz}{0.98} 

\setsymbol{LFI:systematic:effects:4k:p2p:uncertainty:70GHz:units}{1.30\muKCMB} 
\setsymbol{LFI:systematic:effects:4k:p2p:uncertainty:44GHz:units}{9.73\muKCMB} 
\setsymbol{LFI:systematic:effects:4k:p2p:uncertainty:30GHz:units}{9.76\muKCMB} 

\setsymbol{LFI:systematic:effects:4k:p2p:uncertainty:70GHz}{1.30} 
\setsymbol{LFI:systematic:effects:4k:p2p:uncertainty:44GHz}{9.73} 
\setsymbol{LFI:systematic:effects:4k:p2p:uncertainty:30GHz}{9.76} 

\setsymbol{HFI:center:frequency:100GHz:units}{100\,GHz}
\setsymbol{HFI:center:frequency:143GHz:units}{143\,GHz}
\setsymbol{HFI:center:frequency:217GHz:units}{217\,GHz}
\setsymbol{HFI:center:frequency:353GHz:units}{353\,GHz}
\setsymbol{HFI:center:frequency:545GHz:units}{545\,GHz}
\setsymbol{HFI:center:frequency:857GHz:units}{857\,GHz}

\setsymbol{HFI:center:frequency:100GHz}{100}
\setsymbol{HFI:center:frequency:143GHz}{143}
\setsymbol{HFI:center:frequency:217GHz}{217}
\setsymbol{HFI:center:frequency:353GHz}{353}
\setsymbol{HFI:center:frequency:545GHz}{545}
\setsymbol{HFI:center:frequency:857GHz}{857}

\setsymbol{HFI:Ndetectors:100GHz}{8}
\setsymbol{HFI:Ndetectors:143GHz}{11}
\setsymbol{HFI:Ndetectors:217GHz}{12}
\setsymbol{HFI:Ndetectors:353GHz}{12}
\setsymbol{HFI:Ndetectors:545GHz}{3}
\setsymbol{HFI:Ndetectors:857GHz}{4}

\setsymbol{HFI:FWHM:Maps:100GHz:units}{9\parcm88}
\setsymbol{HFI:FWHM:Maps:143GHz:units}{7\parcm18}
\setsymbol{HFI:FWHM:Maps:217GHz:units}{4\parcm87}
\setsymbol{HFI:FWHM:Maps:353GHz:units}{4\parcm65}
\setsymbol{HFI:FWHM:Maps:545GHz:units}{4\parcm72}
\setsymbol{HFI:FWHM:Maps:857GHz:units}{4\parcm39}
\setsymbol{HFI:FWHM:Maps:100GHz}{9.88}
\setsymbol{HFI:FWHM:Maps:143GHz}{7.18}
\setsymbol{HFI:FWHM:Maps:217GHz}{4.87}
\setsymbol{HFI:FWHM:Maps:353GHz}{4.65}
\setsymbol{HFI:FWHM:Maps:545GHz}{4.72}
\setsymbol{HFI:FWHM:Maps:857GHz}{4.39}

\setsymbol{HFI:beam:ellipticity:Maps:100GHz}{1.15}
\setsymbol{HFI:beam:ellipticity:Maps:143GHz}{1.01}
\setsymbol{HFI:beam:ellipticity:Maps:217GHz}{1.06}
\setsymbol{HFI:beam:ellipticity:Maps:353GHz}{1.05}
\setsymbol{HFI:beam:ellipticity:Maps:545GHz}{1.14}
\setsymbol{HFI:beam:ellipticity:Maps:857GHz}{1.19}

\setsymbol{HFI:FWHM:Mars:100GHz:units}{9\parcm37}
\setsymbol{HFI:FWHM:Mars:143GHz:units}{7\parcm04}
\setsymbol{HFI:FWHM:Mars:217GHz:units}{4\parcm68}
\setsymbol{HFI:FWHM:Mars:353GHz:units}{4\parcm43}
\setsymbol{HFI:FWHM:Mars:545GHz:units}{3\parcm80}
\setsymbol{HFI:FWHM:Mars:857GHz:units}{3\parcm67}

\setsymbol{HFI:FWHM:Mars:100GHz}{9.37}
\setsymbol{HFI:FWHM:Mars:143GHz}{7.04}
\setsymbol{HFI:FWHM:Mars:217GHz}{4.68}
\setsymbol{HFI:FWHM:Mars:353GHz}{4.43}
\setsymbol{HFI:FWHM:Mars:545GHz}{3.80}
\setsymbol{HFI:FWHM:Mars:857GHz}{3.67}

\setsymbol{HFI:beam:ellipticity:Mars:100GHz}{1.18}
\setsymbol{HFI:beam:ellipticity:Mars:143GHz}{1.03}
\setsymbol{HFI:beam:ellipticity:Mars:217GHz}{1.14}
\setsymbol{HFI:beam:ellipticity:Mars:353GHz}{1.09}
\setsymbol{HFI:beam:ellipticity:Mars:545GHz}{1.25}
\setsymbol{HFI:beam:ellipticity:Mars:857GHz}{1.03}

\setsymbol{HFI:CMB:relative:calibration:100GHz}{$\lsim 1\%$}
\setsymbol{HFI:CMB:relative:calibration:143GHz}{$\lsim 1\%$}
\setsymbol{HFI:CMB:relative:calibration:217GHz}{$\lsim 1\%$}
\setsymbol{HFI:CMB:relative:calibration:353GHz}{$\lsim 1\%$}
\setsymbol{HFI:CMB:relative:calibration:545GHz}{}
\setsymbol{HFI:CMB:relative:calibration:857GHz}{}

\setsymbol{HFI:CMB:absolute:calibration:100GHz}{$\lsim 2\%$}
\setsymbol{HFI:CMB:absolute:calibration:143GHz}{$\lsim 2\%$}
\setsymbol{HFI:CMB:absolute:calibration:217GHz}{$\lsim 2\%$}
\setsymbol{HFI:CMB:absolute:calibration:353GHz}{$\lsim 2\%$}
\setsymbol{HFI:CMB:absolute:calibration:545GHz}{}
\setsymbol{HFI:CMB:absolute:calibration:857GHz}{}

\setsymbol{HFI:FIRAS:gain:calibration:accuracy:statistical:100GHz}{}
\setsymbol{HFI:FIRAS:gain:calibration:accuracy:statistical:143GHz}{}
\setsymbol{HFI:FIRAS:gain:calibration:accuracy:statistical:217GHz}{}
\setsymbol{HFI:FIRAS:gain:calibration:accuracy:statistical:353GHz}{2.5\%}
\setsymbol{HFI:FIRAS:gain:calibration:accuracy:statistical:545GHz}{1\%}
\setsymbol{HFI:FIRAS:gain:calibration:accuracy:statistical:857GHz}{0.5\%}

\setsymbol{HFI:FIRAS:gain:calibration:accuracy:systematic:100GHz}{}
\setsymbol{HFI:FIRAS:gain:calibration:accuracy:systematic:143GHz}{}
\setsymbol{HFI:FIRAS:gain:calibration:accuracy:systematic:217GHz}{}
\setsymbol{HFI:FIRAS:gain:calibration:accuracy:systematic:353GHz}{}
\setsymbol{HFI:FIRAS:gain:calibration:accuracy:systematic:545GHz}{7\%}
\setsymbol{HFI:FIRAS:gain:calibration:accuracy:systematic:857GHz}{7\%}

\setsymbol{HFI:FIRAS:zero:point:accuracy:100GHz:units}{0.8\MJysr}
\setsymbol{HFI:FIRAS:zero:point:accuracy:143GHz:units}{}
\setsymbol{HFI:FIRAS:zero:point:accuracy:217GHz:units}{}
\setsymbol{HFI:FIRAS:zero:point:accuracy:353GHz:units}{1.4\MJysr}
\setsymbol{HFI:FIRAS:zero:point:accuracy:545GHz:units}{2.2\MJysr}
\setsymbol{HFI:FIRAS:zero:point:accuracy:857GHz:units}{1.7\MJysr}

\setsymbol{HFI:FIRAS:zero:point:accuracy:100GHz}{0.8}
\setsymbol{HFI:FIRAS:zero:point:accuracy:143GHz}{}
\setsymbol{HFI:FIRAS:zero:point:accuracy:217GHz}{}
\setsymbol{HFI:FIRAS:zero:point:accuracy:353GHz}{1.4}
\setsymbol{HFI:FIRAS:zero:point:accuracy:545GHz}{2.2}
\setsymbol{HFI:FIRAS:zero:point:accuracy:857GHz}{1.7}

\setsymbol{HFI:unit:conversion:100GHz:units}{0.2415\MJysrmK}
\setsymbol{HFI:unit:conversion:143GHz:units}{0.3694\MJysrmK}
\setsymbol{HFI:unit:conversion:217GHz:units}{0.4811\MJysrmK}
\setsymbol{HFI:unit:conversion:353GHz:units}{0.2883\MJysrmK}
\setsymbol{HFI:unit:conversion:545GHz:units}{0.05826\MJysrmK}
\setsymbol{HFI:unit:conversion:857GHz:units}{0.002238\MJysrmK}

\setsymbol{HFI:unit:conversion:100GHz}{0.2415}
\setsymbol{HFI:unit:conversion:143GHz}{0.3694}
\setsymbol{HFI:unit:conversion:217GHz}{0.4811}
\setsymbol{HFI:unit:conversion:353GHz}{0.2883}
\setsymbol{HFI:unit:conversion:545GHz}{0.05826}
\setsymbol{HFI:unit:conversion:857GHz}{0.002238}

\setsymbol{HFI:colour:correction:alpha=-2:V1.01:100GHz}{0.9893}
\setsymbol{HFI:colour:correction:alpha=-2:V1.01:143GHz}{0.9759}
\setsymbol{HFI:colour:correction:alpha=-2:V1.01:217GHz}{1.0007}
\setsymbol{HFI:colour:correction:alpha=-2:V1.01:353GHz}{1.0028}
\setsymbol{HFI:colour:correction:alpha=-2:V1.01:545GHz}{1.0019}
\setsymbol{HFI:colour:correction:alpha=-2:V1.01:857GHz}{0.9889}

\setsymbol{HFI:colour:correction:alpha=0:V1.01:100GHz}{1.0008}
\setsymbol{HFI:colour:correction:alpha=0:V1.01:143GHz}{1.0148}
\setsymbol{HFI:colour:correction:alpha=0:V1.01:217GHz}{0.9909}
\setsymbol{HFI:colour:correction:alpha=0:V1.01:353GHz}{0.9888}
\setsymbol{HFI:colour:correction:alpha=0:V1.01:545GHz}{0.9878}
\setsymbol{HFI:colour:correction:alpha=0:V1.01:857GHz}{1.0014}

%% file: AuthorList_P02_LFI_Processing_authors_and_institutes.tex
\author{\small
Planck Collaboration:
N.~Aghanim\inst{61}
\and
C.~Armitage-Caplan\inst{92}
\and
M.~Arnaud\inst{75}
\and
M.~Ashdown\inst{72, 6}
\and
F.~Atrio-Barandela\inst{18}
\and
J.~Aumont\inst{61}
\and
C.~Baccigalupi\inst{86}
\and
A.~J.~Banday\inst{95, 9}
\and
R.~B.~Barreiro\inst{68}
\and
E.~Battaner\inst{97}
% ...
}
\institute{\small
APC, AstroParticule et Cosmologie, Universit\'{e} Paris Diderot, CNRS/IN2P3, CEA/lrfu, Observatoire de Paris, Sorbonne Paris Cit\'{e}, 10, rue Alice Domon et L\'{e}onie Duquet, 75205 Paris Cedex 13, France\\
\and
Aalto University Mets\"{a}hovi Radio Observatory, Mets\"{a}hovintie 114, FIN-02540 Kylm\"{a}l\"{a}, Finland\\
\and
African Institute for Mathematical Sciences, 6-8 Melrose Road, Muizenberg, Cape Town, South Africa\\
\and
Agenzia Spaziale Italiana Science Data Center, Via del Politecnico snc, 00133, Roma, Italy\\
\and
Agenzia Spaziale Italiana, Viale Liegi 26, Roma, Italy\\
\and
Astrophysics Group, Cavendish Laboratory, University of Cambridge, J J Thomson Avenue, Cambridge CB3 0HE, U.K.\\
\and
CITA, University of Toronto, 60 St. George St., Toronto, ON M5S 3H8, Canada\\
\and
CNR - ISTI, Area della Ricerca, via G. Moruzzi 1, Pisa, Italy\\
\and
CNRS, IRAP, 9 Av. colonel Roche, BP 44346, F-31028 Toulouse cedex 4, France\\
\and
California Institute of Technology, Pasadena, California, U.S.A.\\
\and
Centre for Theoretical Cosmology, DAMTP, University of Cambridge, Wilberforce Road, Cambridge CB3 0WA, U.K.\\
\and
Centro de Estudios de F\'{i}sica del Cosmos de Arag\'{o}n (CEFCA), Plaza San Juan, 1, planta 2, E-44001, Teruel, Spain\\
\and
Computational Cosmology Center, Lawrence Berkeley National Laboratory, Berkeley, California, U.S.A.\\
\and
Consejo Superior de Investigaciones Cient\'{\i}ficas (CSIC), Madrid, Spain\\
\and
DSM/Irfu/SPP, CEA-Saclay, F-91191 Gif-sur-Yvette Cedex, France\\
\and
DTU Space, National Space Institute, Technical University of Denmark, Elektrovej 327, DK-2800 Kgs. Lyngby, Denmark\\
\and
D\'{e}partement de Physique Th\'{e}orique, Universit\'{e} de Gen\`{e}ve, 24, Quai E. Ansermet,1211 Gen\`{e}ve 4, Switzerland\\
\and
Departamento de F\'{\i}sica Fundamental, Facultad de Ciencias, Universidad de Salamanca, 37008 Salamanca, Spain\\
\and
Departamento de F\'{\i}sica, Universidad de Oviedo, Avda. Calvo Sotelo s/n, Oviedo, Spain\\
\and
Departamento de Matem\'{a}ticas, Estad\'{\i}stica y Computaci\'{o}n, Universidad de Cantabria, Avda. de los Castros s/n, Santander, Spain\\
\and
Department of Astronomy and Astrophysics, University of Toronto, 50 Saint George Street, Toronto, Ontario, Canada\\
\and
Department of Astrophysics/IMAPP, Radboud University Nijmegen, P.O. Box 9010, 6500 GL Nijmegen, The Netherlands\\
\and
Department of Electrical Engineering and Computer Sciences, University of California, Berkeley, California, U.S.A.\\
\and
Department of Physics \& Astronomy, University of British Columbia, 6224 Agricultural Road, Vancouver, British Columbia, Canada\\
\and
Department of Physics and Astronomy, Dana and David Dornsife College of Letter, Arts and Sciences, University of Southern California, Los Angeles, CA 90089, U.S.A.\\
\and
Department of Physics and Astronomy, University College London, London WC1E 6BT, U.K.\\
\and
Department of Physics, Florida State University, Keen Physics Building, 77 Chieftan Way, Tallahassee, Florida, U.S.A.\\
\and
Department of Physics, Gustaf H\"{a}llstr\"{o}min katu 2a, University of Helsinki, Helsinki, Finland\\
\and
Department of Physics, Princeton University, Princeton, New Jersey, U.S.A.\\
\and
Department of Physics, University of California, One Shields Avenue, Davis, California, U.S.A.\\
\and
Department of Physics, University of California, Santa Barbara, California, U.S.A.\\
\and
Department of Physics, University of Illinois at Urbana-Champaign, 1110 West Green Street, Urbana, Illinois, U.S.A.\\
\and
Dipartimento di Fisica e Astronomia G. Galilei, Universit\`{a} degli Studi di Padova, via Marzolo 8, 35131 Padova, Italy\\
\and
Dipartimento di Fisica e Scienze della Terra, Universit\`{a} di Ferrara, Via Saragat 1, 44122 Ferrara, Italy\\
\and
Dipartimento di Fisica, Universit\`{a} La Sapienza, P. le A. Moro 2, Roma, Italy\\
\and
Dipartimento di Fisica, Universit\`{a} degli Studi di Milano, Via Celoria, 16, Milano, Italy\\
\and
Dipartimento di Fisica, Universit\`{a} degli Studi di Trieste, via A. Valerio 2, Trieste, Italy\\
\and
Dipartimento di Fisica, Universit\`{a} di Roma Tor Vergata, Via della Ricerca Scientifica, 1, Roma, Italy\\
\and
Discovery Center, Niels Bohr Institute, Blegdamsvej 17, Copenhagen, Denmark\\
\and
Dpto. Astrof\'{i}sica, Universidad de La Laguna (ULL), E-38206 La Laguna, Tenerife, Spain\\
\and
European Space Agency, ESAC, Planck Science Office, Camino bajo del Castillo, s/n, Urbanizaci\'{o}n Villafranca del Castillo, Villanueva de la Ca\~{n}ada, Madrid, Spain\\
\and
European Space Agency, ESTEC, Keplerlaan 1, 2201 AZ Noordwijk, The Netherlands\\
\and
Haverford College Astronomy Department, 370 Lancaster Avenue, Haverford, Pennsylvania, U.S.A.\\
\and
Helsinki Institute of Physics, Gustaf H\"{a}llstr\"{o}min katu 2, University of Helsinki, Helsinki, Finland\\
\and
INAF - Osservatorio Astrofisico di Catania, Via S. Sofia 78, Catania, Italy\\
\and
INAF - Osservatorio Astronomico di Padova, Vicolo dell'Osservatorio 5, Padova, Italy\\
\and
INAF - Osservatorio Astronomico di Roma, via di Frascati 33, Monte Porzio Catone, Italy\\
\and
INAF - Osservatorio Astronomico di Trieste, Via G.B. Tiepolo 11, Trieste, Italy\\
\and
INAF Istituto di Radioastronomia, Via P. Gobetti 101, 40129 Bologna, Italy\\
\and
INAF/IASF Bologna, Via Gobetti 101, Bologna, Italy\\
\and
INAF/IASF Milano, Via E. Bassini 15, Milano, Italy\\
\and
INFN, Sezione di Bologna, Via Irnerio 46, I-40126, Bologna, Italy\\
\and
INFN, Sezione di Roma 1, Universit\`{a} di Roma Sapienza, Piazzale Aldo Moro 2, 00185, Roma, Italy\\
\and
IPAG: Institut de Plan\'{e}tologie et d'Astrophysique de Grenoble, Universit\'{e} Joseph Fourier, Grenoble 1 / CNRS-INSU, UMR 5274, Grenoble, F-38041, France\\
\and
ISDC Data Centre for Astrophysics, University of Geneva, ch. d'Ecogia 16, Versoix, Switzerland\\
\and
IUCAA, Post Bag 4, Ganeshkhind, Pune University Campus, Pune 411 007, India\\
\and
Imperial College London, Astrophysics group, Blackett Laboratory, Prince Consort Road, London, SW7 2AZ, U.K.\\
\and
Infrared Processing and Analysis Center, California Institute of Technology, Pasadena, CA 91125, U.S.A.\\
\and
Institut N\'{e}el, CNRS, Universit\'{e} Joseph Fourier Grenoble I, 25 rue des Martyrs, Grenoble, France\\
\and
Institut Universitaire de France, 103, bd Saint-Michel, 75005, Paris, France\\
\and
Institut d'Astrophysique Spatiale, CNRS (UMR8617) Universit\'{e} Paris-Sud 11, B\^{a}timent 121, Orsay, France\\
\and
Institut d'Astrophysique de Paris, CNRS (UMR7095), 98 bis Boulevard Arago, F-75014, Paris, France\\
\and
Institute for Space Sciences, Bucharest-Magurale, Romania\\
\and
Institute of Astronomy and Astrophysics, Academia Sinica, Taipei, Taiwan\\
\and
Institute of Astronomy, University of Cambridge, Madingley Road, Cambridge CB3 0HA, U.K.\\
\and
Institute of Theoretical Astrophysics, University of Oslo, Blindern, Oslo, Norway\\
\and
Instituto de Astrof\'{\i}sica de Canarias, C/V\'{\i}a L\'{a}ctea s/n, La Laguna, Tenerife, Spain\\
\and
Instituto de F\'{\i}sica de Cantabria (CSIC-Universidad de Cantabria), Avda. de los Castros s/n, Santander, Spain\\
\and
Istituto di Fisica del Plasma, CNR-ENEA-EURATOM Association, Via R. Cozzi 53, Milano, Italy\\
\and
Jet Propulsion Laboratory, California Institute of Technology, 4800 Oak Grove Drive, Pasadena, California, U.S.A.\\
\and
Jodrell Bank Centre for Astrophysics, Alan Turing Building, School of Physics and Astronomy, The University of Manchester, Oxford Road, Manchester, M13 9PL, U.K.\\
\and
Kavli Institute for Cosmology Cambridge, Madingley Road, Cambridge, CB3 0HA, U.K.\\
\and
LAL, Universit\'{e} Paris-Sud, CNRS/IN2P3, Orsay, France\\
\and
LERMA, CNRS, Observatoire de Paris, 61 Avenue de l'Observatoire, Paris, France\\
\and
Laboratoire AIM, IRFU/Service d'Astrophysique - CEA/DSM - CNRS - Universit\'{e} Paris Diderot, B\^{a}t. 709, CEA-Saclay, F-91191 Gif-sur-Yvette Cedex, France\\
\and
Laboratoire Traitement et Communication de l'Information, CNRS (UMR 5141) and T\'{e}l\'{e}com ParisTech, 46 rue Barrault F-75634 Paris Cedex 13, France\\
\and
Laboratoire de Physique Subatomique et de Cosmologie, Universit\'{e} Joseph Fourier Grenoble I, CNRS/IN2P3, Institut National Polytechnique de Grenoble, 53 rue des Martyrs, 38026 Grenoble cedex, France\\
\and
Laboratoire de Physique Th\'{e}orique, Universit\'{e} Paris-Sud 11 \& CNRS, B\^{a}timent 210, 91405 Orsay, France\\
\and
Lawrence Berkeley National Laboratory, Berkeley, California, U.S.A.\\
\and
Max-Planck-Institut f\"{u}r Astrophysik, Karl-Schwarzschild-Str. 1, 85741 Garching, Germany\\
\and
McGill Physics, Ernest Rutherford Physics Building, McGill University, 3600 rue University, Montr\'{e}al, QC, H3A 2T8, Canada\\
\and
MilliLab, VTT Technical Research Centre of Finland, Tietotie 3, Espoo, Finland\\
\and
Niels Bohr Institute, Blegdamsvej 17, Copenhagen, Denmark\\
\and
Observational Cosmology, Mail Stop 367-17, California Institute of Technology, Pasadena, CA, 91125, U.S.A.\\
\and
SB-ITP-LPPC, EPFL, CH-1015, Lausanne, Switzerland\\
\and
SISSA, Astrophysics Sector, via Bonomea 265, 34136, Trieste, Italy\\
\and
School of Physics and Astronomy, Cardiff University, Queens Buildings, The Parade, Cardiff, CF24 3AA, U.K.\\
\and
School of Physics and Astronomy, University of Nottingham, Nottingham NG7 2RD, U.K.\\
\and
Space Sciences Laboratory, University of California, Berkeley, California, U.S.A.\\
\and
Special Astrophysical Observatory, Russian Academy of Sciences, Nizhnij Arkhyz, Zelenchukskiy region, Karachai-Cherkessian Republic, 369167, Russia\\
\and
Stanford University, Dept of Physics, Varian Physics Bldg, 382 Via Pueblo Mall, Stanford, California, U.S.A.\\
\and
Sub-Department of Astrophysics, University of Oxford, Keble Road, Oxford OX1 3RH, U.K.\\
\and
Theory Division, PH-TH, CERN, CH-1211, Geneva 23, Switzerland\\
\and
UPMC Univ Paris 06, UMR7095, 98 bis Boulevard Arago, F-75014, Paris, France\\
\and
Universit\'{e} de Toulouse, UPS-OMP, IRAP, F-31028 Toulouse cedex 4, France\\
\and
Universities Space Research Association, Stratospheric Observatory for Infrared Astronomy, MS 232-11, Moffett Field, CA 94035, U.S.A.\\
\and
University of Granada, Departamento de F\'{\i}sica Te\'{o}rica y del Cosmos, Facultad de Ciencias, Granada, Spain\\
% ...
}

%% file: 00_abstract.tex
\abstract{
 We describe the data processing pipeline of the \Planck\ Low
  Frequency Instrument (LFI) data processing center (DPC) to
  create and characterize full-sky maps based on the first 15.5 months of operations at 30, 44, and 70\,GHz.
  In particular, we discuss the various
  steps involved in reducing the data, from telemetry
  packets through to the production of cleaned, calibrated timelines
  and calibrated frequency maps. Data are continuously calibrated
  using the modulation induced on the mean temperature of the
  cosmic microwave background radiation by the proper motion of
  the spacecraft. Sky signals other than the
  dipole are removed by an iterative procedure based on simultaneous
  fitting of calibration parameters and sky maps. Noise properties are
  estimated from time-ordered data after the sky signal has been removed,
  using a generalized least squares map-making algorithm. A destriping code ({\tt Madam})
  is employed to combine radiometric data and pointing information into sky maps,
  minimizing the variance of correlated noise.  Noise covariance matrices, required to
  compute statistical uncertainties on LFI and \Planck\ products, are also produced. Main
  beams are estimated down to the $\approx -20$\,dB level using Jupiter
  transits, which are also used for the  geometrical
  calibration of the focal plane.}

%% file: 01_introduction.tex
This paper, one of a set associated with the 2013 release of data
from the \Planck\footnote{\Planck\
(\url{http://www.esa.int/Planck}) is a project of the European
Space Agency (ESA) with instruments provided by two scientific
consortia funded by ESA member states (in particular the lead
countries France and Italy), with contributions from NASA (USA)
and telescope reflectors provided by a collaboration between ESA
and a scientific consortium led and funded by Denmark.}  mission
\citep{planck2013-p01}, describes the Low Frequency Instrument
(LFI) data processing that supports the first \Planck\
cosmological release based on the nominal \Planck\ survey
(15.5\,months of observation). This paper represents an updated
version of the LFI data processing description
\citep{planck2011-1.6} that was part of the first wave of
astrophysical results published in early 2011 (Planck
Collaboration VIII--XXVI 2011). This work describes the overall
data flow of the pipeline implemented at the LFI DPC, from
instrument scientific telemetry and housekeeping data to frequency
maps, as well as the test plan applied to validate the data
products. Detailed descriptions of critical aspects of the data
analysis and products, including justifications for choices of
algorithms used in the pipeline, are given in three companion
papers: \citet{planck2013-p02a} discusses systematic effects and
gives the overall error budget; \citet{planck2013-p02d} describes
determination of main beams and uncertainties from in-flight
planet-crossing measurements; and \citet{planck2013-p02b}
describes photometric calibration, including methods and related
uncertainties.  The main results and reference tables in these
three areas are summarized in this paper.  \citet{planck2013-p28}
provides detailed descriptions of the products delivered.
\all2013resultspapers

%% file: 02-00_inflight_behav_intro.tex
The \Planck\ LFI instrument is described in \citet{
bersanelli2010} and \citet{mennella2010}. It comprises eleven
radiometer chain assemblies (RCAs), two at 30\,GHz, three at
44\,GHz, and six at 70\,GHz, each composed of two independent
pseudo-correlation radiometers sensitive to orthogonal linear
polarizations. Each radiometer has two independent square-law
diodes for detection, integration, and conversion from radio
frequency signals into DC voltages. The focal plane is
cryogenically cooled to 20\,K, while the pseudo-correlation design
uses internal, blackbody, reference loads cooled to 4.5\,K.  The
radiometer timelines are produced by taking differences between
the signals from the sky, $V_{\rm sky}$, and from the reference
loads, $V_{\rm ref}$. Radiometer balance is optimized by
introducing a gain modulation factor, typically stable within
0.04\,\% throughout the mission, which greatly reduces $1/f$ noise
and improves immunity to a wide class of systematic effects
\citep{planck2011-1.4}. During the entire nominal survey, the
behavior of all 22 LFI radiometers was stable, with $1/f$ knee
frequencies unchanging within 10\,\% and white noise levels within
0.5\,\%.

%% file: 02-01_operations.tex
During the period of observations, no changes have been applied on
the satellite \citep{planck2013-p01}, with a single exception.
Three months before the end of the nominal mission it was
necessary to switch from the nominal to the redundant sorption
cooler. This operation, described below, was visible in the LFI
scientific data, but the effect on the temperature power spectrum
was negligible (Sect.~\ref{sec_operations_switchover}).

\subsection{Switchover from nominal to redundant sorption cooler}
\label{sec_operations_switchover}

The 20\,K cooling on \Planck\ is provided by the sorption cooler
system. This cooler uses six metal hydride compressor elements to
produce high-pressure hydrogen that expands through a
Joule-Thomson valve to provide 1\,W of cooling at 20\,K. Gas
compression is achieved by heating a single compressor element to
440\,K and a pressure of 30\,bar.  After expansion through the
Joule-Thomson valve, the gas is recovered by three compressor
elements at 270\,K and 0.3\,bar.  To reduce power consumption,
gas-gap heat switches are used to isolate the compressor elements
from the radiator while the heating elements are powered.  Two
sorption coolers were flown on \Planck\ to meet mission lifetime
requirements.  A switchover procedure was developed to change
between the operating cooler and the redundant cooler.  In early
August of 2010, one of the gas-gap heat switches for a compressor
element failed on the active cooler.  Although the sorption cooler
can operate with as few as four compressor elements, it was
decided to implement the switchover procedure and activate the
redundant cooler.  On 11 August 2010 at 17:30 UTC, the working
cooler was commanded off and the redundant cooler was switched on.
Adequate cryogenic cooling was restored in about 1\,hour;  return
to thermal stability took 48\,hours.  After thermal stability of
the cooler was restored, anomalous temperature fluctuations were
observed on the LFI focal plane. These excess fluctuations are
thought to be due to sloshing of liquid hydrogen remaining at the
cold end of the cooler that had been switched off. It had been
thought that essentially all of the hydrogen in the system would
be absorbed in the metal hydride beds after the cooler was
switched off.  It seems, however, that the normal loss of storage
capacity during operations left enough hydrogen in the piping to
form liquid at the cold end. While these fluctuations produced a
measurable effect in the LFI data, their propagation to the
temperature power spectrum is more than two orders of magnitude
below the cosmic microwave background (CMB) signal
\citep{planck2013-p02a}. Furthermore, by the end of the nominal
mission in February 2011, these fluctuations reduced to a much
lower level. More details of these issues will be discussed in a
future paper.

%% file: 02-02_instrupdate_summary.tex
Table~\ref{tab_summary_performance} gives a top-level summary of
instrument performance parameters measured in flight during the
nominal data period. Optical properties have been successfully
reconstructed using Jupiter transits \citep{planck2013-p02d}, and
the main parameters are in agreement with pre-launch and early
estimates \citep{planck2011-1.4}. The white noise sensitivity and
parameters describing the 1/$f$ noise component are in line with
ground measurements \citep{mennella2010}, and agree with the values
in \citet{planck2011-1.4}.  Photometric
calibration based on the CMB dipole yields an overall statistical
uncertainty of 0.25\,\% \citep{planck2013-p02b}. Variations due to
slow instrumental changes are traced by the calibration
pipeline, yielding an overall uncertainty between 0.1\,\%
and 0.2\,\%. The residual systematic uncertainty varies between 21 and 6\,$\mu\mathrm{K}_\mathrm{CMB}$
\citep{planck2013-p02a}.

\begin{table*}
\begingroup
\newdimen\tblskip \tblskip=5pt
\caption{LFI performance parameters.}
\label{tab_summary_performance}
\nointerlineskip
\vskip -3mm
\footnotesize
\setbox\tablebox=\vbox{
   \newdimen\digitwidth
   \setbox0=\hbox{\rm 0}
   \digitwidth=\wd0
   \catcode`*=\active
   \def*{\kern\digitwidth}
   \newdimen\signwidth
   \setbox0=\hbox{+}
   \signwidth=\wd0
   \catcode`!=\active
   \def!{\kern\signwidth}
\halign{\hbox to 2.7in{#\leaderfil}\tabskip=3em&
        \hfil#\hfil&
        \hfil#\hfil&
        \hfil#\hfil\tabskip=0pt\cr
\noalign{\doubleline}
\omit\hfil Parameter\hfil&30\,GHz&44\,GHz&70\,GHz\cr
\noalign{\vskip 3pt\hrule\vskip 5pt}
Center frequency [GHz]&28.4&44.1&70.4\cr
\noalign{\vskip 3pt}
Scanning beam FWHM$^{\rm a}$ [arcmin]&33.16&28.09&13.08\cr
\noalign{\vskip 3pt}
Scanning beam ellipticity$^{\rm a}$&1.37&1.25&1.27\cr
\noalign{\vskip 3pt}
Effective beam FWHM$^{\rm b}$ [arcmin]&32.34&27.12&13.31\cr
\noalign{\vskip 3pt}
White noise level in map$^{\rm c}$ [\muKCMB]&**9.2&*12.5&*23.2\cr
\noalign{\vskip 3pt}
White noise level in timelines$^{\rm d}$ [\muKCMBs]&148.5&173.2&151.9\cr
\noalign{\vskip 3pt}
$f_{\rm knee}$$^{\rm d}$ [mHz]&114.5&45.7&20.2\cr
\noalign{\vskip 3pt}
$1/f$ slope$^{\rm d}$&\llap{$-$}0.92&\llap{$-$}0.90&\llap{$-$}1.13\cr
\noalign{\vskip 3pt}
Overall calibration uncertainty$^{\rm e}$ [\%]&0.82&0.55&0.62\cr
\noalign{\vskip 3pt}
Systematic effects uncertainty$^{\rm f}$ [\muKCMB]&21.02&5.61&7.87\cr
\noalign{\vskip 5pt\hrule\vskip 3pt}}}
\endPlancktablewide
\tablenote a Determined by fitting Jupiter observations
directly in the timelines.\par 
\tablenote b Calculated from the main beam solid angle of the effective beam, $\Omega_{\rm eff} =
\hbox{mean}(\Omega)$ (Sect.~\ref{sec_effectivebeam}). These values
are used in the source extraction pipeline
~\citep{planck2013-p05}.\par \tablenote c White noise per pixel
computed from half-ring difference maps. These values are within
1\% of the white noise sensitivity computed directly on the
timelines, taking into account the actual integration time
represented in the maps.\par \tablenote d Values derived
from fitting noise spectra (Sect.~\ref{sec_noise}).\par
\tablenote e Sum of the error on the estimation of the
calibration constant (0.25\,\%) and the square root of the squared
sum of the following errors: beam uncertainty; sidelobe
convolution effect; and unknown systematics as measured from the
power spectrum at $50 < \ell < 250$ (see
\citealt{planck2013-p02b}).\par \tablenote f Peak-to-peak
difference between 99\,\% and 1\,\% quantiles in the pixel value
distributions from simulated maps (see
\citealt{planck2013-p02a}).\par

\endgroup
\end{table*}

%% file: 03-00_data_proc_over.tex
\begin{figure*} [th]
\centering
\includegraphics[width=18cm]{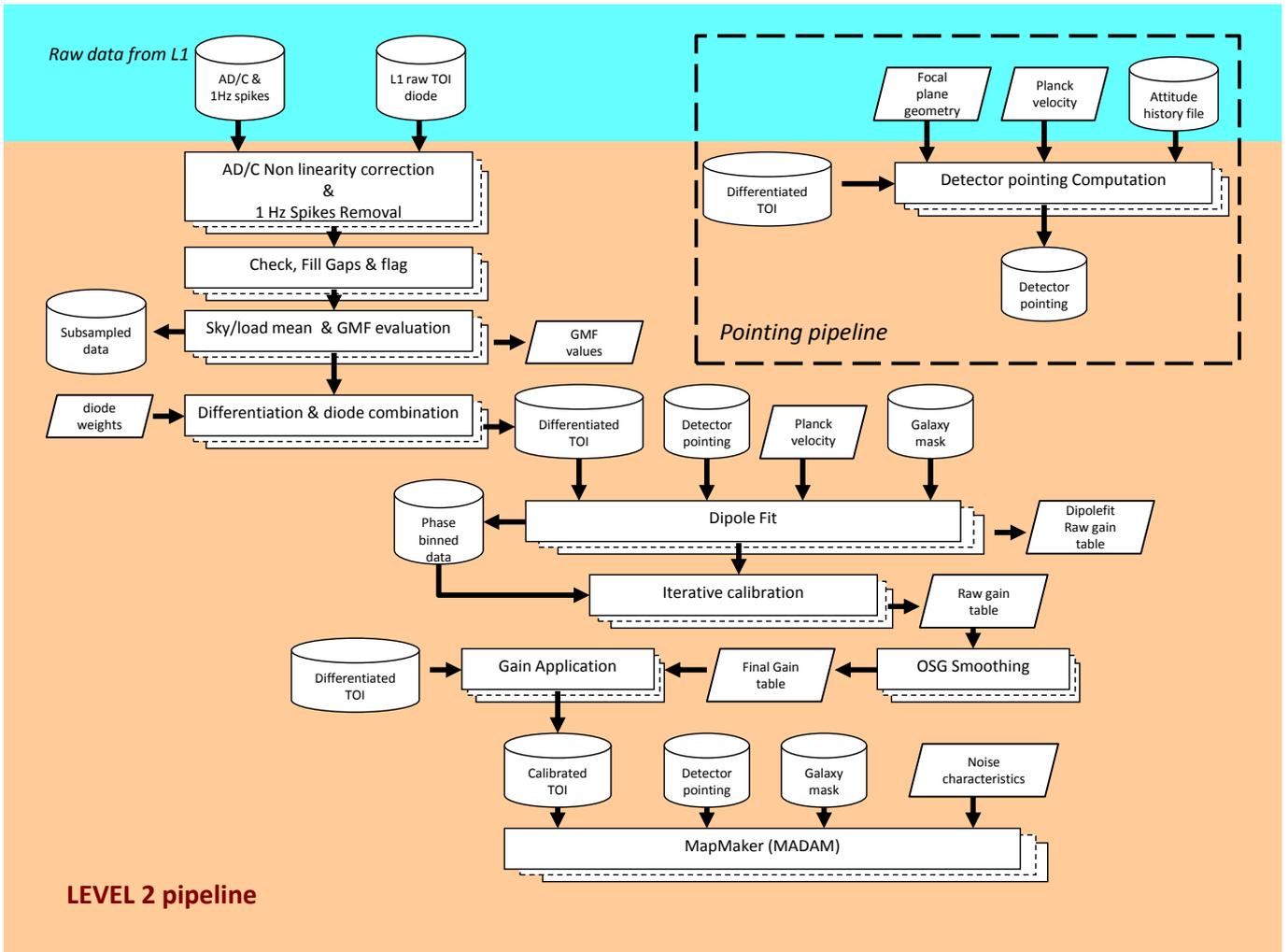}
\caption{Schematic representation of the Level~2 and pointing pipelines of the LFI DPC}
\label{dpcpipeline}
\end{figure*}

The processing of LFI data is divided into levels shown
schematically in Fig.~\ref{dpcpipeline}. Processing  starts at
Level~1, which retrieves all necessary information from packets
and auxiliary data received each day from the Mission Operation
Center, and transforms the scientific packets and housekeeping
data into a form manageable by Level~2.  Level~2 uses scientific
and housekeeping information to:

\begin{itemize}

\item build the LFI reduced instrument model (RIMO), which contains the main characteristics of the instrument;

\item remove analogue-to-digital converter (ADC) non-linearities and 1\,Hz spikes at diode level (see Sects.~\ref{sec_adc_nonlinearity} and ~\ref{sec_electronic_spikes});

\item compute and apply the gain modulation factor to minimize $1/f$ noise (see Sect.~\ref{sec_gain_modulation});

\item combine signals from the diodes (see Sect.~\ref{sec_comb_diodes});

\item compute corresponding detector pointing for each sample, based on auxiliary data and beam information (see Sect.~\ref{sec_pointing});

\item calibrate the scientific timelines to physical units ($\mathrm{K}_\mathrm{CMB}$), fitting the dipole convolved with the $4\pi$ beam representation (see Sect.~\ref{sec_calibration});

\item remove the dipole convolved with the $4\pi$ beam representation from the scientific calibrated timeline;

\item combine the calibrated TOIs into aggregate products such as maps at each frequency (see Sect.~\ref{sec_mmaking_intro}).

\end{itemize}

Level~3 collects Level~2 outputs from both HFI
\citep{planck2013-p03} and LFI and derives various products such
as component-separated maps of astrophysical foregrounds,
catalogues of various classes of sources, and the likelihood of
various cosmological and astrophysical models given the frequency
maps.

%% file: 04-00_toi_processing.tex
The Level~1 pipeline receives telemetry data as a stream of
packets that are handled automatically in several steps:

\begin{itemize}

\item uncompress the retrieved packets;

\item de-quantize and de-mix the uncompressed packets to retrieve
the original signal in analog to digital units (ADU);

\item transform ADU data into volts using a conversion factor stored in the packet header;

\item cross-correlate time information to time stamp each sample uniquely;

\item store the resulting timelines in a database interface to the Level 2 pipeline.

\end{itemize}

We made no change in Level~1 software during the mission.
Detailed information on how each of the steps listed above was applied is provided in
\citet{planck2011-1.6}.   To avoid strong gradients in
the signal and signals that do not project correctly in the maps,
we established the procedure to flag a single scientific sample
described in Sect.~\ref{sec_flagging}.

%% file: 04-01_flagging.tex
For each sample we define a 32-bit flag mask to identify potential
inconsistencies in the data and to enable the pipeline to skip
that sample or handle it differently. The TOI from all LFI
detectors are archived in the Level~1 database, and regularly
checked to identify and flag events that can affect the scientific
analysis. These events include missing or anomalous data, and data
acquired during the manoeuvres regularly performed to repoint the
telescope according to the \Planck\ scanning strategy.
Table~\ref{tab_data_flags_percentage} summarizes the percentage of
time associated with these events for the nominal mission. The
table also reports the total percentage of Level~1 TOIs usable in
the scientific analysis. Most of the missing data are from
telemetry packets in which the arithmetic compression performed by
the Science Processing Unit (SPU) is incorrect, causing a
decompression error. They are rare, and have negligible impact on
the scientific analysis. For instance, for the entire 70\,GHz
channel, the total amount of missing data corresponds to 130 lost
seconds in 15\,months. The instrument team performs a daily check
of the data retrieved during the daily telecommunication period
with the satellite; the data cover an entire operational day (OD).
Part of this analysis consists of identifying, for each detector,
time windows where either the total power signal or the
differentiated signal shows anomalous fluctuations or jumps.
Depending on the characteristics of the anomaly identified, a time
window can be flagged as unusable for science. Currently, the
criteria defined to flag time windows as unusable include:

\begin{itemize}

\item gain changes in the data acquisition electronics that cause saturation of the sky or reference load signals;

\item abrupt changes in voltage output with slow recovery ($>$1\,min), caused by gain fluctuations in the back-end module amplifier, induced by electrical or thermal variations, which generate discontinuities in the differentiated signal;

\item short, abrupt changes in voltage output caused by fluctuations in the low noise amplifiers in the front-end module, which produce asymmetries between the sky and reference load signals and possibly first order effects in the differentiated signal;

\item permanent changes in the voltage output caused by a permanent change at the front-end module (amplifier bias or focal plane unit temperature) or back-end module (temperature or HEMT gain variations) -- in such cases, only a small time window around the discontinuity is flagged as unusable;

\item ``popcorn'' noise on the total power signal of one or both detectors due to variations in the back end diode or in the front end low noise amplifiers, causing short time windows ($<$1\,m) of unusable data.

\end{itemize}

In Table~\ref{tab_data_flags_percentage}, the row labelled
``Anomalies'' reports the percentage of observation time flagged
as unusable for these reasons in the scientific analysis. The almost 1.0\,\% shown for
the 44\,GHz channel corresponds to a total time of 113 hours.
Finally, the times of manoeuvres and stable pointing periods are recovered from the
attitude history files provided by the \Planck\ flight dynamics
team. Detector samples corresponding to manoeuvres are flagged
so they can be ignored in subsequent steps.

\begin{table}[tmb]
  \begingroup
  \newdimen\tblskip \tblskip=5pt
  \caption{Percentage of LFI observation time lost due to missing or unusable data, and to manoeuvres$^{\rm a}$.}
  \label{tab_data_flags_percentage}
  \nointerlineskip
  \vskip -3mm
  \footnotesize
  \setbox\tablebox=\vbox{
    \newdimen\digitwidth
    \setbox0=\hbox{\rm 0}
    \digitwidth=\wd0
    \catcode`*=\active
    \def*{\kern\digitwidth}
    \newdimen\signwidth
    \setbox0=\hbox{+}
    \signwidth=\wd0
    \catcode`!=\active
    \def!{\kern\signwidth}
    \halign{\hbox to 1.31in{#\leaderfil}\tabskip=1em&
      \hfil#\hfil&
      \hfil#\hfil&
      \hfil#\hfil\tabskip=0pt\cr                            
      \noalign{\doubleline}
      \omit\hfil Category\hfil& 30\,GHz& 44\,GHz& 70\,GHz\cr
      \noalign{\vskip 3pt\hrule\vskip 5pt}
        Missing [\%]& *0.00014& *0.00023& *0.00032\cr
      Anomalies [\%]& *0.82220& *0.99763& *0.82925\cr
      Manoeuvres [\%]& *8.07022& *8.07022& *8.07022\cr
\noalign{\vskip 3pt}
      Usable [\%]& 91.10744& 90.93191& 91.10021\cr
      \noalign{\vskip 5pt\hrule\vskip 3pt}
    }}
  \endPlancktable
  \par \tablenote a The remaining percentage is used in scientific analysis. \par

  \endgroup
\end{table}

Tasks within the Level~2 pipelines both fill gaps in the data with
artificial noise and flag them properly. Other tasks locate
transits of planets and other moving objects within the solar
system, again flagging samples affected by such observations.

%% file: 04-02_adc.tex
The ADCs convert the analogue detector voltages to numbers, which
are then processed on-board by the radiometer electronics box
assembly.  Since they are directly involved with the signal power,
their linearity is as important as that of the receivers and
detectors, with any departure appearing as a distortion in the
system power response curve.  In differential measurements such as
those carried out by the \Planck\ LFI instrument, small localized
distortions in this curve can have a large impact, since the
calibration factor depends on the gradient of the response curve
at the point at which the differential measurements are made. This
effect is described in detail in \citet{planck2013-p02a}; its
impact on calibration is described in \citet{planck2013-p02b}.

\begin{figure}
\includegraphics[width=9cm]{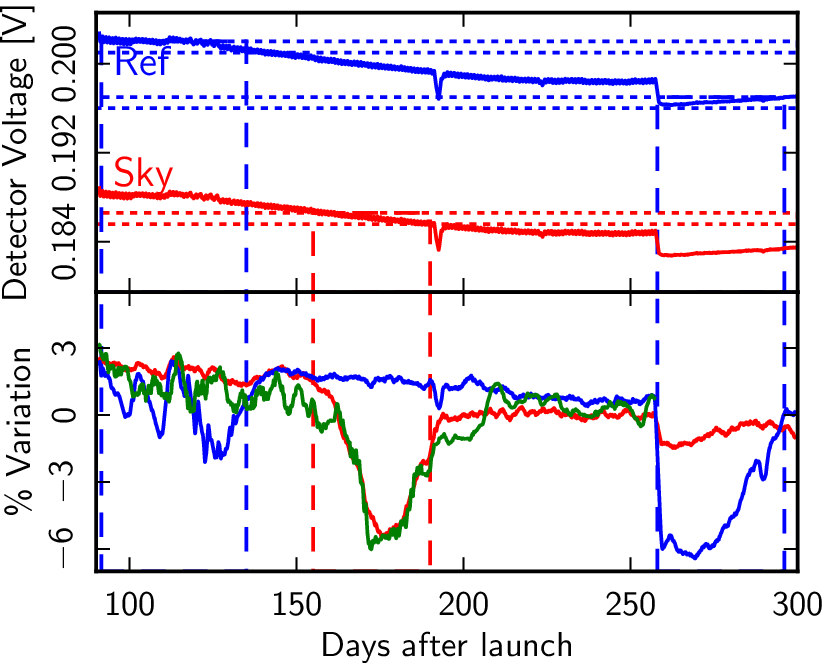}
\caption{Effect of ADC non-linearities on time-ordered data of one
44\,GHz diode. The upper plot shows the recorded detector voltages
for sky (red) and reference (blue).  Voltage ranges affected by
ADC non-linearities are marked by horizontal dotted lines. Time
ranges affected are marked by vertical dashed lines. The lower
plot shows the percentage variation of the inverse of the gain
factor from the dipole gain (green) and the ``white noise''
estimates on the sky and reference voltages (sky red, reference
blue). The gain estimates have been smoothed by a three-day moving
mean, the noise by a one-day moving mean.}
\label{fig_adc_gain_deltav_effect}
\end{figure}

The effect is observed in some LFI radiometer data, appearing as
gain variations seen at particular detector voltages. This is
shown for the most affected channel, RCA2501, in
Fig.~\ref{fig_adc_gain_deltav_effect}, where the upper plot shows
the measured voltages of the sky and reference loads and the lower
plot shows the percentage variations of gain and noise in the sky
and reference voltages.  The range of the upper plot is matched to
that of the lower plot, so for normal gain variations the same
pattern should be seen for both. That is clearly not the case.
When the sky signal is near 0.186\,V, marked by horizontal dotted
lines, both the inverse gain and the sky ``white noise" estimates
show anomalies (the time interval affected is indicated by
vertical dashed lines). The same anomalous behavior of the
reference white noise signal and inverse gain is seen in two
intervals when the reference signal is near 0.197\,V and 0.202\,V.
Outside of these limited ranges, the variations in all plotted
signals track one another, such as the feature at day 192 in the
sky voltage, or the drop at day 257 when the transponder was
turned on permanently.

The response curves can be reconstructed by tracking how the noise
amplitude varies with the apparent detector voltage in the TOI.
The radiometers are assumed to be stable and the intrinsic thermal
noise can be taken to be constant in terms of temperature, so any
voltage variations are then assumed to be due to both gain drift
and ADC effects. The method for this correction is set out in
appendix A of \citet{planck2013-p02a}.

%% file: 04-03_spikes.tex
Electronic spikes in the signal are caused by an interaction between the housekeeping
electronics clock and the scientific data line in the on-board
data acquisition system. The spikes are synchronous with the
on-board time, with no changes in phase over the entire acquisition
period, allowing the construction of dedicated templates that are
then removed from the timelines.  Spikes are present in
all frequencies, but are significant only at 44\,GHz due to
the high gain of these detectors.  Consequently, electronic spikes
are removed only in this channel. This process and the evaluation
of the effect at map level are described in \citet{planck2013-p02a}.

%% file: 04-04_gain_modulation.tex
Each diode switches at 4096\,Hz \citep{mennella2010} between the
sky and the 4\,K reference load.  Voltages $V_{\rm sky}$ and $V_{\rm load}$
are dominated by $1/f$ noise, with knee frequencies of tens of hertz.
This noise is highly correlated between the two streams, a result of the
pseudo-correlation design \citep{bersanelli2010}, and differencing
the streams results in a dramatic reduction of the $1/f$ noise. To
force the mean of the difference to zero, the load signal is
multiplied by the gain modulation factor (GMF in Fig.~\ref{dpcpipeline}) $R$, which can be
computed in several ways \citep{mennella2003}. The simplest
method, and the one implemented in the processing pipeline, is to
take the ratio of DC levels from sky and load outputs obtained by
averaging the two time streams, i.e., $R = \langle V_{\rm
sky}\rangle/\langle V_{\rm load}\rangle$.  Then
\begin{equation}
\Delta V(t) = V_{\rm sky}(t) - \frac{\langle V_{\rm sky}\rangle}{\langle V_{\rm load}\rangle} V_{\rm load}(t)\, .
\label{requation}
\end{equation}
$R$ is computed from unflagged data for each pointing period and
then applied to create the differenced timelines. The $R$ factor
has been stable over the mission so far, with overall variations
of 0.03--0.04\,\%.  A full discussion regarding the theory of this
value is reported in \citet{planck2011-1.4}.

%% file: 04-05_comb_diodes.tex
The receiver architecture is symmetric, with two complementary
detector diodes providing output for each receiver channel. As
described in \citet{seiffert2002} and \citet{mennella2010},
imperfect matching of components limits the isolation between the
complementary diodes of the receivers to between $-10$ and
$-15$\,dB. This imperfect isolation leads to a small
anticorrelated component in the white noise. We perform a weighted
average of the time-ordered data from the two diodes of each
receiver just before the differentiation. This avoids the
complication of tracking the anticorrelated white noise throughout
the subsequent analysis. We treat the combined diode data as the
raw data, and calibration, noise estimation, mapmaking etc. are
performed on these combined data. We use inverse noise weights
determined from an initial estimate of the calibrated noise for
each detector. The weights, reported in
Table~\ref{tab_diode_weights}, are kept fixed for the entire
mission.

\begin{table}[tmb]
  \begingroup
  \newdimen\tblskip \tblskip=5pt
  \caption{Weights used in combining diodes$^{\rm a}$.}
  \label{tab_diode_weights}
  \nointerlineskip
  \vskip -3mm
  \footnotesize
  \setbox\tablebox=\vbox{
    \newdimen\digitwidth
    \setbox0=\hbox{\rm 0}
    \digitwidth=\wd0
    \catcode`*=\active
    \def*{\kern\digitwidth}
    \newdimen\signwidth
    \setbox0=\hbox{+}
    \signwidth=\wd0
    \catcode`!=\active
    \def!{\kern\signwidth}
    \halign{\hbox to 1.0in{#\leaderfil}\tabskip=2em&
      \hfil#\hfil&
            \hfil#\hfil&
      \hfil#\hfil&
      \hfil#\hfil\tabskip=0pt\cr                            
      \noalign{\doubleline}
      \omit&\multispan4\hfil Diode\hfil\cr
      \noalign{\vskip -3pt}
      \omit&\multispan4\hrulefill\cr
      \noalign{\vskip 2pt}
      \omit\hfil Radiometer\hfil& M-00& M-01& S-10& S-11\cr
      \noalign{\vskip 3pt\hrule\vskip 5pt}
      LFI-18& 0.567& 0.433& 0.387& 0.613\cr
      LFI-19& 0.502& 0.498& 0.551& 0.449\cr
      LFI-20& 0.523& 0.477& 0.477& 0.523\cr
      LFI-21& 0.500& 0.500& 0.564& 0.436\cr
      LFI-22& 0.536& 0.464& 0.554& 0.446\cr
      LFI-23& 0.508& 0.492& 0.362& 0.638\cr
      LFI-24& 0.602& 0.398& 0.456& 0.544\cr
      LFI-25& 0.482& 0.518& 0.370& 0.630\cr
      LFI-26& 0.593& 0.407& 0.424& 0.576\cr
      LFI-27& 0.520& 0.480& 0.485& 0.515\cr
      LFI-28& 0.553& 0.447& 0.468& 0.532\cr
      \noalign{\vskip 5pt\hrule\vskip 3pt}
    }}
  \endPlancktable
  \par \tablenote a A perfect instrument would have weights of 0.500 for both
  diodes. \par
  \endgroup
\end{table}

%% file: 05-00_pointing.tex
\def\beamfwhm{b}
\def\Pointing{\hat{\vec{P}}}
\def\PointingRad{\hat{\vec{P}}_{\mathrm{rad}}}
\def\RotEclBody{\mathcal{R}_{\mathrm{Ecl},\mathrm{Body}}}
\def\RotBodyBeam{\mathcal{R}_{\mathrm{Body},\mathrm{rad}}}
\def\eVersorZ{\hat{\vec{e}}_{\mathrm z}}
\def\APointing{\hat{\vec{P}}'}
\def\deflection{\delta_{\mathrm P}}
\def\Vel{\mathbf{v}_{\mathrm{Planck}}}
\def\GalLon{l_{\mathrm{Gal}}}
\def\GalLat{b_{\mathrm{Gal}}}

\def\psiOne{\psi_1}
\def\psiTwo{\psi_2}
\def\psiThree{\psi_3}
\def\FpsiTwo{\tilde{\psi}_{2}}
\def\RotEclSTR{\mathcal{R}_{\mathrm{Ecl},\mathrm{STR}}}
\def\RotPABody{\mathcal{R}_{\mathrm{PA},\mathrm{Body}}}
\def\RotEclPA{\mathcal{R}_{\mathrm{Ecl},\mathrm{PA}}}
\def\RotBodySTR{\mathcal{R}_{\mathrm{Body},\mathrm{STR}}}

Proper pointing reconstruction is critical and has a direct
impact in the determination of an accurate photometric
calibration. The pointing for each radiometer $\PointingRad(t)$ at
time $t$ is given by
\begin{equation}
\PointingRad(t)=\RotEclBody(t)\,\RotBodyBeam\eVersorZ.
\label{eq:pointing}
\end{equation}

\noindent The $\RotBodyBeam$ matrix encodes the orientation of the
beam pattern with respect to the body reference frame defined by
the spacecraft structure.  We adopt the convention that in the
reference frame of the beam, the optical axis is aligned with
$\eVersorZ$.  $\RotBodyBeam$ is parameterized by a set of rotation
angles in the RIMO derived from flight data and ground-based
measurements. $\RotEclBody(t)$ is derived by time interpolation of
quaternions distributed in the attitude history files, it encodes
the orientation of the spacecraft body with respect to the
reference frame . The spacecraft attitude is determined from
\Planck\ star tracker data, and during periods of stability
between maneuvers is sampled at 8\,Hz, much lower than the LFI
sampling frequency. Equation~\ref{eq:pointing} incorporates a
large amount of information on the satellite and a long chain of
transformations between reference frames, each one being a
possible source of systematic error. Indeed, even a small
aberration compared to the beam size can introduce significant
photometric effects if the gradient of the temperature field is
large enough. The two most important sources of aberration
identified and corrected are stellar aberration and the apparent
change in wobble angles likely produced by thermal deformations of
the star tracker support.

\subsection{Stellar aberration}

The star tracker system is the basis for the reconstructed
astrometric attitude of the \Planck\ spacecraft in the solar
system barycentric reference frame; however, the effective
pointing direction is affected by stellar aberration due to the
orbital motion of \Planck\ and the finite speed of light.  In the
non-relativistic case, stellar aberration is given by
\begin{equation}
  \APointing = (\Pointing + \Vel/c)/\left|\Pointing + \Vel/c\right|,
\end{equation}
where $\APointing$ is the aberrated pointing direction, $\Vel$ is
the orbital velocity of \Planck\ in the solar system barycentric
frame, and $c$ is the speed of light.  From this formula, the
deflection angle $\deflection = \arccos(\APointing\cdot\Pointing)$
can be derived. \Planck\ moves at about 30\,km\,s$^{-1}$ in the
ecliptic plane, and scan circles are nearly normal to it.
Therefore $\deflection \le 20.6$\arcs, and  the greatest
deflection occurs near the ecliptic poles.  If left uncorrected,
this aberration would distort the maps, producing a seasonal shift
near the equator and a blurring near the ecliptic poles.  Accurate
simulations show that the distortion radius is maximal at the
ecliptic poles, $(\GalLon,\GalLat)=(96\pdeg384, 29\pdeg811)$ and
$(276\pdeg384, -29\pdeg811)$, and that it decreases towards the
ecliptic down to a minimum of about 0\pdeg1 on the ecliptic. The
boundary of the region in which the distortion radius is at least
half the polar value is roughly a ring centred on the poles, with
radius about $60\deg$. There are some variations in the radius and
in the longitudinal shape of the boundary, both smaller than a few
degrees, due to the scanning strategy, and also to the different
angular distances from the spin axis of the various feedhorns.

\subsection{Wobble angles}

Wobble angles describe the unavoidable misalignment of the body
reference frame with respect to the reference frame defined by the
satellite principal inertial axis.
The nominal spin axis for the satellite is nearly 0\pdeg5
away from the principal moment of inertia, and the effective scan circles are about half
a degree smaller than the nominal ones \citep{planck2011-1.1}.
Wobble angles and their variations in time, either real or apparent,
are measured by careful modelling of the observed
\Planck\ attitude dynamics included in the attitude history files.
\citet{planck2011-1.1} reported an apparent
variation of the wobble angles likely produced by thermoelastic deformations
that change the relative orientation of the star tracker with respect to
the body reference frame.  The change was detected in scans of Jupiter.
Since this variation is rigidly transported by the rotations of
spacecraft  body, its effect will be largely averaged out near the
poles and will be maximal near the ecliptic, the opposite of the
stellar aberration effect.

Of the three angles that describe the wobble, $\psiOne$ has
largely negligible effects and $\psiThree$ is badly determined, so
the LFI pipeline corrects only for variations in $\psiTwo$, whose
effect is apparent changes of the angular distance between the
telescope and the spin axis. Typical changes of this angle are
equivalent to apparent changes of scan circle radii of
$\pm0\parcm1$, giving equivalent displacements in pointing between
consecutive surveys of $0\parcm2$.

%% file: 06-00_beamrecovery.tex
The profiles and locations of the beams are determined from the
four observations of Jupiter listed in Table~\ref{tab:ods},
following the procedure described in \citet{planck2011-1.6} and
\citet{planck2011-1.4}.  Details are given in
\citet{planck2013-p02d}.  The origin of the focal plane is the
optical line of sight defined in \cite{tauber2010b}. The LFI beam
centers are given by four numbers, $\theta_{uv}$, $\phi_{uv}$,
$\psi_{\mathrm  uv}$, and $\psi_{\rm pol}$ (see
\citealt{planck2013-p28} for the definitions of these angles).
Only $\theta_{uv}$ and $\phi_{uv}$, which are the beam pointing in
spherical coordinates referred to the line of sight, can be
determined with Jupiter observations.  The polarization
orientation of the beams, defined by $\psi_{uv} + \psi_{\rm pol}$,
is not estimated from flight data but is derived from main beam
simulations based on ground measurements.

\begin{table}[tmb]
  \begingroup
  \newdimen\tblskip \tblskip=5pt
  \caption{Approximate dates of the Jupiter observations$^{\rm a}$.}
  \label{tab:ods}
  \nointerlineskip
  \vskip -3mm
  \footnotesize
  \setbox\tablebox=\vbox{
    \newdimen\digitwidth
    \setbox0=\hbox{\rm 0}
    \digitwidth=\wd0
    \catcode`*=\active
    \def*{\kern\digitwidth}
    \newdimen\signwidth
    \setbox0=\hbox{+}
    \signwidth=\wd0
    \catcode`!=\active
    \def!{\kern\signwidth}
    \halign{\hbox to 1.0in{#\leaderfil}\tabskip=1em&
      \hfil#\hfil&
      \hfil#\hfil&
      \hfil#\hfil\tabskip=0pt\cr                            
      \noalign{\doubleline}
      \omit\hfil Jupiter transit \hfil& Date& OD\cr
      \noalign{\vskip 3pt\hrule\vskip 5pt}
      Scan 1 (J1)& 21/10/2009 -- 05/11/2009 & 161 -- 176\cr
      Scan 2 (J2)& 27/06/2010 -- 12/07/2010 & 410 -- 425\cr
      Scan 3 (J3)& 03/12/2010 -- 18/12/2010 & 569 -- 584\cr
      Scan 4 (J4)& 30/07/2011 -- 08/08/2011 & 808 -- 817\cr
      \noalign{\vskip 5pt\hrule\vskip 3pt}
    }}
  \endPlancktable
  \par \tablenote a The periods include the scan by the entire LFI field of view. \par
  \endgroup
\end{table}

For each beam, the pointing is determined by the location of the maximum of an
elliptical Gaussian fit to that beam. This was done for each beam in each
single scan.  Results are reported, with errors, in
\cite{planck2013-p02d}.

\begin{figure}[!hb]
\centering\includegraphics[width=8.8cm]{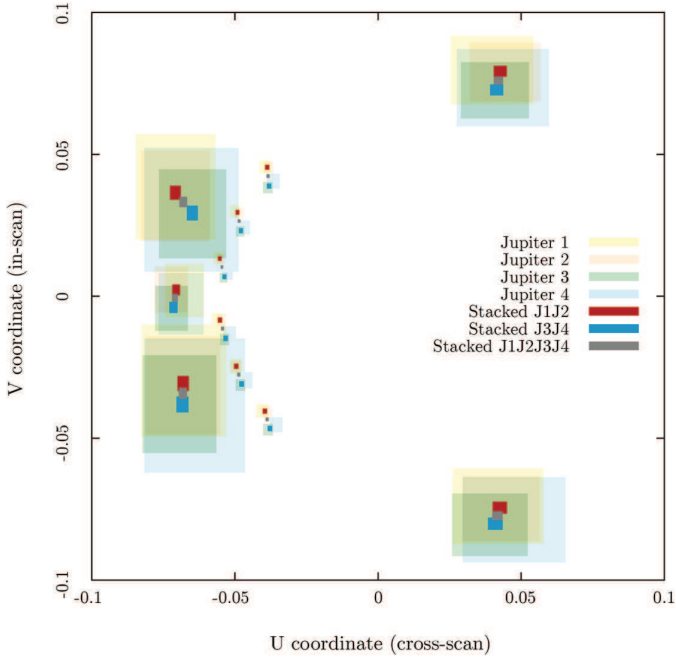}
\caption{Main beam pointing directions measured with the first
four Jupiter crossings.  Single scans are yellow, light red,
green, and light blue. First and second stacked scans are red,
third and fourth stacked scans are blue, and four stacked scans
are grey. The colored boxes refer to the measured uncertainties
magnified by a factor of 100. The differences in pointing were
normalized to the J1 measurements, and were magnified by the same
factor of 100. The $U$ and $V$ axis are defined as
$U=\sin(\theta)\cos(\phi)$ and $V=\sin(\theta)\sin(\phi)$, where
$\theta$ and $\phi$ are the angle respect the LOS (line of sight)
defined in \cite{tauber2010b}.} \label{fig:fpg}
\end{figure}

In addition, the beams are stacked in pairs (J1J2 and J3J4) and
all together (J1J2J3J4) in order to improve the signal-to-noise
ratio of the measurements. Before the stacking, each beam is
artificially repointed along the direction given by the arithmetic
average of the center of each beam to be stacked. Then a fit is
performed again on the stacked beams and the resulting parameters
recorded. For single scans it has been found that there is an
agreement within $2$\arcs\ in the pointing direction between J1
and J2. The same agreement occurs between J3 and J4. In contrast,
a $\sim15$\arcs\ systematic deviation of the beam center was
detected when comparing J1J2 to J3J4. Figure~\ref{fig:fpg} shows
the reconstructed beam positions and errors in the line-of-sight
frame magnified by a factor of 100. The shift is evident for the
70\,GHz beams, as well as in all the J1J2 and J3J4 stacked beam
centers. The change in the location has been found mainly in the
scan direction (i.e., $v$-coordinate). To account for this
pointing shift, we apply two pointing solutions for LFI. The first
focal plane calibration is valid from OD91 to OD540 and is based
on the J1J2 beam pointing determination. The second calibration is
valid from OD541 to OD563 and is based on the J3J4 beam pointing
calibration. The reconstructed angles are reported in Table
\ref{tab_fpg}.

\begin{table*}[tmb]
  \begingroup
  \newdimen\tblskip \tblskip=5pt
  \caption{Focal plane geometry}
  \label{tab_fpg}
  \nointerlineskip
  \vskip -3mm
  \footnotesize
  \setbox\tablebox=\vbox{
    \newdimen\digitwidth
    \setbox0=\hbox{\rm 0}
    \digitwidth=\wd0
    \catcode`*=\active
    \def*{\kern\digitwidth}
    \newdimen\signwidth
    \setbox0=\hbox{+}
    \signwidth=\wd0
    \catcode`!=\active
    \def!{\kern\signwidth}
    \halign{\hbox to 1.31in{#\leaderfil}\tabskip=2em&
      \hfil#\hfil\tabskip=1em&
      \hfil#\hfil\tabskip=2.8em&
      \hfil#\hfil\tabskip=1em&
      \hfil#\hfil\tabskip=2em&
      \hfil#\hfil\tabskip=1em&
      \hfil#\hfil\tabskip=0pt\cr                            
      \noalign{\doubleline}
      \omit\hfil Radiometer\hfil& $\theta_{uv}$\rlap{$^{\rm a}$}& $\phi_{uv}$\rlap{$^{\rm a}$}&
                                  $\theta_{uv}$\rlap{$^{\rm b}$}& $\phi_{uv}$\rlap{$^{\rm b}$}&
                                  $\psi_{uv}$\rlap{$^{\rm c}$}& $\psi_{\rm pol}$\rlap{$^{\rm c}$}\cr
      \noalign{\vskip 3pt\hrule\vskip 5pt}
      LFI-18S& 3.334& $-$131.803& 3.335& $-$131.752&   !*22.2& *0.0\cr
      LFI-18M& 3.333& $-$131.812& 3.335& $-$131.759&   !*22.2& 90.2\cr
      LFI-19S& 3.208& $-$150.472& 3.209& $-$150.408&   !*22.4& *0.0\cr
      LFI-19M& 3.208& $-$150.467& 3.209& $-$150.402&   !*22.4& 90.0\cr
      LFI-20S& 3.183& $-$168.189& 3.183& $-$168.121&   !*22.4& *0.0\cr
      LFI-20M& 3.183& $-$168.178& 3.183& $-$168.109&   !*22.4& 89.9\cr
      LFI-21S& 3.184&   !169.265& 3.182&   !169.324& *$-$22.4& *0.0\cr
      LFI-21M& 3.184&   !169.274& 3.183&   !169.336& *$-$22.4& 90.1\cr
      LFI-22S& 3.172&   !151.352& 3.170&   !151.405& *$-$22.4& *0.1\cr
      LFI-22M& 3.172&   !151.345& 3.170&   !151.398& *$-$22.4& 90.1\cr
      LFI-23S& 3.280&   !132.255& 3.277&   !132.287& *$-$22.1& *0.0\cr
      LFI-23M& 3.280&   !132.234& 3.277&   !132.274& *$-$22.1& 89.7\cr
      LFI-24S& 4.070& $-$179.506& 4.069& $-$179.449&   !**0.0& *0.0\cr
      LFI-24M& 4.070& $-$179.538& 4.071& $-$179.488&   !**0.0& 90.0\cr
      LFI-25S& 4.984&   !*61.105& 4.981&   !*61.084& $-$113.2& *0.0\cr
      LFI-25M& 4.985&   !*61.065& 4.981&   !*61.051& $-$113.2& 89.5\cr
      LFI-26S& 5.037& *$-$61.662& 5.040& *$-$61.669&   !113.2& *0.0\cr
      LFI-26M& 5.037& *$-$61.649& 5.040& *$-$61.676&   !113.2& 90.5\cr
      LFI-27S& 4.343&   !153.958& 4.343&   !154.033& *$-$22.3& *0.0\cr
      LFI-27M& 4.345&   !153.981& 4.341&   !154.010& *$-$22.3& 89.7\cr
      LFI-28S& 4.374& $-$153.413& 4.376& $-$153.369&   !*22.3& *0.0\cr
      LFI-28M& 4.374& $-$153.419& 4.376& $-$153.371&   !*22.3& 90.3\cr
      \noalign{\vskip 5pt\hrule\vskip 3pt}
    }}
  \endPlancktablewide
  \tablenote a Beam pointing reconstructed using the first two Jupiter transits (J1 and J2). \par
  \tablenote b Beam pointing reconstructed using the last two Jupiter transits (J3 and J4). \par
  \tablenote c Polarization orientation of the beam measured during ground test. \par
  \endgroup
\end{table*}

%% file: 06-01_scanningbeam.tex
Scanning beams are defined as the beams measured in flight on
planets. The scanning beam derives from the optical beam coupled
with the radiometer response, and smeared by the satellite motion.
With four Jupiter transit measurements we were able to reconstruct
the beam shape down to $-20$\,dB from the peak at $30$ and
$44$\,GHz, and down to $-25$\,dB at $70$\,GHz. From the beam shape
we estimated the main beam parameters using a bivariate Gaussian
fit on the four stacked beams (J1J2J3J4). The fitting procedure,
described in \citet{planck2011-1.6}, was slightly modified to
correct for offsets in the data and to avoid noise contamination.
We refer to the companion paper on LFI beams
\citep{planck2013-p02d} for details on procedures and results.
Table~\ref{tab_beam-parameters} gives the average values of the
FWHM and ellipticity, with errors.

  \begin{table}
  \begingroup
  \newdimen\tblskip \tblskip=5pt
  \caption{LFI beam FWHM and ellipticity measured in flight from four Jupiter passes$^{\rm a}$.}
  \label{tab_beam-parameters}
  \nointerlineskip
  \vskip -3mm
  \footnotesize
  \setbox\tablebox=\vbox{
    \newdimen\digitwidth
    \setbox0=\hbox{\rm 0}
    \digitwidth=\wd0
    \catcode`*=\active
    \def*{\kern\digitwidth}
    \newdimen\signwidth
    \setbox0=\hbox{+}
    \signwidth=\wd0
    \catcode`!=\active
    \def!{\kern\signwidth}
  \halign{\hbox to 1.2in{#\leaderfil}\tabskip=2.5em&
      \hfil$#$\hfil&
      \hfil$#$\hfil\tabskip=0pt\cr
  \noalign{\doubleline}
  \omit&\omit\hfil FWHM$^{\rm b}$\hfil&\cr
  \omit\hfil Beam\hfil&[\rm arcmin]&\omit\hfil Ellipticity$^{\rm c}$\hfil\cr
  \noalign{\vskip 3pt\hrule\vskip 5pt}
  {\bf 70\,GHz mean}&\getsymbol{LFI:FWHM:70GHz}&\getsymbol{LFI:ellipticity:70GHz}\cr
  \noalign{\vskip 4pt}
  \hglue 2em  LFI-18 &\getsymbol{LFI:FWHM:LFI18}\pm\getsymbol{LFI:FWHM:uncertainty:LFI18}&\getsymbol{LFI:ellipticity:LFI18}\pm\getsymbol{LFI:ellipticity:uncertainty:LFI18}\cr
  \hglue 2em  LFI-19 &\getsymbol{LFI:FWHM:LFI19}\pm\getsymbol{LFI:FWHM:uncertainty:LFI19}&\getsymbol{LFI:ellipticity:LFI19}\pm\getsymbol{LFI:ellipticity:uncertainty:LFI19}\cr
  \hglue 2em  LFI-20 &\getsymbol{LFI:FWHM:LFI20}\pm\getsymbol{LFI:FWHM:uncertainty:LFI20}&\getsymbol{LFI:ellipticity:LFI20}\pm\getsymbol{LFI:ellipticity:uncertainty:LFI20}\cr
  \hglue 2em  LFI-21 &\getsymbol{LFI:FWHM:LFI21}\pm\getsymbol{LFI:FWHM:uncertainty:LFI21}&\getsymbol{LFI:ellipticity:LFI21}\pm\getsymbol{LFI:ellipticity:uncertainty:LFI21}\cr
  \hglue 2em  LFI-22 &\getsymbol{LFI:FWHM:LFI22}\pm\getsymbol{LFI:FWHM:uncertainty:LFI22}&\getsymbol{LFI:ellipticity:LFI22}\pm\getsymbol{LFI:ellipticity:uncertainty:LFI22}\cr
  \hglue 2em  LFI-23 &\getsymbol{LFI:FWHM:LFI23}\pm\getsymbol{LFI:FWHM:uncertainty:LFI23}&\getsymbol{LFI:ellipticity:LFI23}\pm\getsymbol{LFI:ellipticity:uncertainty:LFI23}\cr
  \noalign{\vskip 5pt}
  {\bf 44\,GHz mean}&\getsymbol{LFI:FWHM:44GHz}&\getsymbol{LFI:ellipticity:44GHz}\cr
  \noalign{\vskip 4pt}
  \hglue 2em  LFI-24&\getsymbol{LFI:FWHM:LFI24}\pm\getsymbol{LFI:FWHM:uncertainty:LFI24}&\getsymbol{LFI:ellipticity:LFI24}\pm\getsymbol{LFI:ellipticity:uncertainty:LFI24}\cr
  \hglue 2em  LFI-25&\getsymbol{LFI:FWHM:LFI25}\pm\getsymbol{LFI:FWHM:uncertainty:LFI25}&\getsymbol{LFI:ellipticity:LFI25}\pm\getsymbol{LFI:ellipticity:uncertainty:LFI25}\cr
  \hglue 2em  LFI-26&\getsymbol{LFI:FWHM:LFI26}\pm\getsymbol{LFI:FWHM:uncertainty:LFI26}&\getsymbol{LFI:ellipticity:LFI26}\pm\getsymbol{LFI:ellipticity:uncertainty:LFI26}\cr
  \noalign{\vskip 5pt}
  {\bf 30\,GHz mean}&\getsymbol{LFI:FWHM:30GHz}&\getsymbol{LFI:ellipticity:30GHz}\cr
  \noalign{\vskip 4pt}
  \hglue 2em  LFI-27&\getsymbol{LFI:FWHM:LFI27}\pm\getsymbol{LFI:FWHM:uncertainty:LFI27}&\getsymbol{LFI:ellipticity:LFI27}\pm\getsymbol{LFI:ellipticity:uncertainty:LFI27}\cr
  \hglue 2em  LFI-28&\getsymbol{LFI:FWHM:LFI28}\pm\getsymbol{LFI:FWHM:uncertainty:LFI28}&\getsymbol{LFI:ellipticity:LFI28}\pm\getsymbol{LFI:ellipticity:uncertainty:LFI28}\cr
  \noalign{\vskip 5pt\hrule\vskip 3pt}}}
  \endPlancktable
  \tablenote a Uncertainties are the standard deviation of the mean of the $1\sigma$ statistical uncertainties of the fit.  A small difference is expected between the {\tt M} and {\tt S} beams, caused by optics and receiver
  non-idealities.\par
  \tablenote b The square root of the product of the major axis and minor axis FWHMs of the individual horn beams, averaged between {\tt M} and {\tt S} radiometers.\par
  \tablenote c Ratio of the major and minor axes of the fitted elliptical Gaussian.\par
  \endgroup
  \end{table}

%% file: 06-02_effectivebeam.tex
The effective beam at a given pixel in a map of the sky is the
average of all scanning beams that observed that pixel during the
observing period of the map given the \Planck\ scan strategy. We
compute the effective beam at each LFI frequency scanning beam and
scan history in real space using the {\tt FEBeCoP}
\citep{mitra2010} method. Details of the application of {\tt
FEBeCoP} to \Planck\ data will be discussed in a future paper.
Effective beams were used to calculate the effective beam window
function as reported in \cite{planck2013-p02d} and in the source
detection pipeline necessary to generate the PCCS catalogue
~\citep{planck2013-p05}. Table~\ref{tab_eff} lists the mean and
rms variation across the sky of the main parameters computed with
{\tt FEBeCoP}. Note that the FWHM and ellipticity in
Table~\ref{tab_eff} differ slightly from the values reported in
Table~\ref{tab_beam-parameters}.  This results from the different
way in which the Gaussian fit was applied. The scanning beam fit
was determined by fitting the profile of Jupiter on timelines and
limiting the fit to the data with signal-to-noise ratio greater
than 3, while the fit of the effective beam was computed on {\tt
GRASP} maps projected in several positions of the sky
~\citep{planck2013-p02d}.  The latter are less affected by the
noise.

\begin{table*}[tmb]
  \begingroup
  \newdimen\tblskip \tblskip=5pt
  \caption{Mean and rms variation across the sky of FWHM, ellipticity, orientation, and solid angle of the {\tt FEBeCop} effective beams computed with the {\tt GRASP} beam fitted scanning beams$^{\rm a}$.}
  \label{tab_eff}
  \nointerlineskip
  \vskip -3mm
  \footnotesize
  \setbox\tablebox=\vbox{
    \newdimen\digitwidth
    \setbox0=\hbox{\rm 0}
    \digitwidth=\wd0
    \catcode`*=\active
    \def*{\kern\digitwidth}
    \newdimen\signwidth
    \setbox0=\hbox{+}
    \signwidth=\wd0
    \catcode`!=\active
    \def!{\kern\signwidth}
    \halign{\hbox to 0.9 in{#\leaderfil}\tabskip=2em&
      \hfil$#$\hfil&
      \hfil$#$\hfil&
      \hfil$#$\hfil&
      \hfil$#$\hfil&
      \hfil#\hfil\tabskip=0pt\cr                            
      \noalign{\doubleline}
\omit\hfil Frequency\hfil&\omit\hfil FWHM [arcmin]\hfil& e& \psi\,[\rm deg]& \Omega\,[\hbox{arcmin}^2]& FWHM$_{\mathrm {eff}}$\cr
      \noalign{\vskip 3pt\hrule\vskip 5pt}
      70& 13.252\pm0.033& 1.223\pm0.026& !0.587\pm55.066& *200.742\pm*1.027& 13.31\cr
      44& 27.005\pm0.552& 1.034\pm0.033& !0.059\pm53.767& *832.946\pm31.774& 27.12\cr
      30& 32.239\pm0.013& 1.320\pm0.031& -0.304\pm55.349& 1189.513\pm*0.842& 32.24\cr
      \noalign{\vskip 5pt\hrule\vskip 3pt}
    }}
  \endPlancktablewide
  \tablenote a FWHM$_{\mathrm {eff}}$ is the effective FWHM estimated from the main beam solid angle of the effective beam, $\Omega_{\mathrm {eff}} =
  {\mathrm{mean}}(\Omega)$.\par
  \endgroup
\end{table*}

%% file: 07-00_calibration.tex
Conversion of time-ordered streams of voltages into time-ordered streams of thermodynamic temperatures is modelled by
\begin{equation}
\label{eq:calibrationEquation}
V = G \times \bigl(T_\mathrm{sky} + T_\mathrm{noise}\bigr),
\end{equation}
where $V$ is the voltage measured by the ADC, $T_\mathrm{sky}$ is obtained by convolving the sky
temperature with the beam response of the instrument at a given
time, and $T_\mathrm{noise}$ is the noise temperature of the
radiometer.  In general, we are interested in $K = G^{-1}$, as the
purpose of the calibration is to convert $V$ back into a
temperature. As described in \citet{planck2013-p02b}, two
different algorithms are used for calibrating the LFI radiometers
in this data release:

\begin{enumerate}

\item For the 44 and 70\,GHz radiometers, we use a technique
called optimal search of gain, which is similar to the one used by
{\tt WMAP} \citep{hinshaw2009}.  It is based on fitting the
radiometric signal to the expected dipolar anisotropy induced by
the motion of the spacecraft with respect to the CMB rest frame.

\item For the 30\,GHz radiometers, we use a
technique that combines the knowledge of the dipolar anisotropy
(as above), then additionally takes into account the observed
fluctuations in the measurement of the signal of the 4\,K
reference loads.

\end{enumerate}

The overall accuracy in the calibration is reported in
Table~\ref{tab_summary_performance}. The reasons why we used two
different algorithms are discussed in \citet{planck2013-p02b}.
We describe the algorithms in the following sections.

%% file: 07-01_iterative_calib.tex
The main features of the iterative calibration algorithm used for 44 and 70\,GHz are the following:

\begin{enumerate}
\item We combine the speed of the spacecraft with respect to the Sun,
  $\vec{v}_\mathrm{Planck}$, and the speed of the Sun with respect
  to the CMB, $\vec{v}_\mathrm{Sun}$. The angle between the velocity vector and the axis of the relevant beam is $\theta$. The dipole is then evaluated considering the
  relativistic correction
  \begin{equation}
  \label{eq:dipoleAmplitudeKCMB}
    \Delta T =
  T_\mathrm{CMB} \left( \frac{1}{\gamma (1
  - \beta  \cos  \theta)} -1 \right),
  \end{equation}
where $T_\mathrm{CMB} = 2.7255\,\mathrm{K}$.
We produce discrete time ordered data (TOD) of the expected
overall dipole signal for each sample in a pointing period.

\item Using pointing information, we project both $V_i$ and
$\Delta T_i$ on a {\tt HEALPix} map ($N_{\rm side}=256$). Multiple
hits on the same pixels are averaged. The result is
a pair of maps, $V^\mathrm{map}_k$ and $\Delta T^\mathrm{map}_k$,
with $k$ being the pixel index\footnote{Most of the pixels in the
maps are not set, as during one pointing period the beam paints a
thin circle in the sky. We assume hereafter that the index $k$
runs only through the pixels which have been hit at least once.}.

\item We use weighted least squares to estimate $K = G\mo$ in
Eq.~\ref{eq:calibrationEquation} from the correlation between the
signal in volts, $V^\mathrm{sky}_k$, and $\Delta T^\mathrm{sky}_k$:
    \begin{equation}
    V^\mathrm{map}_k = K^\mathrm{dip}\,\Delta T^\mathrm{map}_k + \epsilon,
    \end{equation}
where $K$ and $\epsilon$ are the parameters used in the fit. Each
sample $k$ is weighted according to the number of hits per pixel.
In computing the fit, we use a frequency-dependent mask to avoid
those pixels where a strong non-Gaussian signal other than the
dipole is expected, i.e., point sources and the Galaxy.

\item The main source of uncertainties in the fit using the dipole
is the cosmological CMB signal itself. To improve the result, we
calibrate the data using $K_i$ and $\epsilon_i$, remove the dipole
convolved with the beam, and make a map, which represents an
estimation of the cosmological signal. To reduce the effect of
noise, we combine data streams from both radiometers of the same
horn. Then we remove the estimated cosmological signal from the data, make
a map using a simplified destriping algorithm, and use the results
to refine the values of $K_i$ and $\epsilon_i$. We iterate the procedure
until convergence. The result of this process is a set of gains,
$K^\mathrm{iter}_i$, and offsets, $\epsilon^\mathrm{iter}_i$.

\item An adaptive low-pass filter based on wavelets is applied to
the vectors $K^\mathrm{iter}_i$ and $\epsilon^\mathrm{iter}_i$ to reduce high-frequency noise, particularly near the
regions where the spacecraft is unfavorably aligned with the
dipole.
\end{enumerate}

%% file: 07-02_4k_calib.tex
To calibrate the 30\,GHz radiometers, we used a
different calibration scheme based on the signal measuring the
temperature of the 4\,K reference loads. This calibration has the
advantage of being less dependent on optical systematics such as
far sidelobes \citep{planck2013-p02a}, at the expense of being
more sensitive to systematics in the radiometers such as ADC
non-linearities \citep{planck2013-p02b}. The algorithm can be
summarized as follows:

\begin{enumerate}
\item For each pointing period $i$, a set of gains
$K^\mathrm{iter}_i$ is estimated using the iterative procedure
described in Sect.~\ref{sec_iterative_calib}.

\item The values of $K^\mathrm{iter}_i$ are used to estimate the value of the constant $K_0$ in the equation
\begin{equation}
K^\mathrm{iter}_i = K_0 \times \left(2 - \frac{V^\mathrm{ref}_i}{V^\mathrm{ref}_0}\right),
\end{equation}
where $V^\mathrm{ref}_i$ is the average value of the 4\,K reference load signal (in volts) over the $i$-th pointing period, and $V^\mathrm{ref}_0 = \langle V^\mathrm{ref}_i\rangle$ is a voltage representative of the value of $V^\mathrm{ref}_i$ over the \emph{whole} mission.  The constant $K_0$ is estimated using a weighted, one-parameter, linear least squares fit, where the weights are chosen to be proportional to the expected amplitude of the dipole-like signal in the sky, $\Delta T^\mathrm{dip}_i$, at the $i$-th pointing.

\item Using the value of $K_0$ estimated in the previous point, we extract a new set of gains $K^\mathrm{4,\mathrm{K}}_i$ with the equation
\begin{equation}
K^{4\,{\rm K}}_i \equiv K_0 \times \left(2 - \frac{V^\mathrm{ref}_i}{V^\mathrm{ref}_0}\right).
\end{equation}
\end{enumerate}

The procedure can be modelled by the following GNU R\footnote{\url{http://www.r-project.org/}.} code:
\begin{small}
\begin{verbatim}
data<-data.frame(gain = iterative.dipole.gains,
                 vref = 2 - signal.4K/mean(signal.4K),
                 dipole = dipole.amplitude.KCMB)
fit<-lm(gain ~ vref + 0, data, weights = dipole)
gains.4K <- fit$coefficients[1] * data$dvref
\end{verbatim}
\end{small}
where \texttt{iterative.dipole.gains}, \texttt{signal.4K}, and
\texttt{dipole.amplitude.KCMB} are three vectors containing the
iterative gains $K^\mathrm{iter}_i$ before the smoothing filter,
the 4\,K reference load signal $V^\mathrm{ref}_i$ averaged over
each pointing period, and the values of $\Delta T_i$
(Eq.~\ref{eq:dipoleAmplitudeKCMB}), respectively.

Unlike the procedure in Sect.~\ref{sec_iterative_calib}, in this case there is no need to smooth the stream of gains, as they share the  stability of the voltages $V^\mathrm{ref}_i$.

%% file: 07-03_colcorr.tex
Table~\ref{tbl:colourCorrections} gives color corrections
calculated following the method given in \citet{planck2013-p02b}.
Values for intermediate spectral indices can be derived by
interpolation.  The data release includes the {\tt UcCC} {\tt IDL}
package used by both LFI and HFI \citep{planck2013-p03d} that
calculates color corrections and unit conversions using the
band-averaged bandpass stored in the reduced instrument model
(RIMO) file, which is also included in the data release.

\begin{table*}[tmb]
\begingroup
\newdimen\tblskip \tblskip=5pt
\caption{Multiplicative color corrections cc$(\alpha)$ for
individual LFI Radiometer Chain Assemblies and for the band
average maps.} \label{tbl:colourCorrections} \nointerlineskip
\vskip -3mm \footnotesize \setbox\tablebox=\vbox{
   \newdimen\digitwidth
   \setbox0=\hbox{\rm 0}
   \digitwidth=\wd0
   \catcode`*=\active
   \def*{\kern\digitwidth}
   \newdimen\signwidth
   \setbox0=\hbox{+}
   \signwidth=\wd0
   \catcode`!=\active
   \def!{\kern\signwidth}
\halign{
\hbox to 3.0cm{#\leaderfil}\tabskip 1.0em& 
\hfil$#$\hfil&   
\hfil$#$\hfil&   
\hfil$#$\hfil&   
\hfil$#$\hfil&   
\hfil$#$\hfil&   
\hfil$#$\hfil&   
\hfil$#$\hfil&   
\hfil$#$\hfil&   
\hfil$#$\hfil&   
\hfil$#$\hfil&   
\hfil$#$\hfil&   
\hfil$#$\hfil&   
\hfil$#$\hfil\tabskip=0pt\cr 
\noalign{\doubleline}
\omit&\multispan{13}\hfil Spectral index $\alpha$\hfil\cr
\noalign{\vskip -3pt}
\omit&\multispan{13}\hrulefill\cr
\noalign{\vskip 3pt}
\omit\hfil Horn\hfil&-2.0& -1.5& -1.0& -0.5& 0.0& 0.5& 1.0& 1.5& 2.0& 2.5& 3.0& 3.5& 4.0\cr
\noalign{\vskip 3pt\hrule\vskip 5pt}
{\bf 70 GHz mean}& 0.938& 0.951& 0.963& 0.973& 0.982& 0.988& 0.994& 0.997& 0.999& 0.999& 0.998& 0.995& 0.991\cr
\noalign{\vskip 5pt}
\hglue 2em LFI-18& 0.948& 0.961& 0.972& 0.981& 0.988& 0.994& 0.997& 0.998& 0.997& 0.995& 0.990& 0.983& 0.975\cr
\hglue 2em LFI-19& 0.856& 0.878& 0.899& 0.919& 0.939& 0.957& 0.975& 0.991& 1.006& 1.020& 1.032& 1.043& 1.053\cr
\hglue 2em LFI-20& 0.889& 0.908& 0.925& 0.941& 0.956& 0.970& 0.983& 0.994& 1.003& 1.011& 1.018& 1.023& 1.027\cr
\hglue 2em LFI-21& 0.917& 0.933& 0.947& 0.960& 0.971& 0.981& 0.989& 0.996& 1.001& 1.004& 1.006& 1.006& 1.004\cr
\hglue 2em LFI-22& 1.024& 1.026& 1.027& 1.026& 1.023& 1.018& 1.011& 1.003& 0.993& 0.982& 0.969& 0.955& 0.940\cr
\hglue 2em LFI-23& 0.985& 0.991& 0.996& 0.999& 1.001& 1.002& 1.002& 1.000& 0.997& 0.993& 0.988& 0.982& 0.975\cr
\noalign{\vskip 10pt}
{\bf 44 GHz mean}& 0.968& 0.975& 0.981& 0.986& 0.990& 0.994& 0.997& 0.999& 1.000& 1.000& 0.999& 0.998& 0.995\cr
\noalign{\vskip 5pt}
\hglue 2em LFI-24& 0.978& 0.984& 0.988& 0.993& 0.996& 0.998& 0.999& 1.000& 0.999& 0.998& 0.996& 0.993& 0.989\cr
\hglue 2em LFI-25& 0.967& 0.974& 0.980& 0.985& 0.990& 0.994& 0.996& 0.999& 1.000& 1.000& 1.000& 0.999& 0.997\cr
\hglue 2em LFI-26& 0.957& 0.966& 0.973& 0.980& 0.985& 0.990& 0.995& 0.998& 1.000& 1.001& 1.002& 1.002& 1.000\cr
\noalign{\vskip 10pt}
{\bf 30 GHz mean}& 0.947& 0.959& 0.969& 0.977& 0.985& 0.991& 0.995& 0.998& 1.000& 1.000& 0.998& 0.994& 0.989\cr
\noalign{\vskip 5pt}
\hglue 2em LFI-27& 0.948& 0.959& 0.969& 0.978& 0.985& 0.991& 0.995& 0.998& 1.000& 1.000& 0.998& 0.995& 0.991\cr
\hglue 2em LFI-28& 0.946& 0.958& 0.968& 0.977& 0.985& 0.991& 0.996& 0.998& 1.000& 0.999& 0.997& 0.993& 0.988\cr
\noalign{\vskip 5pt\hrule\vskip 3pt}}}
\endPlancktablewide
\endgroup
\end{table*}

%% file: 08-00_noise.tex
The estimation of noise properties is fundamental in several
aspects of the data analysis. For instance, such measurements are
used in the Monte Carlo simulations of noise necessary for power
spectrum estimation, as well as to determine proper horn weights
to be employed during the map-making process. In addition,
inspection of noise properties throughout the mission lifetime is
of paramount importance in tracking possible variations and
anomalies in instrument performance.  Our noise estimation
pipeline has been improved over the log-periodogram approach used
in \citet{planck2011-1.6} by the implementation of a Markov chain
Monte Carlo (MCMC) approach for the extraction of basic noise
parameters. This allows for an unbiased estimate of the parameters
that characterize the non-white  noise.

\begin{figure*}
\includegraphics[width=6cm]{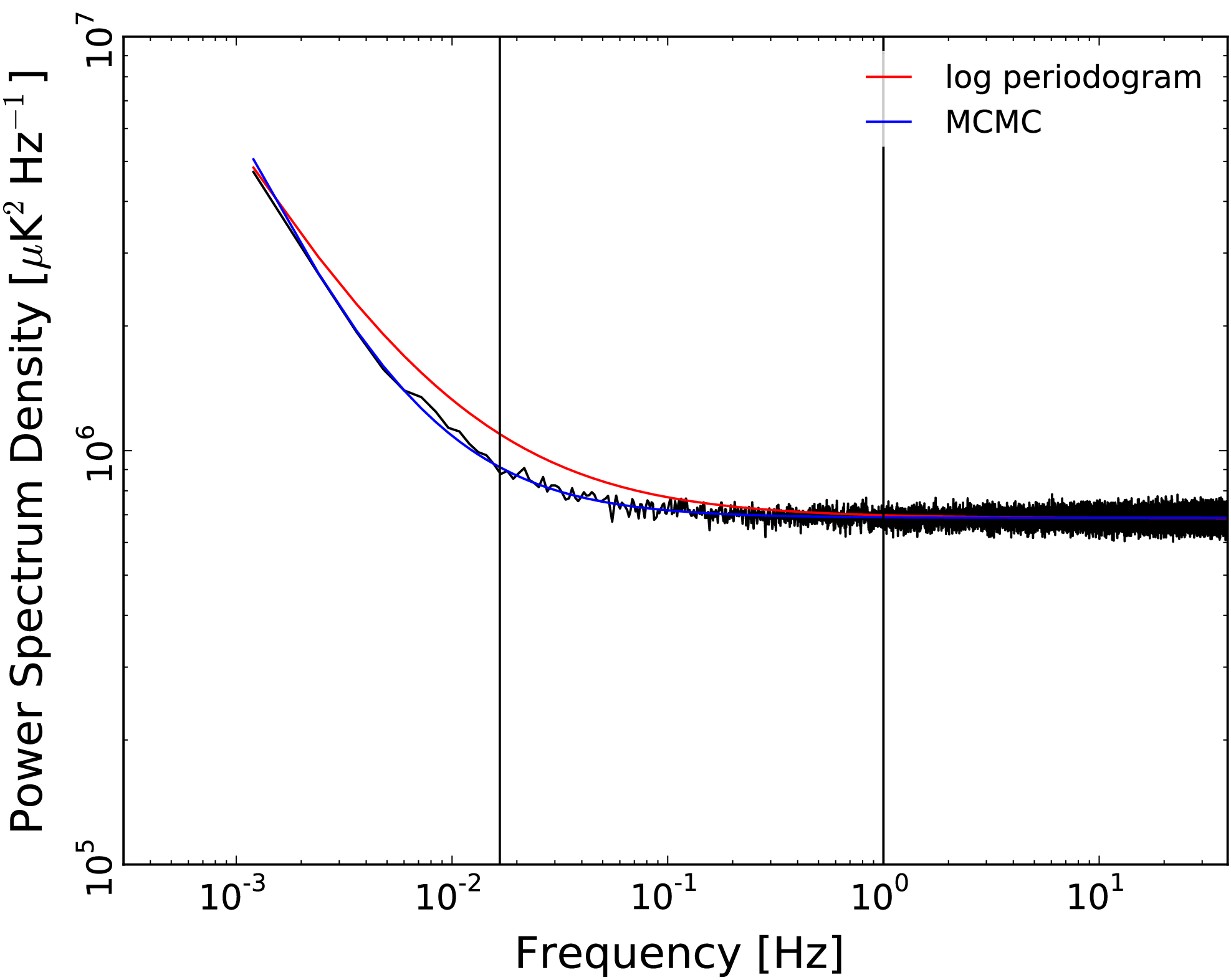}
\includegraphics[width=6cm]{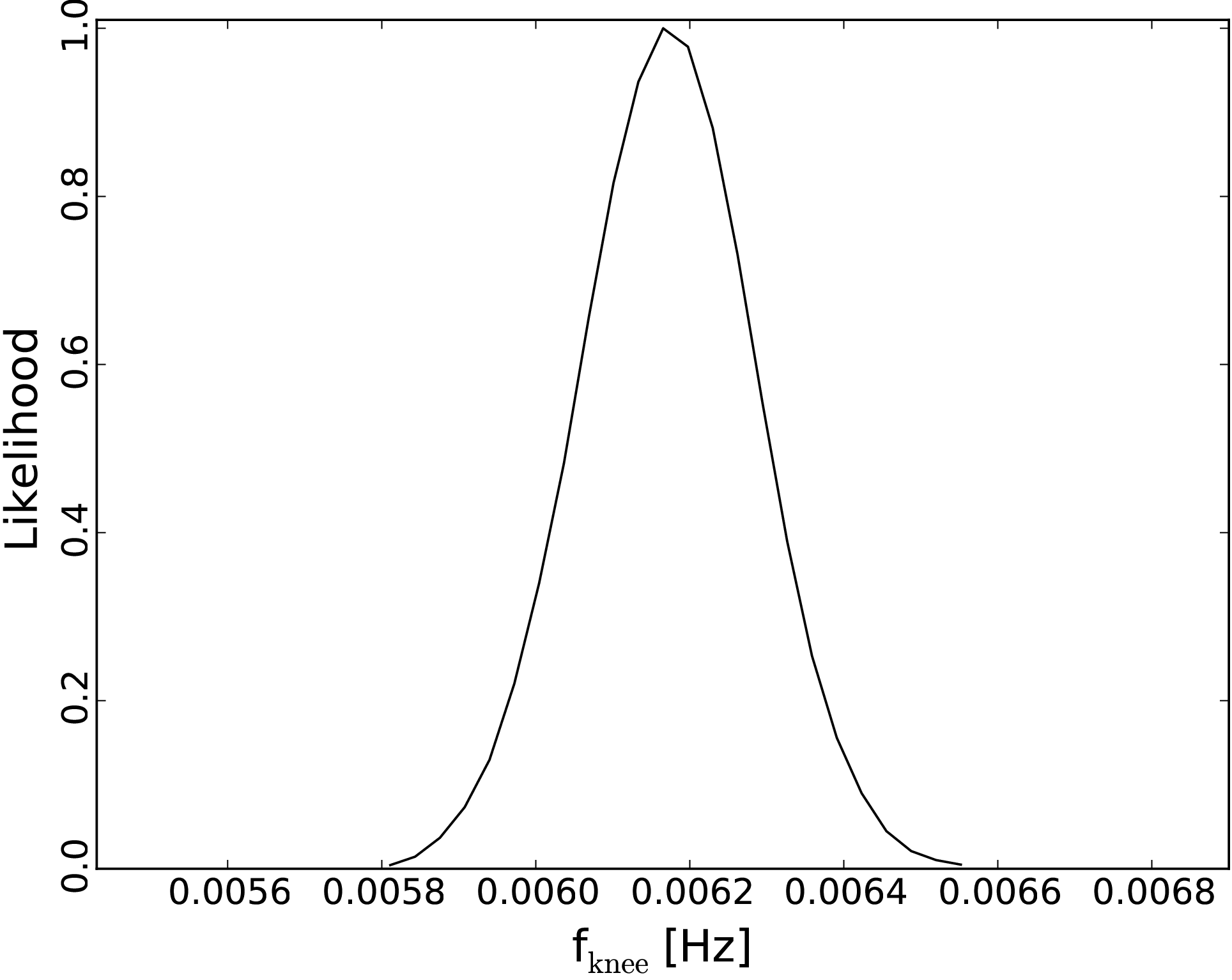}
\includegraphics[width=6cm]{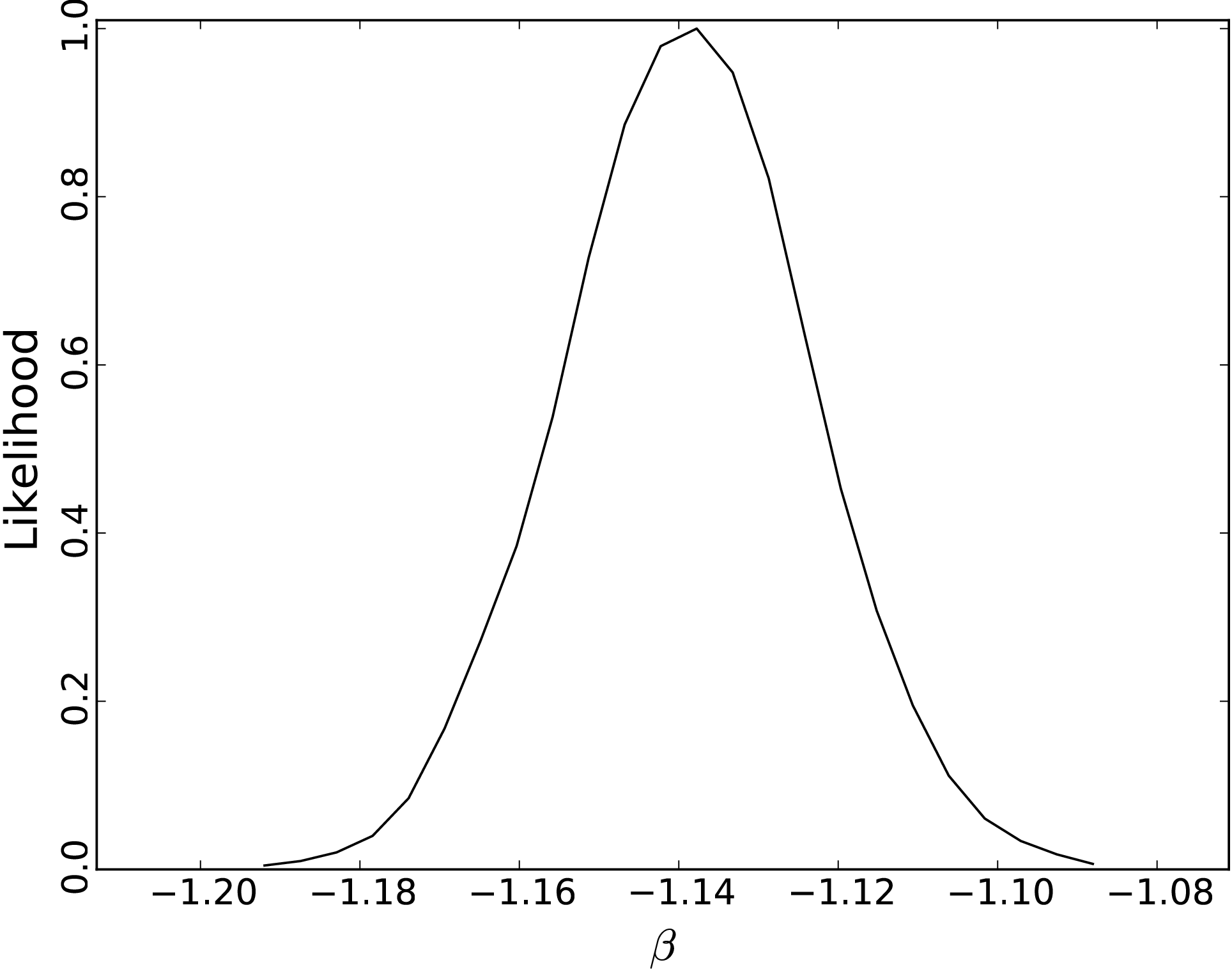}
\caption{{\it Left\/}: Typical noise spectrum at 70\,GHz, with ``old'' log-periodogram fit (red line) and  ``new'' MCMC fit (blue
  line). Vertical lines mark the frequencies corresponding to the spin period ($1/60$\,Hz) and the baseline used in mapmaking ($1$\,Hz). The corresponding distributions of knee-frequency and slope from the MCMC chains are shown in the center and right panels, respectively.
for the example spectrum.}
\label{fig_mcmc_noise}
\end{figure*}

We write the noise spectrum as
\begin{equation}
P(f) = \sigma^2\left[1 + \left(\frac{f}{f_\mathrm{knee}}\right)^\beta\right]\,,
\label{eq:noisemodel}
\end{equation}
where $\sigma^2$ is the white noise level, and $f_{\rm knee}$ and
$\beta$ characterize the non-white noise.  As before, $\sigma^2$
is calculated as the mean of the noise spectrum over the flat,
high-frequency tail (see Figs.~\ref{fig_mcmc_noise} and
\ref{fig_switchover_noise_spectra_rad}), typically over the
highest 10\% of frequency bins shown in the figures.  For the
30\,GHz radiometers, which have $f_{\rm knee} \approx 100$\,mHz, a
smaller percentage must be used to get an unbiased estimation.
Once white noise is computed, the code creates Markov chains for
the other parameters.  We get the expected value and variance of
each noise parameter from the chain distribution, ignoring the
burn-in period.

The left panel of Fig.~\ref{fig_mcmc_noise} shows a typical spectrum at 70\,GHz with the old log-periodogram fit (red line) and the new MCMC-derived spectrum (blue line) superimposed.
The center and right panels show the distributions of knee-frequency and slope from the MCMC chains.

%% file: 08-01_instrupdate_noise.tex
Radiometer noise properties have been evaluated using the new MCMC
just described. We select calibrated radiometer data in periods of
five days, and compute noise spectra with the {\tt roma} iterative
generalized least squares map-making algorithm
\citep{natoli2001,degasperis2005,prunet2001,planck2011-1.6}. The
output is a frequency spectrum to which the new MCMC code is
applied.  Results at radiometer level on white noise sensitivity
are reported in Table~\ref{tab_white_noise_per_radiometer}, while
Table~\ref{tab_one_over_f_noise_per_radiometer} shows $1/f$ noise
parameters.  These are computed taking the median of the ten
estimates made for different time ranges over the nominal mission.

\begin{figure*}[th]
\includegraphics[width=6cm]{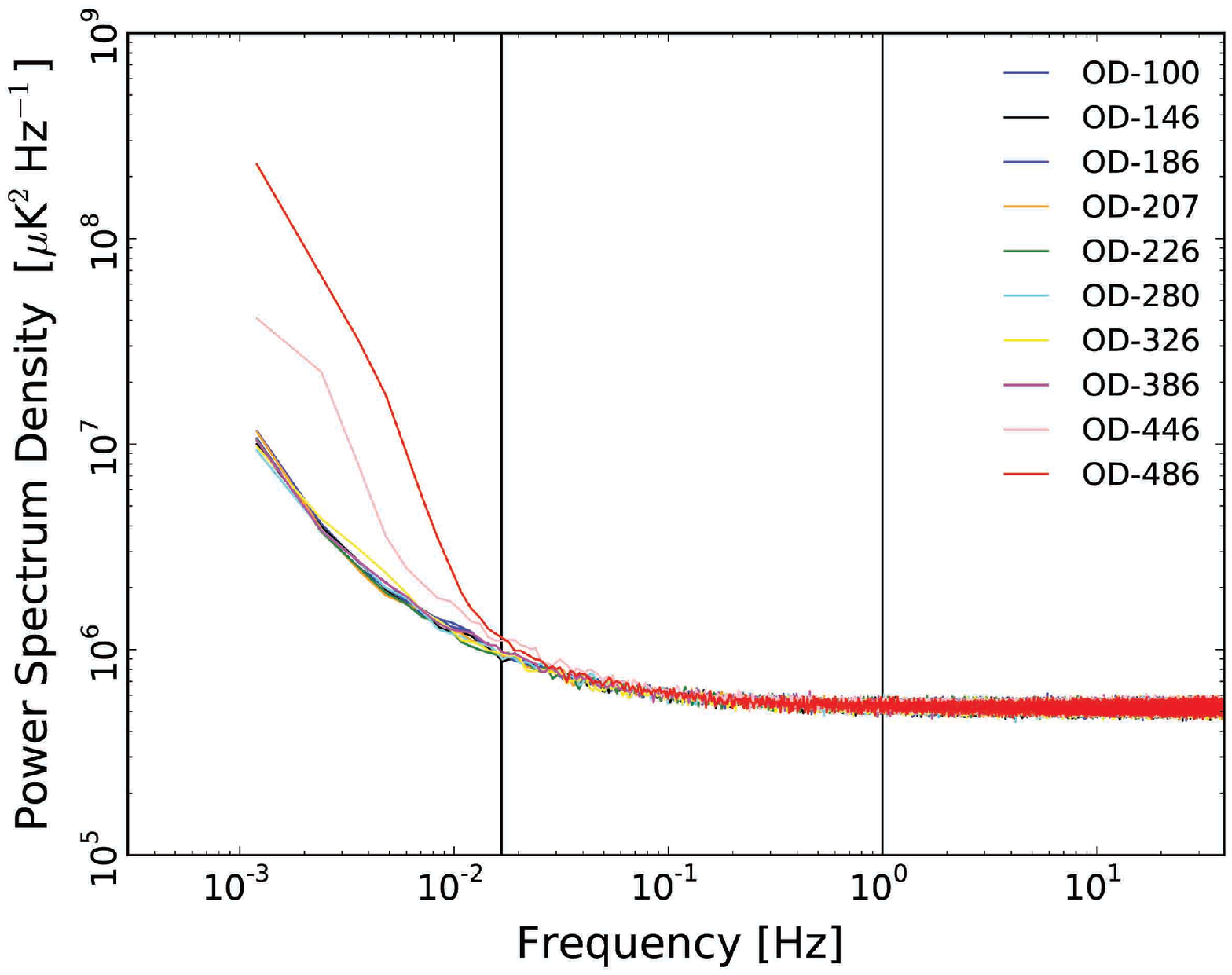}
\includegraphics[width=6cm]{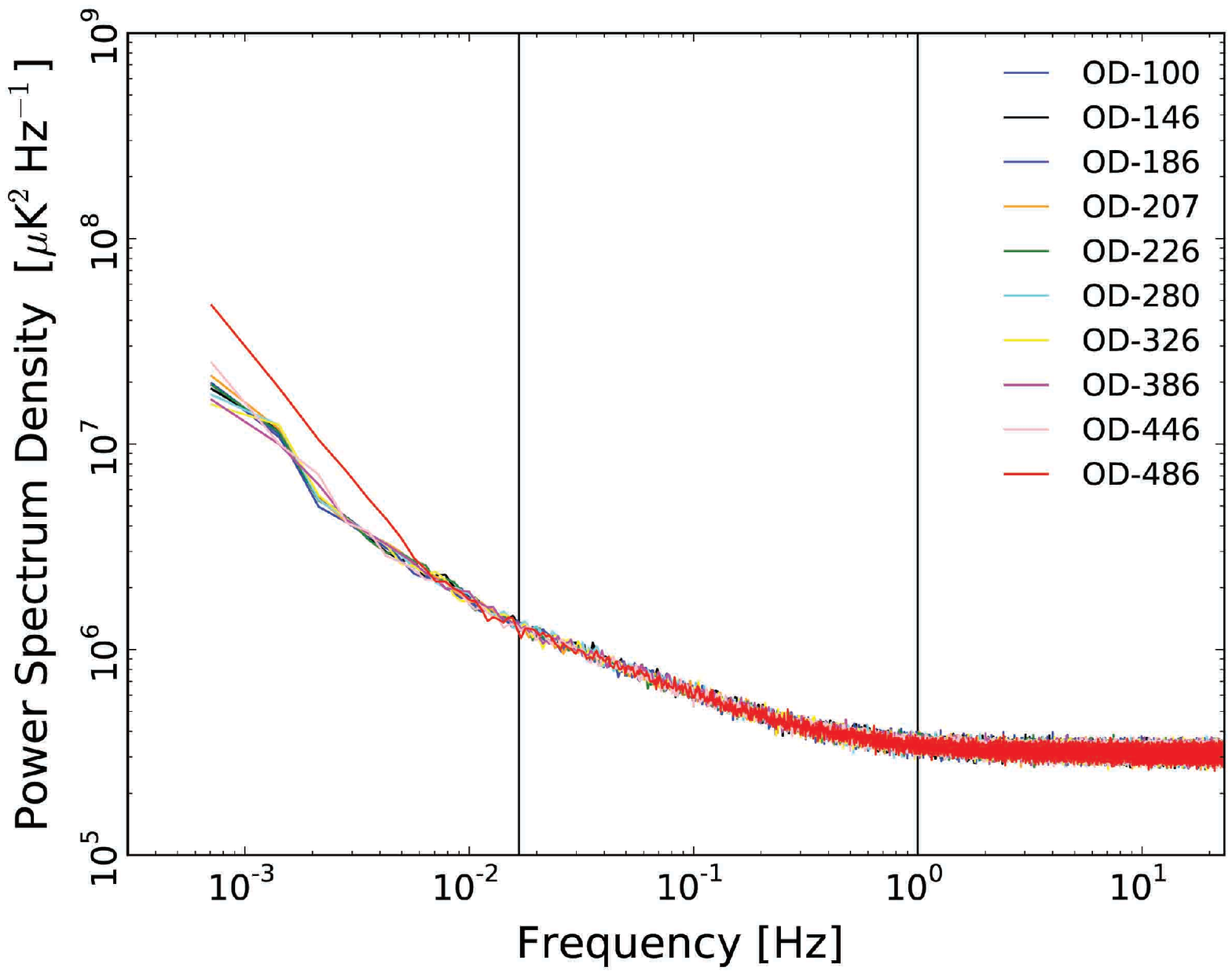}
\includegraphics[width=6cm]{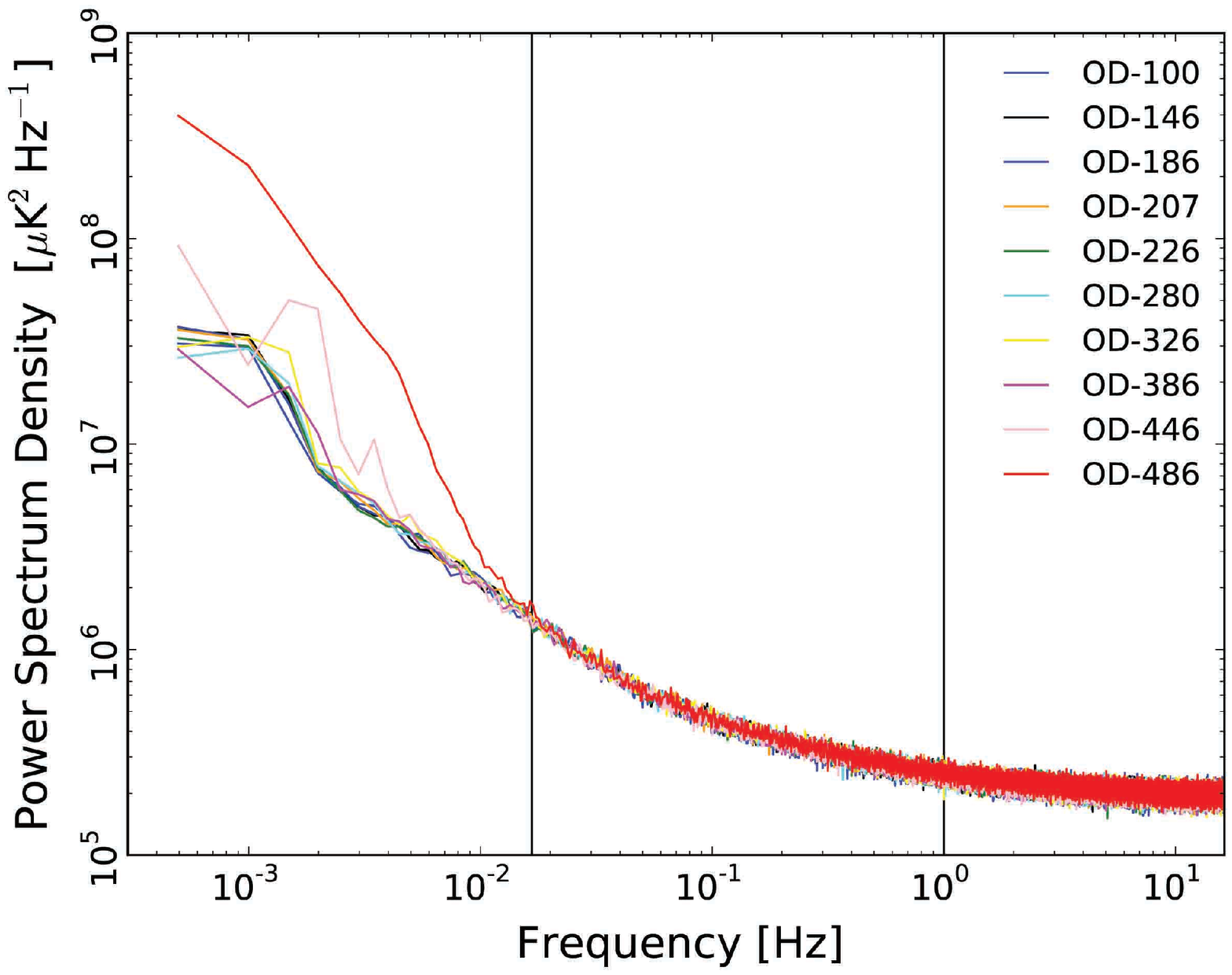}
\caption{Time behaviour of noise spectra on selected periods for
radiometers 18M (70\,GHz, left), 24S (44\,GHz, centre), and 28M
(30\,GHz, right).  White noise and $1/f$ noise are constant within
$0.5$\% until OD 326, after which degradation of the sorption
cooler and the switchover to the redundant cooler introduce higher
thermal noise at the lowest frequencies. Vertical lines mark the
frequencies corresponding to the spin period ($1/60$\,Hz) and the
baseline used in the mapmaking ($1$\,Hz).}
\label{fig_switchover_noise_spectra_rad}
\end{figure*}

Time variations of the noise properties provide a valuable
diagnostic of possible changes in the instrument behavior.  The
switchover between the two sorption coolers provides an example.
Variations in noise properties driven by temperature changes were
expected as the performance of the first cooler degraded with
time, as well as at the switchover to the redundant cooler.
Figure~\ref{fig_switchover_noise_spectra_rad} shows noise
frequency spectra for radiometers LFI28M, LFI24S, and LFI18M for
the nominal mission.  White noise levels are stable within
$0.5$\%. Knee frequencies and slopes are also quite stable until
OD 326, after which the spectra show increasing noise and two
slopes for the low-frequency part. The latter becomes more evident
for spectra around OD~366 and OD~466, when the first cooler starts
to be less effective and produces low-frequency thermal noise.
This behavior is present at some level in all radiometers, but
with different trends, ranging from the small effect shown by
LFI24S to more prominent effects as shown by LFI28M and LFI18M.

  \begin{table}
  \begingroup
  \newdimen\tblskip \tblskip=5pt
  \caption{White noise sensitivities for the LFI radiometers.}
  \label{tab_white_noise_per_radiometer}
  \nointerlineskip
  \vskip -5mm
  \footnotesize
  \setbox\tablebox=\vbox{
  \newdimen\digitwidth
  \setbox0=\hbox{\rm 0}
  \digitwidth=\wd0
  \catcode`*=\active
  \def*{\kern\digitwidth}
  \newdimen\signwidth
  \setbox0=\hbox{+}
  \signwidth=\wd0
  \catcode`!=\active
  \def!{\kern\signwidth}
  \halign{\hbox to 1.3in{#\leaderfil}\tabskip=2em&
      \hfil#\hfil&
      \hfil#\hfil\tabskip=0pt\cr
  \noalign{\doubleline}
  \omit&\multispan2\hfil W{\sc hite} N{\sc oise} S{\sc ensitivity}\hfil\cr
  \noalign{\vskip -4pt}
  \omit&\multispan2\hrulefill\cr
  \omit&Radiometer M&Radiometer S\cr
  \omit\hfil \hfil&[$\,\mu\mathrm{K}_{\rm CMB}\, \mathrm{s}^{1/2}$]&[$\,\mu\mathrm{K}_{\rm CMB}\, \mathrm{s}^{1/2}$]\cr
  \noalign{\vskip 3pt\hrule\vskip 5pt}
  \omit{\bf 70\,GHz}\hfil\cr
  \noalign{\vskip 4pt}
  \hglue 2em LFI-18   &\getsymbol{LFI:white:noise:sensitivity:LFI18:Rad:M}$\,\pm$\getsymbol{LFI:white:noise:sensitivity:uncertainty:LFI18:Rad:M} &\getsymbol{LFI:white:noise:sensitivity:LFI18:Rad:S}$\,\pm$\getsymbol{LFI:white:noise:sensitivity:uncertainty:LFI18:Rad:S}\cr
  \hglue 2em LFI-19   &\getsymbol{LFI:white:noise:sensitivity:LFI19:Rad:M}$\,\pm$\getsymbol{LFI:white:noise:sensitivity:uncertainty:LFI19:Rad:M} &\getsymbol{LFI:white:noise:sensitivity:LFI19:Rad:S}$\,\pm$\getsymbol{LFI:white:noise:sensitivity:uncertainty:LFI19:Rad:S}\cr
  \hglue 2em LFI-20   &\getsymbol{LFI:white:noise:sensitivity:LFI20:Rad:M}$\,\pm$\getsymbol{LFI:white:noise:sensitivity:uncertainty:LFI20:Rad:M} &\getsymbol{LFI:white:noise:sensitivity:LFI20:Rad:S}$\,\pm$\getsymbol{LFI:white:noise:sensitivity:uncertainty:LFI20:Rad:S}\cr
  \hglue 2em LFI-21   &\getsymbol{LFI:white:noise:sensitivity:LFI21:Rad:M}$\,\pm$\getsymbol{LFI:white:noise:sensitivity:uncertainty:LFI21:Rad:M} &\getsymbol{LFI:white:noise:sensitivity:LFI21:Rad:S}$\,\pm$\getsymbol{LFI:white:noise:sensitivity:uncertainty:LFI21:Rad:S}\cr
  \hglue 2em LFI-22   &\getsymbol{LFI:white:noise:sensitivity:LFI22:Rad:M}$\,\pm$\getsymbol{LFI:white:noise:sensitivity:uncertainty:LFI22:Rad:M} &\getsymbol{LFI:white:noise:sensitivity:LFI22:Rad:S}$\,\pm$\getsymbol{LFI:white:noise:sensitivity:uncertainty:LFI22:Rad:S}\cr
  \hglue 2em LFI-23   &\getsymbol{LFI:white:noise:sensitivity:LFI23:Rad:M}$\,\pm$\getsymbol{LFI:white:noise:sensitivity:uncertainty:LFI23:Rad:M} &\getsymbol{LFI:white:noise:sensitivity:LFI23:Rad:S}$\,\pm$\getsymbol{LFI:white:noise:sensitivity:uncertainty:LFI23:Rad:S}\cr
  \noalign{\vskip 5pt}
  \omit{\bf 44\,GHz}\hfil\cr
  \noalign{\vskip 4pt}
  \hglue 2em LFI-24   &\getsymbol{LFI:white:noise:sensitivity:LFI24:Rad:M}$\,\pm$\getsymbol{LFI:white:noise:sensitivity:uncertainty:LFI24:Rad:M} &\getsymbol{LFI:white:noise:sensitivity:LFI24:Rad:S}$\,\pm$\getsymbol{LFI:white:noise:sensitivity:uncertainty:LFI24:Rad:S}\cr
  \hglue 2em LFI-25   &\getsymbol{LFI:white:noise:sensitivity:LFI25:Rad:M}$\,\pm$\getsymbol{LFI:white:noise:sensitivity:uncertainty:LFI25:Rad:M} &\getsymbol{LFI:white:noise:sensitivity:LFI25:Rad:S}$\,\pm$\getsymbol{LFI:white:noise:sensitivity:uncertainty:LFI25:Rad:S}\cr
  \hglue 2em LFI-26   &\getsymbol{LFI:white:noise:sensitivity:LFI26:Rad:M}$\,\pm$\getsymbol{LFI:white:noise:sensitivity:uncertainty:LFI26:Rad:M} &\getsymbol{LFI:white:noise:sensitivity:LFI26:Rad:S}$\,\pm$\getsymbol{LFI:white:noise:sensitivity:uncertainty:LFI26:Rad:S}\cr
  \noalign{\vskip 5pt}
  \omit{\bf 30\,GHz}\hfil\cr
  \noalign{\vskip 4pt}
  \hglue 2em LFI-27   &\getsymbol{LFI:white:noise:sensitivity:LFI27:Rad:M}$\,\pm$\getsymbol{LFI:white:noise:sensitivity:uncertainty:LFI27:Rad:M} &\getsymbol{LFI:white:noise:sensitivity:LFI27:Rad:S}$\,\pm$\getsymbol{LFI:white:noise:sensitivity:uncertainty:LFI27:Rad:S}\cr
  \hglue 2em LFI-28   &\getsymbol{LFI:white:noise:sensitivity:LFI28:Rad:M}$\,\pm$\getsymbol{LFI:white:noise:sensitivity:uncertainty:LFI28:Rad:M} &\getsymbol{LFI:white:noise:sensitivity:LFI28:Rad:S}$\,\pm$\getsymbol{LFI:white:noise:sensitivity:uncertainty:LFI28:Rad:S}\cr
  \noalign{\vskip 5pt\hrule\vskip 3pt}}}
  \endPlancktable
  \endgroup
  \end{table}

  \begin{table*}
  \begingroup
  \newdimen\tblskip \tblskip=5pt
  \caption{Knee frequency and slope for the LFI radiometers.}
  \label{tab_one_over_f_noise_per_radiometer}
  \nointerlineskip
  \vskip -3mm
  \footnotesize
  \setbox\tablebox=\vbox{
  \newdimen\digitwidth
  \setbox0=\hbox{\rm 0}
  \digitwidth=\wd0
  \catcode`*=\active
  \def*{\kern\digitwidth}
  \newdimen\signwidth
  \setbox0=\hbox{+}
  \signwidth=\wd0
  \catcode`!=\active
  \def!{\kern\signwidth}
  \halign{\hbox to 1.3in{#\leaderfil}\tabskip=2em&
      \hfil#\hfil&
      \hfil#\hfil&
      \hfil#\hfil&
      \hfil#\hfil\tabskip=0pt\cr
  \noalign{\doubleline}
  \omit&\multispan2\hfil K{\sc nee} F{\sc requency} $f_{\rm knee}$ [mHz]\hfil&\multispan2\hfil S{\sc lope} $\beta$\hfil\cr
  \noalign{\vskip -4pt}
  \omit&\multispan2\hrulefill&\multispan2\hrulefill\cr
  \omit& Radiometer M&Radiometer S&Radiometer M&Radiometer S\cr
  \noalign{\vskip 3pt\hrule\vskip 5pt}
  \omit{\bf 70\,GHz}\hfil\cr
  \noalign{\vskip 4pt}
  \hglue 2em LFI-18   &*\getsymbol{LFI:knee:frequency:LFI18:Rad:M}$\,\pm\,$\getsymbol{LFI:knee:frequency:uncertainty:LFI18:Rad:M}&
              *\getsymbol{LFI:knee:frequency:LFI18:Rad:S}$\,\pm\,$\getsymbol{LFI:knee:frequency:uncertainty:LFI18:Rad:S}&
              \getsymbol{LFI:slope:LFI18:Rad:M}$\,\pm\,$\getsymbol{LFI:slope:uncertainty:LFI18:Rad:M}&
              \getsymbol{LFI:slope:LFI18:Rad:S}$\,\pm\,$\getsymbol{LFI:slope:uncertainty:LFI18:Rad:S}\cr
  \hglue 2em LFI-19   &*\getsymbol{LFI:knee:frequency:LFI19:Rad:M}$\,\pm\,$\getsymbol{LFI:knee:frequency:uncertainty:LFI19:Rad:M}&
              *\getsymbol{LFI:knee:frequency:LFI19:Rad:S}$\,\pm\,$\getsymbol{LFI:knee:frequency:uncertainty:LFI19:Rad:S}&
              \getsymbol{LFI:slope:LFI19:Rad:M}$\,\pm\,$\getsymbol{LFI:slope:uncertainty:LFI19:Rad:M}&
              \getsymbol{LFI:slope:LFI19:Rad:S}$\,\pm\,$\getsymbol{LFI:slope:uncertainty:LFI19:Rad:S}\cr
  \hglue 2em LFI-20   &**\getsymbol{LFI:knee:frequency:LFI20:Rad:M}$\,\pm\,$\getsymbol{LFI:knee:frequency:uncertainty:LFI20:Rad:M}&
              **\getsymbol{LFI:knee:frequency:LFI20:Rad:S}$\,\pm\,$\getsymbol{LFI:knee:frequency:uncertainty:LFI20:Rad:S}&
              \getsymbol{LFI:slope:LFI20:Rad:M}$\,\pm\,$\getsymbol{LFI:slope:uncertainty:LFI20:Rad:M}&
              \getsymbol{LFI:slope:LFI20:Rad:S}$\,\pm\,$\getsymbol{LFI:slope:uncertainty:LFI20:Rad:S}\cr
  \hglue 2em LFI-21   &*\getsymbol{LFI:knee:frequency:LFI21:Rad:M}$\,\pm\,$\getsymbol{LFI:knee:frequency:uncertainty:LFI21:Rad:M}&
              *\getsymbol{LFI:knee:frequency:LFI21:Rad:S}$\,\pm\,$\getsymbol{LFI:knee:frequency:uncertainty:LFI21:Rad:S}&
              \getsymbol{LFI:slope:LFI21:Rad:M}$\,\pm\,$\getsymbol{LFI:slope:uncertainty:LFI21:Rad:M}&
              \getsymbol{LFI:slope:LFI21:Rad:S}$\,\pm\,$\getsymbol{LFI:slope:uncertainty:LFI21:Rad:S}\cr
  \hglue 2em LFI-22   &*\getsymbol{LFI:knee:frequency:LFI22:Rad:M}$\,\pm\,$\getsymbol{LFI:knee:frequency:uncertainty:LFI22:Rad:M}&
              *\getsymbol{LFI:knee:frequency:LFI22:Rad:S}$\,\pm\,$\getsymbol{LFI:knee:frequency:uncertainty:LFI22:Rad:S}&
              \getsymbol{LFI:slope:LFI22:Rad:M}$\,\pm\,$\getsymbol{LFI:slope:uncertainty:LFI22:Rad:M}&
              \getsymbol{LFI:slope:LFI22:Rad:S}$\,\pm\,$\getsymbol{LFI:slope:uncertainty:LFI22:Rad:S}\cr
  \hglue 2em LFI-23   &*\getsymbol{LFI:knee:frequency:LFI23:Rad:M}$\,\pm\,$\getsymbol{LFI:knee:frequency:uncertainty:LFI23:Rad:M}&
              *\getsymbol{LFI:knee:frequency:LFI23:Rad:S}$\,\pm\,$\getsymbol{LFI:knee:frequency:uncertainty:LFI23:Rad:S}&
              \getsymbol{LFI:slope:LFI23:Rad:M}$\,\pm\,$\getsymbol{LFI:slope:uncertainty:LFI23:Rad:M}&
              \getsymbol{LFI:slope:LFI23:Rad:S}$\,\pm\,$\getsymbol{LFI:slope:uncertainty:LFI23:Rad:S}\cr
  \noalign{\vskip 5pt}
  \omit{\bf 44\,GHz}\hfil\cr
  \noalign{\vskip 4pt}
  \hglue 2em LFI-24   &*\getsymbol{LFI:knee:frequency:LFI24:Rad:M}$\,\pm\,$\getsymbol{LFI:knee:frequency:uncertainty:LFI24:Rad:M}&
              *\getsymbol{LFI:knee:frequency:LFI24:Rad:S}$\,\pm\,$\getsymbol{LFI:knee:frequency:uncertainty:LFI24:Rad:S}&
              \getsymbol{LFI:slope:LFI24:Rad:M}$\,\pm\,$\getsymbol{LFI:slope:uncertainty:LFI24:Rad:M}&
              \getsymbol{LFI:slope:LFI24:Rad:S}$\,\pm\,$\getsymbol{LFI:slope:uncertainty:LFI24:Rad:S}\cr
  \hglue 2em LFI-25   &*\getsymbol{LFI:knee:frequency:LFI25:Rad:M}$\,\pm\,$\getsymbol{LFI:knee:frequency:uncertainty:LFI25:Rad:M}&
              *\getsymbol{LFI:knee:frequency:LFI25:Rad:S}$\,\pm\,$\getsymbol{LFI:knee:frequency:uncertainty:LFI25:Rad:S}&
              \getsymbol{LFI:slope:LFI25:Rad:M}$\,\pm\,$\getsymbol{LFI:slope:uncertainty:LFI25:Rad:M}&
              \getsymbol{LFI:slope:LFI25:Rad:S}$\,\pm\,$\getsymbol{LFI:slope:uncertainty:LFI25:Rad:S}\cr
  \hglue 2em LFI-26   &*\getsymbol{LFI:knee:frequency:LFI26:Rad:M}$\,\pm\,$\getsymbol{LFI:knee:frequency:uncertainty:LFI26:Rad:M}&
              *\getsymbol{LFI:knee:frequency:LFI26:Rad:S}$\,\pm\,$\getsymbol{LFI:knee:frequency:uncertainty:LFI26:Rad:S}&
              \getsymbol{LFI:slope:LFI26:Rad:M}$\,\pm\,$\getsymbol{LFI:slope:uncertainty:LFI26:Rad:M}&
              \getsymbol{LFI:slope:LFI26:Rad:S}$\,\pm\,$\getsymbol{LFI:slope:uncertainty:LFI26:Rad:S}\cr
  \noalign{\vskip 5pt}
  \omit{\bf 30\,GHz}\hfil\cr
  \noalign{\vskip 4pt}
  \hglue 2em LFI-27   &\getsymbol{LFI:knee:frequency:LFI27:Rad:M}$\,\pm\,$\getsymbol{LFI:knee:frequency:uncertainty:LFI27:Rad:M}&
              \getsymbol{LFI:knee:frequency:LFI27:Rad:S}$\,\pm\,$\getsymbol{LFI:knee:frequency:uncertainty:LFI27:Rad:S}&
              \getsymbol{LFI:slope:LFI27:Rad:M}$\,\pm\,$\getsymbol{LFI:slope:uncertainty:LFI27:Rad:M}&
              \getsymbol{LFI:slope:LFI27:Rad:S}$\,\pm\,$\getsymbol{LFI:slope:uncertainty:LFI27:Rad:S}\cr
  \hglue 2em LFI-28   &\getsymbol{LFI:knee:frequency:LFI28:Rad:M}$\,\pm\,$\getsymbol{LFI:knee:frequency:uncertainty:LFI28:Rad:M}&
              *\getsymbol{LFI:knee:frequency:LFI28:Rad:S}$\,\pm\,$\getsymbol{LFI:knee:frequency:uncertainty:LFI28:Rad:S}&
              \getsymbol{LFI:slope:LFI28:Rad:M}$\,\pm\,$\getsymbol{LFI:slope:uncertainty:LFI28:Rad:M}&
              \getsymbol{LFI:slope:LFI28:Rad:S}$\,\pm\,$\getsymbol{LFI:slope:uncertainty:LFI28:Rad:S}\cr
  \noalign{\vskip 5pt\hrule\vskip 3pt}}}
  \endPlancktable
  \endgroup
  \end{table*}

%% file: 09-00_mmaking_intro.tex
The mapmaking pipeline was described in detail in \citet{planck2011-1.6}.  Here we give an overview, reporting significant updates.

%% file: 09-01_madam.tex
Frequency maps were produced by the {\tt Madam} mapmaking code
\citep{keihanen2010}, which takes as input calibrated TOD and corresponding radiometer pointing
data in the form of three Euler angles ($\theta,\phi,\psi$). The output consists of three pixelized Stokes maps
\emph{(T,Q,U)} representing the temperature and polarization anisotropies of the observed sky.

The algorithm is based on the destriping technique, where the correlated noise component
is modelled by a sequence of offsets, or baselines.
The amplitudes of these baselines are determined through maximum-likelihood analysis.
Higher-frequency noise, which is not captured by the baseline model,
is assumed to be white.

The noise model can be written as
\begin{equation}
n' = Fa+n\,,
\end{equation}
where $n'$ is the total noise stream, $n$ is white noise, $a$ is a vector consisting
of the baselines, and $F$ is a matrix of ones and zeros that spreads the baselines
into a time-ordered data stream.

Unlike conventional destriping, {\tt Madam} also uses information
on the known noise properties, in the form of a noise prior. This
allows extension of the destriping approach to shorter
baseline lengths, well below the scanning period of 1~minute.

The baseline length is a key parameter in the destriping
technique.  We chose the baseline length to
be an integral number of samples near 1\,s, specifically, 33, 47, and 79 samples for 30, 44, and 70\,GHz,
respectively, corresponding to 1.0151, 1.0098, and 1.0029\,s.

A baseline length of one second is a reasonable compromise between computational burden
and the quality of the final map.  Shortening the baseline below one second has very
little effect on the residual noise.

When mapmaking is run with a short baseline, the noise prior plays
an important role. There are not enough crossing points between
the one-second data sections to determine the baselines without
additional constraints. In this case, the additional constraints are
noise priors constructed from the parametrized noise
model (Eq.~\ref{eq:noisemodel}) using the values given in Tables~\ref{tab_white_noise_per_radiometer}
and \ref{tab_one_over_f_noise_per_radiometer}.  From these we
compute the expected covariance between the noise baselines,
$C_{a}=\langle a a^{\mathrm T}\rangle$. The exact derivation is given in
\citet{keihanen2010}. Another important role of the noise prior is
to suppress the signal error, which would otherwise increase
rapidly with decreasing baseline length.  Flagged data sections
are handled by setting the white noise variance to
infinity for those samples, but not altering the baseline pattern.
The flagged samples thus do not contribute to the final map in any
way other than serving as place-holders to maintain the time sequence. This
is essential for the noise prior to be applied correctly.

The use of a priori information has the danger of hiding problems in the data,
since the prior may drive the solution to a correct-looking result, even if the data alone are
not in complete agreement with it. To avoid this pitfall, we compute for comparison a subset of the maps with a one-minute baseline
without using a noise prior, and compared the maps visually. No artefacts could be seen in either of the maps.
Although this is not a quantitative test, similar sanity checks had helped to reveal bugs at earlier stages of the development of the data processing pipeline.

Each data sample is assigned entirely to the pixel in which the
center of the beam falls.  We used resolution $N_{\rm side}=1024$
for all frequencies. The average width of one pixel at this
resolution is 3\parcm5, which may be compared to the FWHM of the
beams at each frequency given in Table~\ref{tab_summary_performance}.
For the nominal mission, every pixel is observed
for the 70\,GHz channel; however, at 30\,GHz and 44\,GHz,
where the sampling frequency is lower, there remain individual pixels
that are unobserved, or observed with insufficient polarization angle
coverage to recover the polarization.  These are marked by a special
value in the product maps. In the 30\,GHz map there are 158 such pixels,
while in the 44\,GHz frequency map there are 250 (0.0013\,\% and
0.0020\,\%, respectively).

The maximum likelihood analysis that lies behind our mapmaking algorithm and
the derivation of the destriping solution are presented in \citet{keihanen2010}.
Here we quote the most important formulas for easier reference.

The vector of baseline $a$ is solved from the linear equation
\begin{equation}
\left(F^{\mathrm T} C_{w}^{-1}ZF +C_{a}^{-1}\right)\,a = F^{\mathrm T} C_{w}^{-1}Z\, y\,,
\label{baseeq}
\end{equation}
where
$C_{a}$ and $C_{w}$ are
covariance matrices that represent the a priori known properties of the
baselines and the white noise component, respectively, $y$ is the observed data stream, and
\begin{equation}
Z=I-\left(P^{\mathrm T}C_{w}^{-1}P\right)^{-1}P^{\mathrm T}C_{w}^{-1}\,,
\end{equation}
where $P$ is the pointing matrix, which picks values
from the \emph{T,Q,U} maps and spreads them into time-ordered
data.

The final map is then constructed as
\begin{equation}
  m = \left(P^{\mathrm T}C_{w}^{-1}P\right)^{-1}P^{\mathrm T}C_{w}^{-1}(y-Fa)\,.
\end{equation}
The white noise component is assumed to be uncorrelated, but not necessarily
with uniform variance. Matrix $C_{w}$ is thus diagonal, but not
constant.

Matrices $P$, $F$, and $C_{w}$ are large, but sparse. They are not
constructed explicitly, rather, the operations represented formally as
matrix multiplication above are performed algorithmically. For
instance, multiplication by $P^{T}$ represents an operation where
the time-ordered data are coadded on a sky map.

We solve Eq.~\ref{baseeq} through conjugate gradient iteration.
Convergence is reached typically after 20 to 100 iterations,
depending on the sky coverage and radiometer combination. Full
mission maps, where the whole sky is covered, typically converge
faster than single-survey maps, where sky coverage is incomplete.

The observed signal can be written as
\begin{equation}
y_{i} = T(\omega_{i}) + Q(\omega_{i})\cos(2\psi_{i}) +U(\omega_{i})\sin(2\psi_{i})\,,
\end{equation}
where $\omega_{i}$ is the sky pixel to which the sample is assigned and $\psi_{i}$ defines
the beam orientation. The elements of the pointing matrix $P$ consist of ones and cosine and sine
factors picked from this equation.

When constructing single-horn maps, we include only the
temperature component of the sky into the computation. In the case
of frequency or horn-pair maps, we include the $I$, $Q$, and $U$
Stokes components, although only the $I$ component maps are included in the 2013 data release.

We deviate from the formulation of the original {\tt Madam} paper
in that we have written $C_{w}$ in place of $C_{n}$. This reflects
the fact that the $C_{w}$ matrix is not necessarily the white noise covariance, but rather a
user-defined weighting factor.  We diverge from the maximum-likelihood noise-weighted solution
in order to have better control over polarization systematics. Specifically, the weight for a given
horn is taken to be
\begin{equation}
  C_{w}^{-1} = \frac{2}{\sigma_{\mathrm M}^2 +\sigma_{\mathrm S}^{2}}\,,
\end{equation}
where $\sigma_{M}$ and $\sigma_{S}$ are the white noise sensitivities of the two radiometers
of the horn, computed from the values given in Table~\ref{tab_white_noise_per_radiometer}. The weights are identical for radiometers of same horn.
The formula above is applied for non-flagged samples.  For flagged samples we set $C_{w}^{-1}=0$.
With this horn-uniform weighting,  polarization maps
become dependent solely on the signal difference between M and S
radiometers, apart from a small leakage due to the fact that the
polarization sensitivities are not exactly at 90\deg\ from each
other. Many systematic effects, which are equal or strongly
correlated within a horn, cancel out in the difference. Horn-uniform weighting has the benefit of reducing
spurious polarization signals arising from beam shape mismatch,
since the beam shapes of a radiometer pair sharing a horn are
typically quite similar (though not identical). Complete
cancellation also requires that the same flags be applied to both
data streams. Therefore, if a sample for one radiometer is
flagged, we discard the corresponding sample for the other
radiometer as well.

The covariance of residual white noise in the map solution is obtained from
\begin{equation}
C_{\mathrm {wn}} = \left(P^{\mathrm T}C_{w}^{-1}P\right)^{-1}P^{\mathrm T}C_{w}^{-1} C_{n} C_{w}^{-1} P\left(P^{\mathrm T}C_{w}^{-1}P\right)^{-1}\,,
\end{equation}
where now $C_{n}$ is the actual white noise covariance of the time-ordered data.
For a given $C_{n}$, the residual noise would be minimized when $C_{w}=C_{n}$, in which case
$C_{\mathrm {wn}}=\left(P^{\mathrm T}C_{n}^{-1}P\right)^{-1}$.
In using a different weighting, we accept slightly higher noise in return for better removal of systematics.

Altering matrix $C_{w}$ does not bias the solution, as may be verified by inserting $y=Pm$ in the solution above.
It merely affects the level of residual noise.

The white noise covariance only takes into account the
uncorrelated component of the noise. The computation of a full
noise covariance matrix, which also captures the residual
correlated noise, is discussed in Sect.~\ref{subsubsec:NCVM}.

The destriping solution assumes that the sky signal is uniform within one pixel.
This is not strictly true, which gives rise to the signal error:
signal differences within a pixel are falsely interpreted as noise, which leads to spurious striping,
especially in the vicinity of point sources or in other regions where the signal gradient is large.
Most of this effect is seen where Galactic emission is strong.

Similarly, mismatch in frequency response between radiometers gives rise to spurious striping,
as different radiometers record slightly different signals from the same source.
In this case also, the main effect is seen where Galactic emission is strong.

To reduce these undesired effects, we mask the strongest Galactic
region and compact sources in the destriping process and use
crossing points of rings only outside the masked region to solve
the noise baselines. The masks for 30, 44, and 70\,GHz leave
78.7\%, 89.4\%, and 89.7\% of the sky, respectively, included in
the analysis. Some of the baselines fall completely inside the
masked region, but can still be recovered with reasonable accuracy
with help of combined information of the neighbouring baselines
and the noise prior.

Table~\ref{tab_maps_released} lists the delivered maps.  All have {\tt HEALPix} resolution
$N_{\rm side}=1024$.

\begin{table*}[tmb]
  \begingroup
  \newdimen\tblskip \tblskip=5pt
  \caption{Released LFI maps.}
  \label{tab_maps_released}
  \nointerlineskip
  \vskip -3mm
  \footnotesize
  \setbox\tablebox=\vbox{
    \newdimen\digitwidth
    \setbox0=\hbox{\rm 0}
    \digitwidth=\wd0
    \catcode`*=\active
    \def*{\kern\digitwidth}
    \newdimen\signwidth
    \setbox0=\hbox{+}
    \signwidth=\wd0
    \catcode`!=\active
    \def!{\kern\signwidth}
    \halign{\hbox to 1.5in{#\leaderfil}\tabskip=2em&
      \hfil#\hfil&
      \hfil#\hfil&
      \hfil#\hfil&
      \hfil#\hfil\tabskip=0pt\cr
      \noalign{\doubleline}
      \omit\hfil Map\hfil& Horns& OD range& Baseline (s)& Sky coverage (\%)\cr
      \noalign{\vskip 3pt\hrule\vskip 5pt}
      30 GHz nominal&  27, 28&                 *91--563& 1.0151& *99.999\cr
      30 GHz survey 1& 27, 28&                 *91--270& 1.0151& *97.205\cr
      30 GHz survey 2& 27, 28&                 270--456& 1.0151& *97.484\cr
      \noalign{\vskip 5pt}
      44 GHz nominal&  24, 25, 26&             *91--563& 1.0098& *99.998\cr
      44 GHz survey 1& 24, 25, 26&             *91--270& 1.0098& *93.934\cr
      44 GHz survey 2& 24, 25, 26&             270--456& 1.0098& *93.310\cr
      \noalign{\vskip 5pt}
      70 GHz nominal&  18, 19, 20, 21, 22, 23& *91--563& 1.0029& 100.000\cr
      70 GHz survey 1& 18, 19, 20, 21, 22, 23& *91--270& 1.0029& *97.938\cr
      70 GHz survey 2& 18, 19, 20, 21, 22, 23& 270--456& 1.0029& *97.474\cr
      \noalign{\vskip 5pt\hrule\vskip 3pt}
    }}
  \endPlancktable
  \endgroup
\end{table*}

%% file: 09-02_low_cov.tex
To fully exploit the information contained in the large scale
structure of the microwave sky, pixel-pixel covariances are needed
in the maximum likelihood estimation of the CMB power spectrum.
Full covariance matrices are impossible to employ at the
native map resolution, because of resource limitations. A
low-resolution dataset is therefore required for the low-$\ell$
analysis. This dataset consists of low-resolution maps plus
descriptions of residual noise present in those maps given by
pixel-pixel noise covariance matrices (NCVMs).  At present, the
low-resolution dataset can be used efficiently only at resolution
$N_{\rm side}=16$ or lower.
All the low-resolution data products are produced at this target resolution.

We first discuss production of the low-resolution maps, and then discuss the NCVMs.

\subsubsection{Low-resolution maps}

We construct low-resolution maps by downgrading the
high-resolution maps to the target resolution, $N_{\rm side}=16$,
using a noise-weighted downgrading scheme (see
\citealt{keskitalo2009} for a discussion of other schemes and
their advantages and disadvantages). Specifically, we apply to a
high-resolution map the operation
$$
 m_{\mathrm l} = \left(P_{\mathrm l}^{\mathrm T}C_{w}^{-1}P_{\mathrm l}\right)^{-1}X \left(P_{\mathrm h}^{\mathrm T}C_{w}^{-1}P_{\mathrm h}\right)\, m_{\mathrm h} \equiv D\,m_{\mathrm h}\,,
$$
where
$$
X_{qp} = \begin{cases} 1, \; p \;  \mathrm{subpixel} \; \mathrm{of} \;  q\\
0, \; \mathrm{otherwise} \end{cases}
$$
sums high-resolution pixels to low-resolution pixels.
Here subscripts h and l refer to the high ($N_{\rm side}=1024$) and
low ($N_{\rm side}=16$) resolution versions of the pointing
matrix. The same matrix $X$  downgrades the pointing matrix,
$P_{\mathrm l}=P_{\mathrm h}X^{\mathrm T}$. The resulting map is identical to the one we
would get by solving the baselines at the higher resolution,
then binning the map to a lower resolution.

By using the noise-weighted downgrading scheme, we get adequate
control over signal and noise in the resulting map. If we were to
calculate a low-resolution map directly at the target resolution,
the signal error would be larger due to sub-pixel structures,
while the noise level would be lower.

After downgrading, the temperature component is smoothed with a
symmetric Gaussian window function with $\mbox{FWHM}=440$\arcm,
while the polarization components are left unsmoothed. Smoothing
is applied to alleviate aliasing due to high frequency power in
the map. The polarization components need to be treated
differently, since the cosmological information contained in
polarization has several orders of magnitude poorer
signal-to-noise ratio. Compared to the approach proposed in
\citet{keskitalo2009}, we intentionally changed the order of
downgrading and smoothing to better deal with noise. The aliased
power is negligible at scales unaffected by the smoothing
operator.

\subsubsection{Noise covariance matrices}
\label{subsubsec:NCVM}

The statistical description of the residual noise present in a low-resolution map is given in the form of a pixel-pixel noise covariance matrix, as described in \citet{keskitalo2009}.
We must apply the same processing steps used in downgrading the maps to the NCVMs for consistency.
Some approximations are involved, either inherent to the {\tt Madam} NCVM method or
introduced to speed up performance.

The pixel-pixel noise covariance matrix for generalized destriping is
$$
N = \left[P^{\mathrm T}\left(C_{w}+FC_{a}F^{\mathrm T}\right)^{-1}P\right]^{-1},
$$
which can be written in a dimensionally reduced form as
\begin{equation}
N^{-1}=P^{\mathrm T}C_{w}^{-1}P - P^{\mathrm T}C_{w}^{-1}F\,\left(F^{\mathrm T}C_{w}^{-1}F+C_{a}^{-1}\right)^{-1}F^{\mathrm T}C_{w}^{-1}P\,.
\label{eq_NCVM}
\end{equation}
Applying Eq.~\ref{eq_NCVM} in practice requires
inversion of a symmetric $3N_{\mathrm {pix}} \times 3N_{\mathrm {pix}}$
matrix in a later analysis step.  Because the inverse NCVMs are
additive, we divide the computations into a number of small
chunks to save computational resources. We first calculate,
using Eq.~\ref{eq_NCVM}, one inverse NCVM per radiometer per survey
at the highest possible resolution  permitted by
computer resources ($N_{\rm side}=32$).  Later, we combine the individual inverse
matrices to produce the full inverse matrix.

To obtain the noise covariance from its inverse, the matrices are
inverted using the eigendecomposition of a matrix. The monopole of
the temperature map cannot be resolved by the mapmaker, and thus
the matrix becomes singular. Therefore this ill-determined mode is
left out of the analysis.

These intermediate-resolution matrices are then downgraded to the target resolution. The downgrading operator is the same $D$ as for the map downgrading, but $P_{\mathrm h}$ is replaced with $P_{\mathrm i}$, i.e., with the intermediate resolution pointing matrix. The downgraded matrix is
\begin{equation}
N_{l} = DND^{\mathrm T}.
\end{equation}

As a final step, the same smoothing operator is applied to the
temperature component of the matrices as was applied to the
low-resolution maps.

The noise covariance matrices are calculated with two different
sets of noise parameters. One set covers the entire mission under
consideration (values given in Tables~\ref{tab_white_noise_per_radiometer}
and \ref{tab_one_over_f_noise_per_radiometer}), while the other has
individual noise parameters for each survey.

Equation~\ref{eq_NCVM} describes the noise correlations when
the noise baselines are solved at the resolution
of the final map. For an exact description of the correlations, we
should construct the matrix at resolution $N_{\rm side}=1024$, and
downgrade the inverted matrix to the target resolution $N_{\rm
side}=16$. This is not feasible, due to the size of the matrix. We
therefore construct the matrix at the highest possible resolution
$N_{\rm side}=32$, and downgrade it to the target resolution.

The same formula inherently assumes that individual detectors are
weighted according to their white noise levels, as suggested by
maximum likelihood analysis. In map-making, however, we apply
horn-uniform weighting, to have better control over systematics,
as explained in Sect.~\ref{sec_madam}. Also, the formulation does
not take into account the effect of the destriping mask. As
a result of these idealizations, the covariance matrix is an
approximate description of the residual noise correlations. We
have performed $\chi^2$ tests to assess the effect of each
idealization individually.

\begin{figure} [!htb]
\centering
\includegraphics[width=8.5cm]{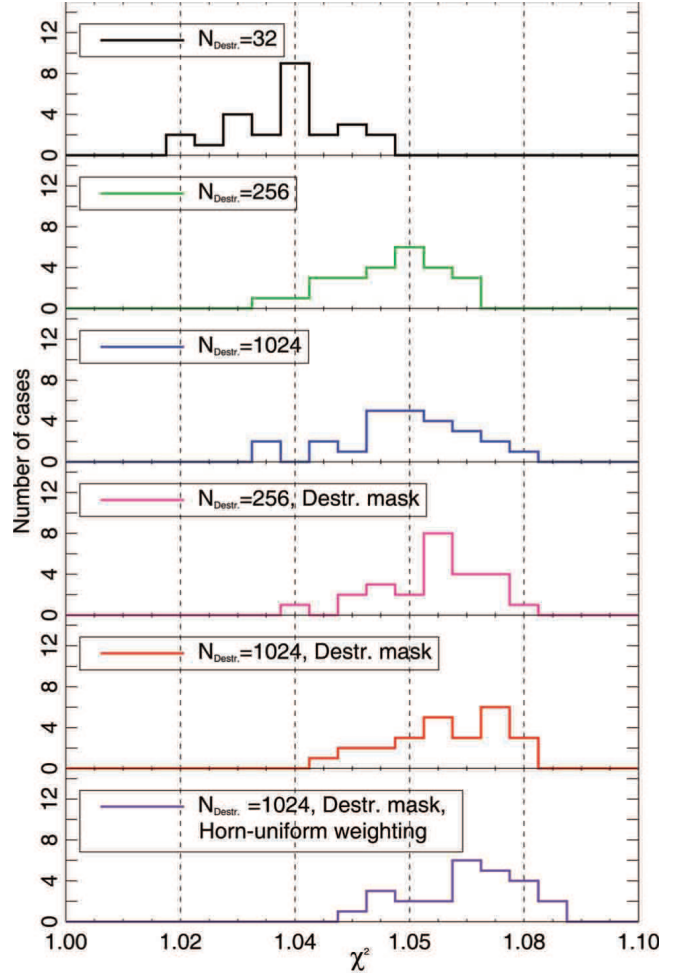}
\caption{Reduced $\chi^{2}$ statistics from 25 noise-only maps for the
30\,GHz 2013 delivery.  The NCVM was calculated using 0.25\,s
baselines, while the simulations were made with 0.5\,s baselines.
The number of idealizations in the noise-only simulations
decreases from top to bottom. The first set of simulations
(plotted in black) contains the same approximations that are made
in the NCVM calculation. The last set of simulations (plotted in
purple) corresponds to the standard mapmaking options: the horns
are weighted uniformly; destriping resolution, $N_{\rm Destr.}$,
is 1024; and a destriping mask is applied.} \label{fig_chi2_2}
\end{figure}

Figure~\ref{fig_chi2_2} illustrates the effect the
approximations inherent in the NCVM computation: horn-uniform
weighting; destriping resolution $N_{\rm Destr.}$ equal to
the map resolution; and masking in the destriping phase. Each
non-ideal factor increases the discrepancy seen in the $\chi^{2}$
test.

\begin{figure} [!htb]
\centering
\includegraphics[width=8.5cm]{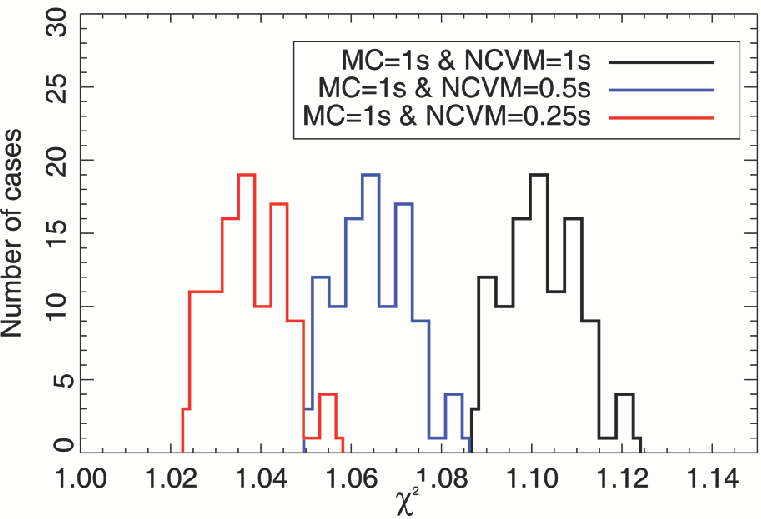}
\includegraphics[width=8.5cm]{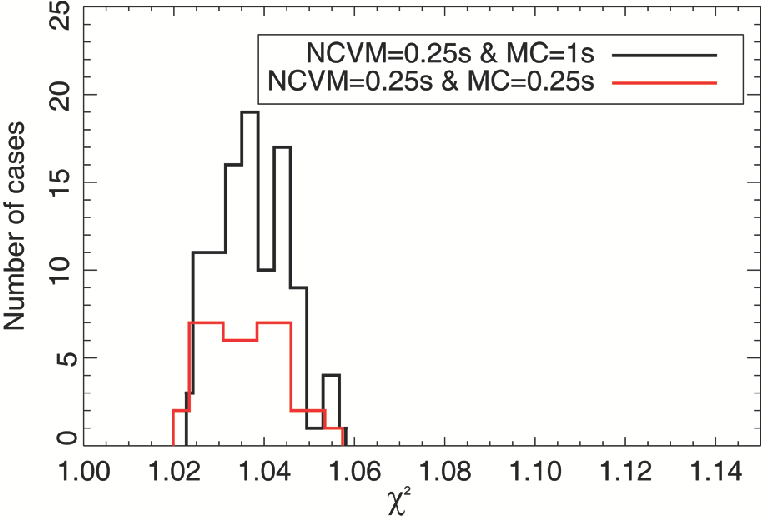}
\caption{Reduced $\chi^{2}$ statistics from noise-only maps
in the 30\,GHz 2013 delivery. \emph{Upper}: The noise-only simulation
set is fixed, while the NCVM baseline length changes. Three baseline
lengths were chosen, 1\,s (black), 0.5\,s (blue), and 0.25\,s (red).
\emph{Lower}: the NCVM is fixed, while the noise-only simulation
varies. Two baseline lengths were chosen, 1\,s (plotted in black)
and 0.25\,s (red).} \label{fig_chi2_1}
\end{figure}

We chose to use shorter baselines in the NCVM production than in
the map-making. They were 0.25\,s (8~samples), 0.5\,s (24~samples),
and 0.5\,s (39~samples) for 30\,GHz, 44\,GHz, and 70\,GHz,
respectively. Since many of the knee frequencies were higher than
anticipated prior to launch, short baselines model the noise
better \citep{keskitalo2009}. We additionally found out that
reducing baseline length in the NCVM calculation affects the
$\chi^{2}$ statistics more than changing baseline length in the
mapmaking (see Fig.~\ref{fig_chi2_1}).

%% file: 09-03_halfring.tex
To estimate the noise directly at the map level and in
the angular power spectra, we produce half-ring maps
($h_1$ and $h_2$) with the same pipeline as
described in Sect.~\ref{sec_madam}, but using data only from the
first or the second half of each stable pointing period. These
half-ring maps contain the same sky signal, since they result from
the same scanning pattern on the sky. Therefore the difference of
maps $h_1$ and $h_2$ captures any noise whose frequency is greater than 
that corresponding to half of the duration of the pointing period, 
i.e., noise whose frequency is $f \gtrsim 1/20\,\mbox{min} =
0.85\,$mHz. The procedure of calculating the half-ring maps and
their hit-count-weighted difference maps is described in more
detail in \citet{planck2011-1.6} and \citet{ planck2013-p28}.

The use of half-ring maps in the validation of data and noise
estimates will be explained in Sect.~\ref{sec_dataval_intro}. In
addition, the half-ring maps were an integral part of the
component separation process ~\citep{planck2013-p06} and
likelihood codes ~\citep{planck2013-p08}.

%% file: 09-04_noise_simulations.tex
Simulated noise timelines are produced according to the
three-parameter noise model (white noise sensitivity, knee
frequency, slope), using the estimated parameter values 
given in Tables~\ref{tab_white_noise_per_radiometer} and
~\ref{tab_one_over_f_noise_per_radiometer}. Maps are made from
these noise timelines using reconstructed flight pointing and the
same {\tt Madam} parameter settings as used for the flight
maps. These steps are repeated to produce 
1000 realizations of noise maps for different radiometer
combinations, including frequency maps and 70\,GHz horn-pair maps.

This Monte Carlo (MC) work is done in two stages, with two
partially different pipelines, first an LFI MC in close connection
with the map-making from the LFI flight data, and then as a part
of the joint LFI/HFI full-focal plane ``FFP6'' simulations.\footnote{ 
\url{http://www.sciops.esa.int/wikiSI/planckpla/index.php?title=Simulation_data&instance=Planck_Public_PLA}.}  
This work is divided between two supercomputing centers, the CSC-IT
Center for Science in Finland and the National Energy Research
Scientific Computing Center (NERSC) in the U.S.

The noise MC maps provide a statistical distribution of noise maps
that can be compared to the half-ring noise maps
(Sec.~\ref{sec_halfring}) to see how well maps from the noise
model match the real flight noise in the half-ring noise maps.
Note, however, that half-ring noise maps cannot represent
properly the noise in the flight maps for timescales of half the pointing period or
longer.

For low-resolution studies, the maps are downgraded to $N_{\rm
side}=32$ and $N_{\rm side}=16$ {\tt HEALPix} resolution, using
the same procedure as for the flight maps. These can be compared
to the low-resolution noise covariance matrices discussed in
Sec.~\ref{sec_low_cov}, which were generated from the same noise
model, but are based on some approximations (see Fig.~\ref{fig_chi2_2} 
and Fig.~\ref{fig_chi2_1}).  This comparison reveals the
effect of these approximations on the NCVM.


In addition, the noise MC maps were used in power spectrum
estimation, component separation \citep{planck2013-p06}, and in
non-Gaussianity estimation (\citealt{planck2013-p09} and
\citealt{planck2013-p09}).

%% file: 09-05_Overview_maps.tex
\begin{figure*}
\begin{centering}
\includegraphics[width=0.8\textwidth]{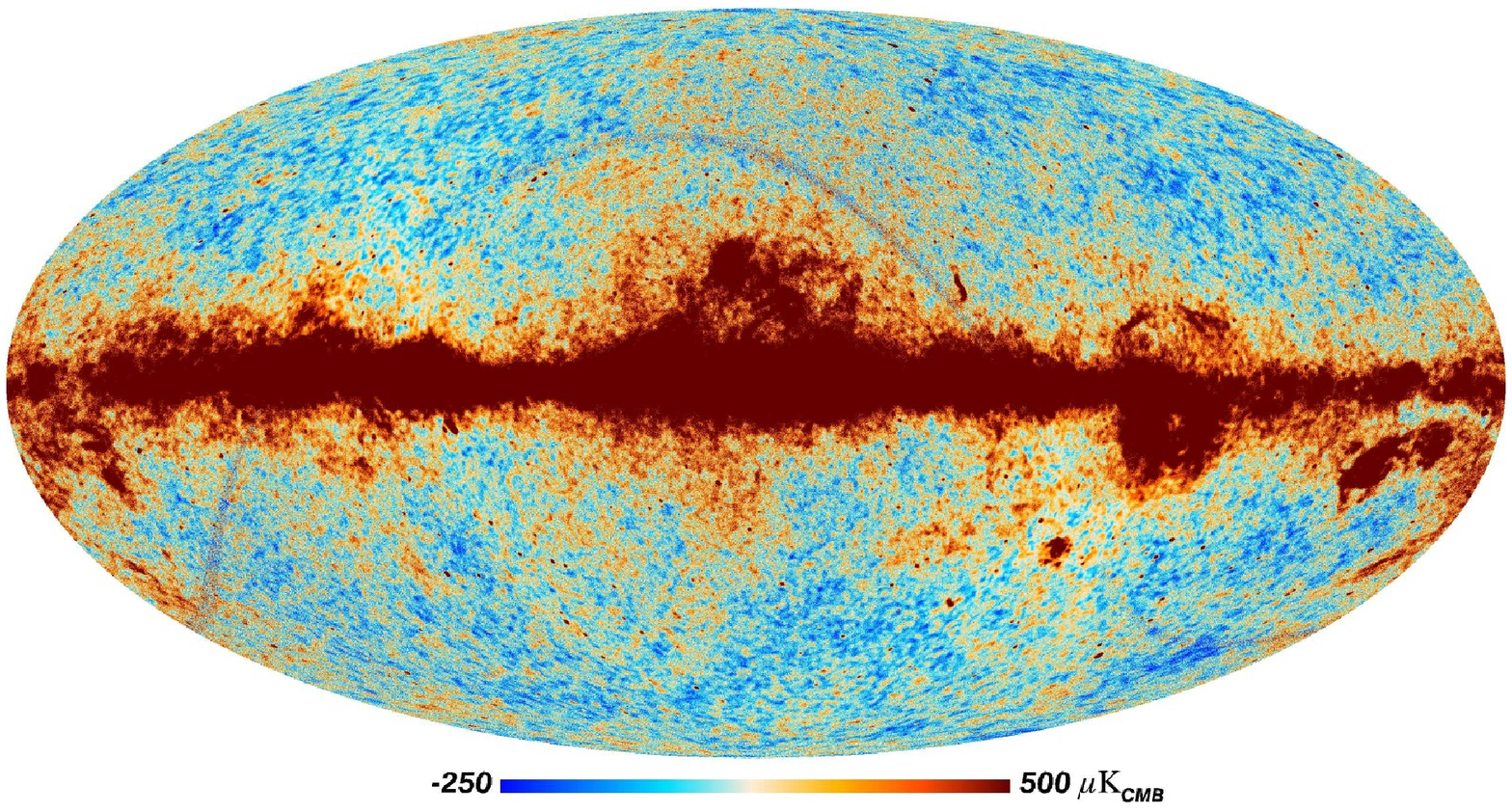}
\includegraphics[width=0.8\textwidth]{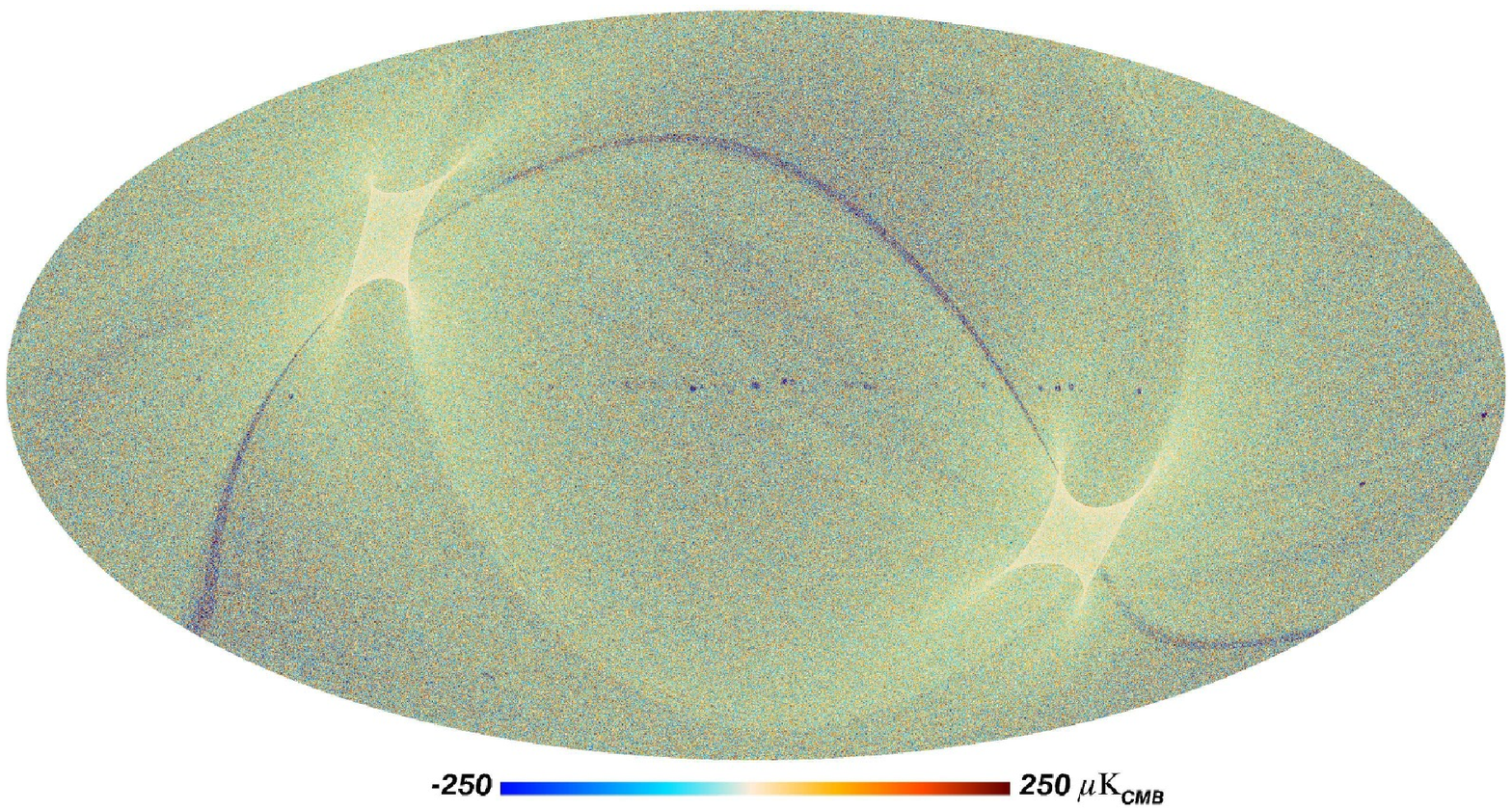}
\includegraphics[width=0.8\textwidth]{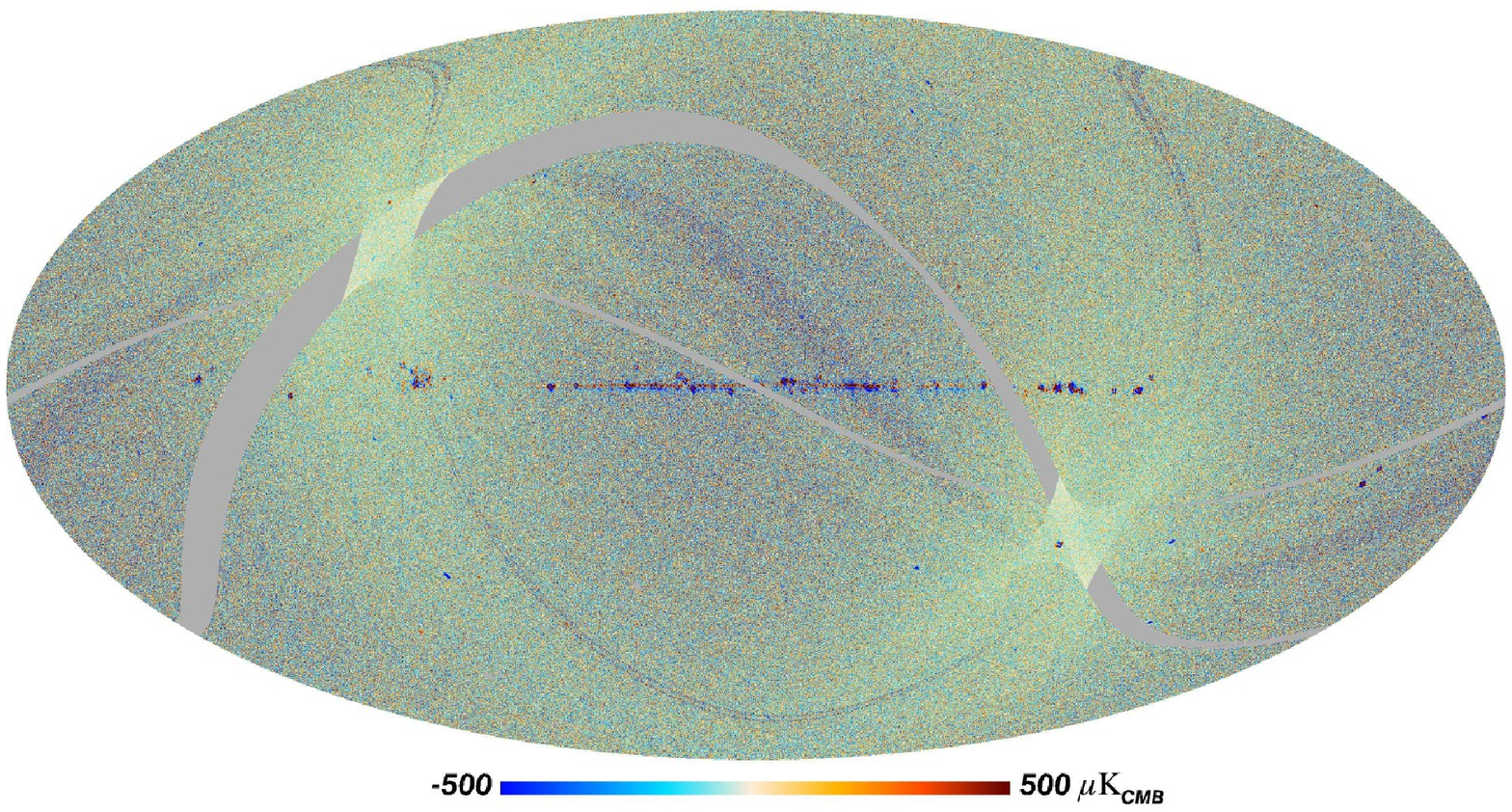}
\caption{\label{fig:Imaps-030}LFI maps at 30\,GHz.  \,\,{\it Top}: Intensity $I$. \,\,{\it Middle}: Half-ring difference between maps made of the first and the second half of each stable pointing period.  \,\,{\it Bottom}: Survey1 minus Survey2.  Smoothed versions of the half-ring and survey difference maps can be found in \citet{planck2013-p02a}.}
\end{centering}
\end{figure*}
\begin{figure*}
\begin{centering}
\includegraphics[width=0.8\textwidth]{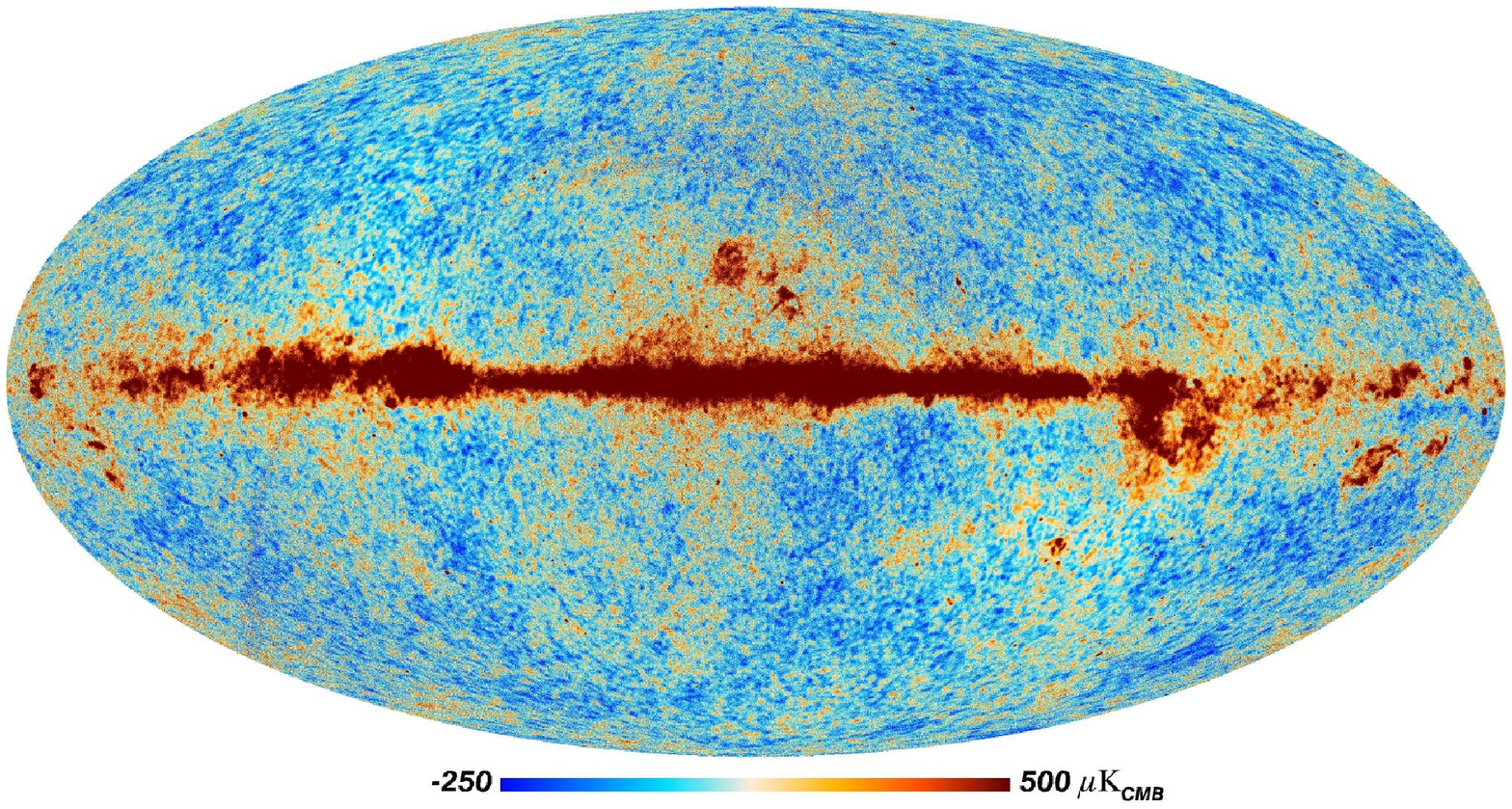}
\includegraphics[width=0.8\textwidth]{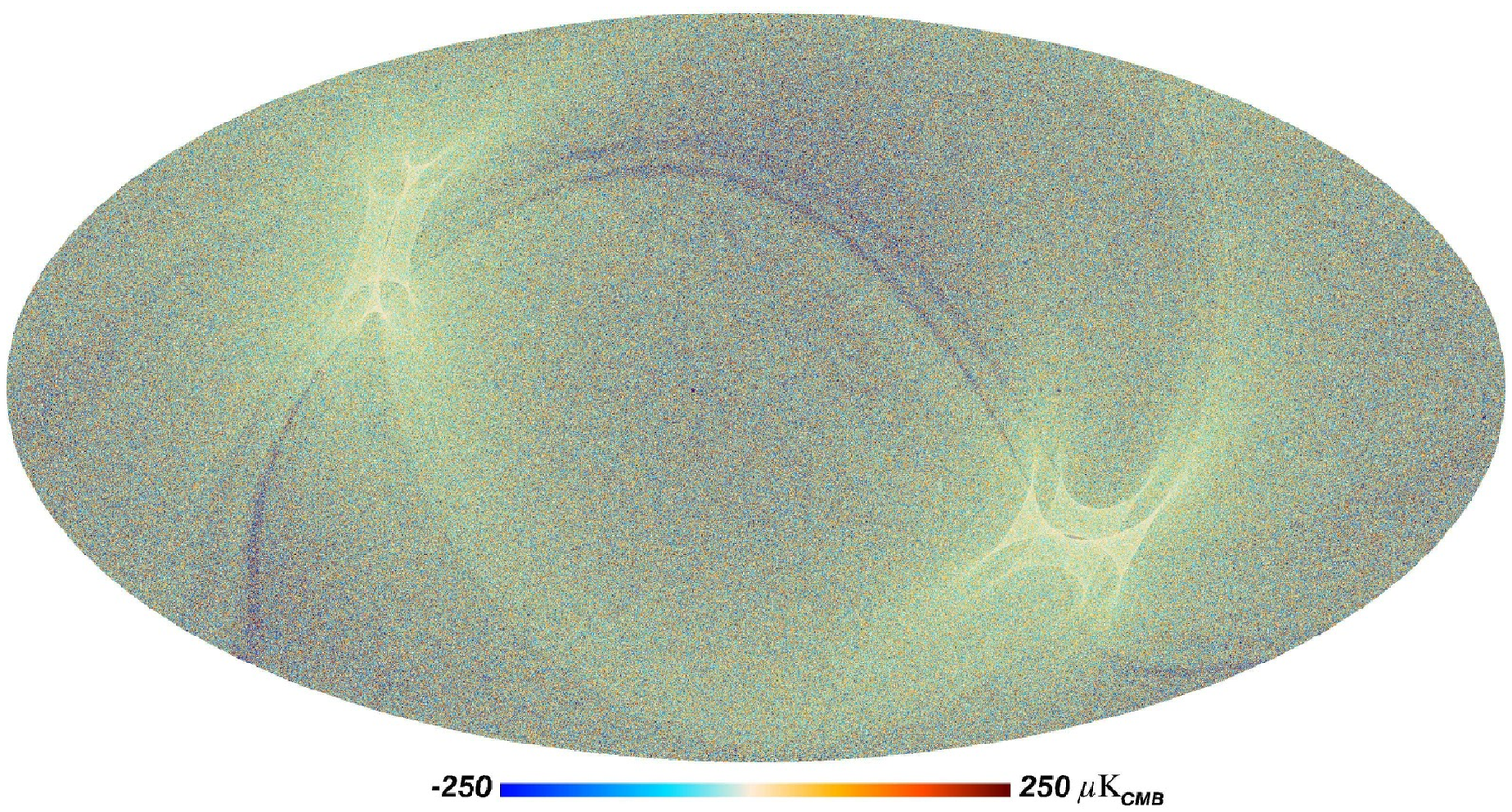}
\includegraphics[width=0.8\textwidth]{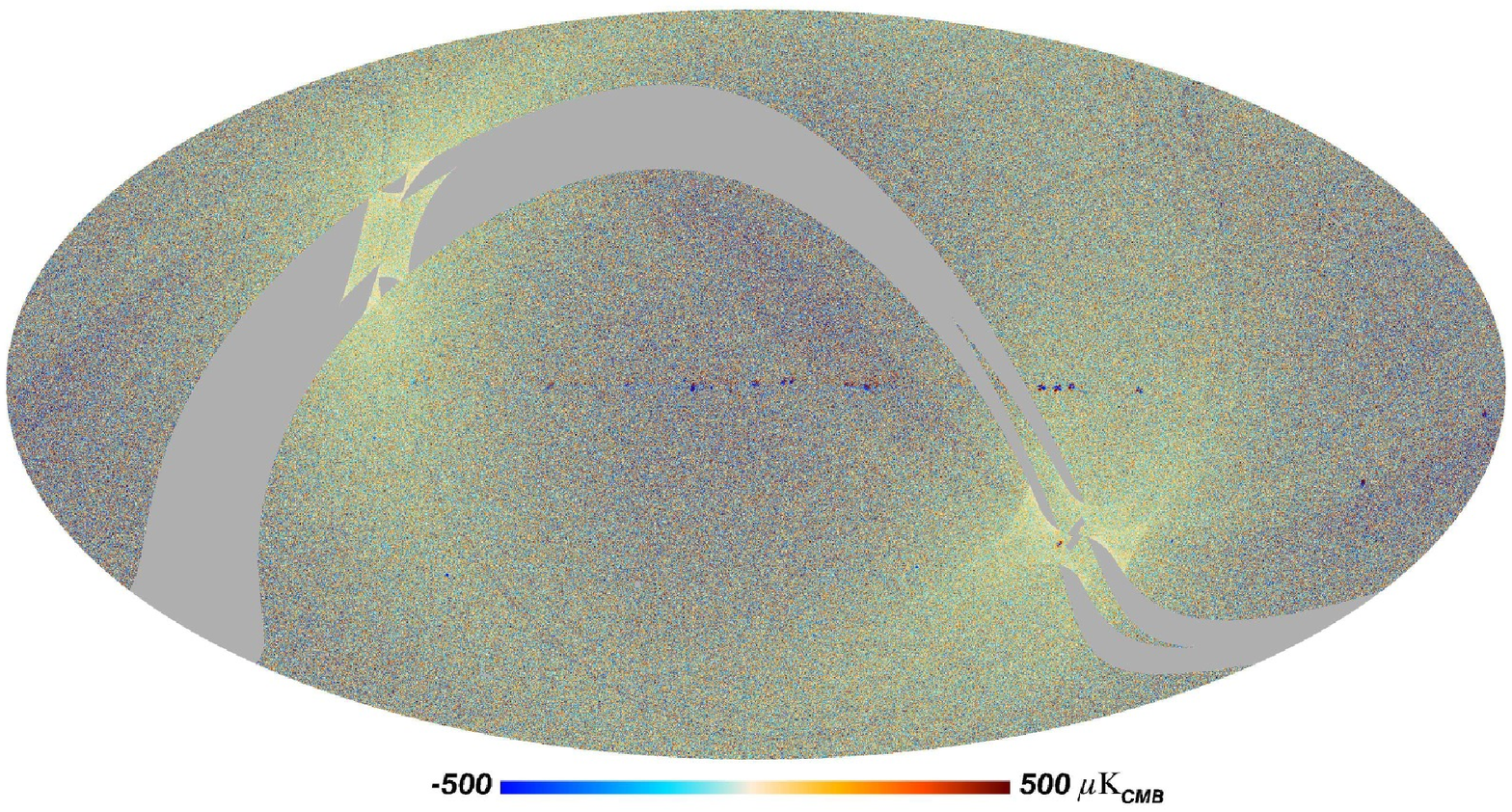}
\caption{Same as Fig.~\ref{fig:Imaps-030} for 44\,GHz.}
\label{fig:Imaps-044}
\end{centering}
\end{figure*}
\begin{figure*}
\begin{centering}
\includegraphics[width=0.8\textwidth]{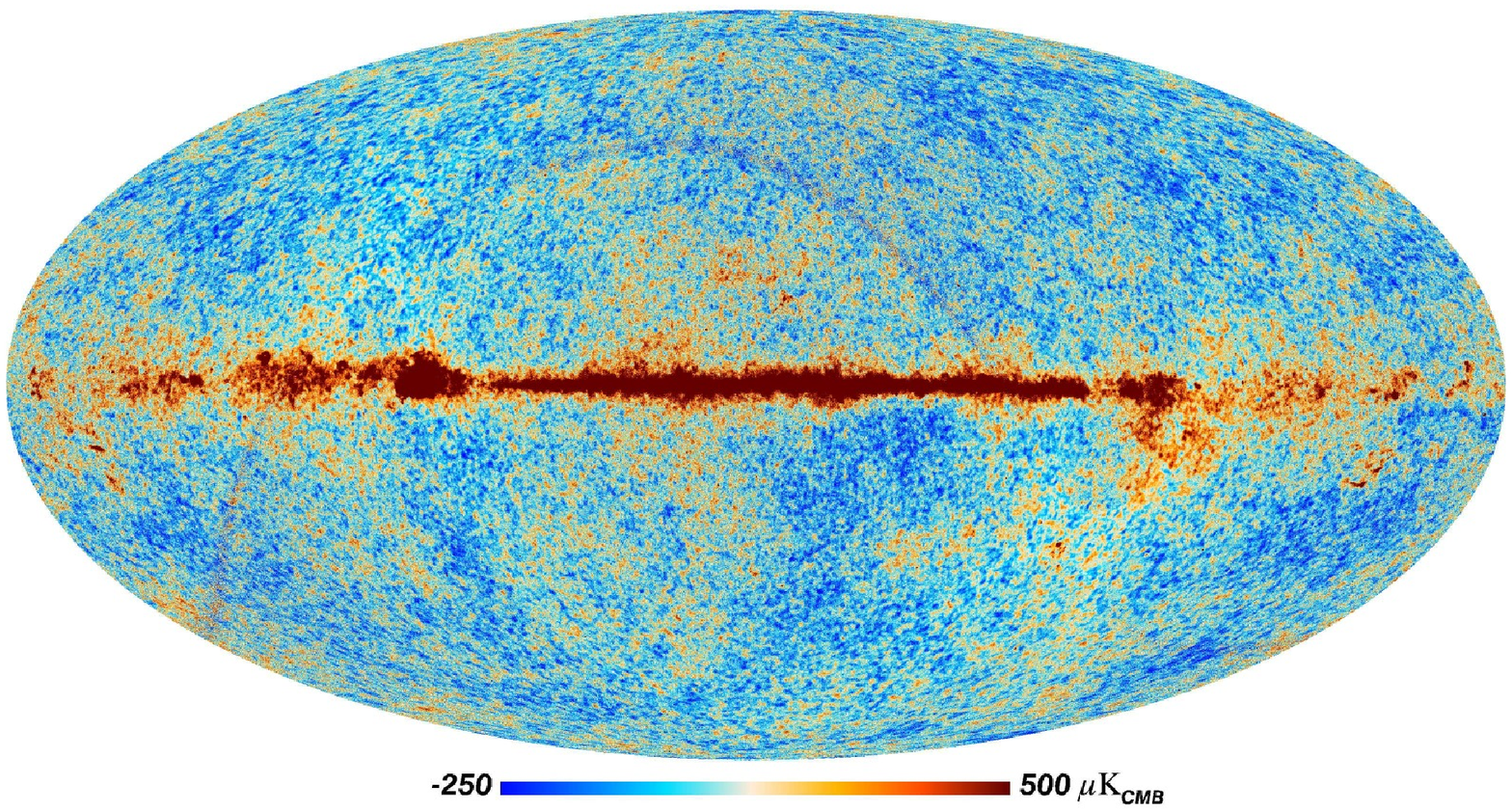}
\includegraphics[width=0.8\textwidth]{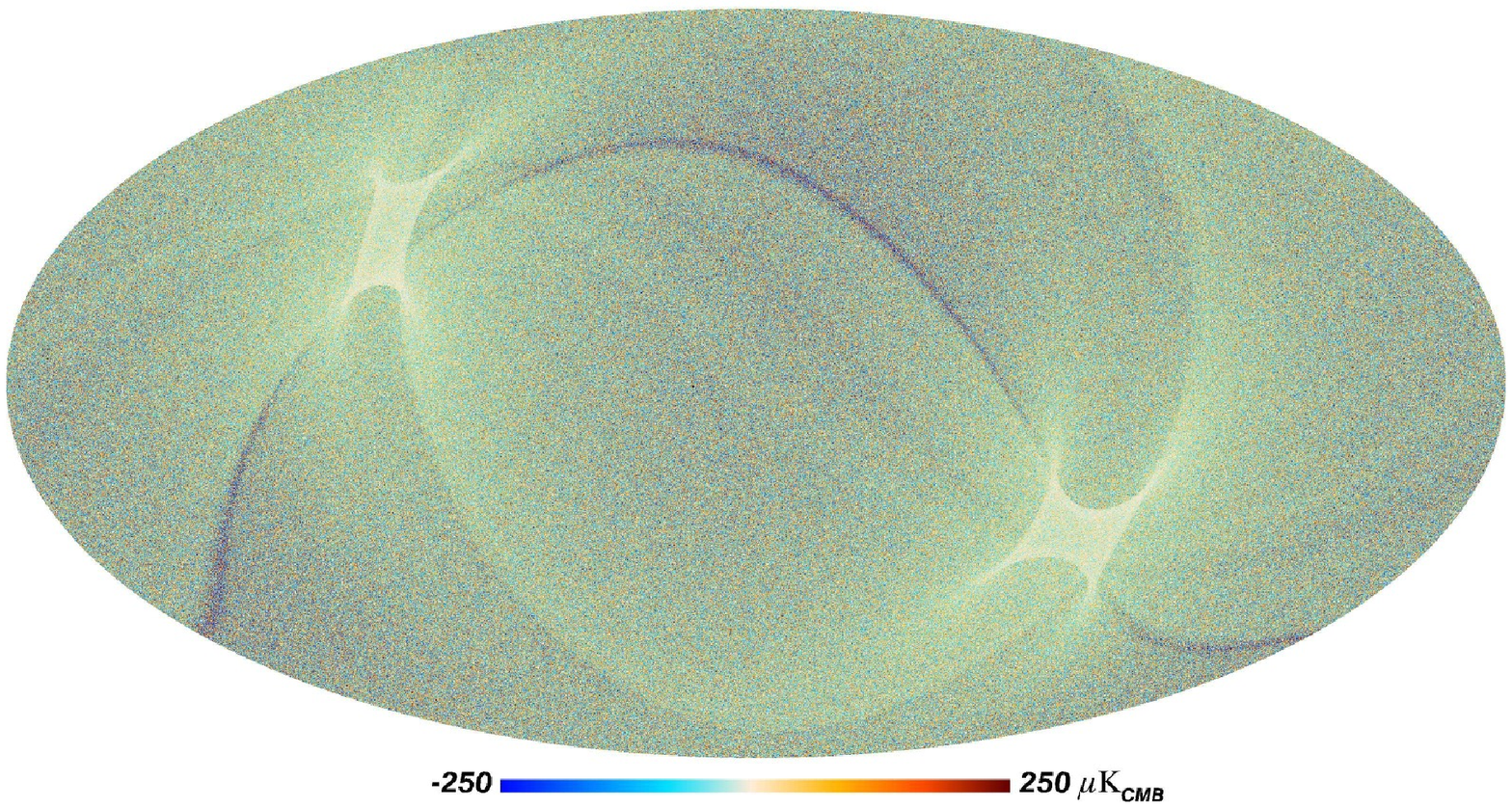}
\includegraphics[width=0.8\textwidth]{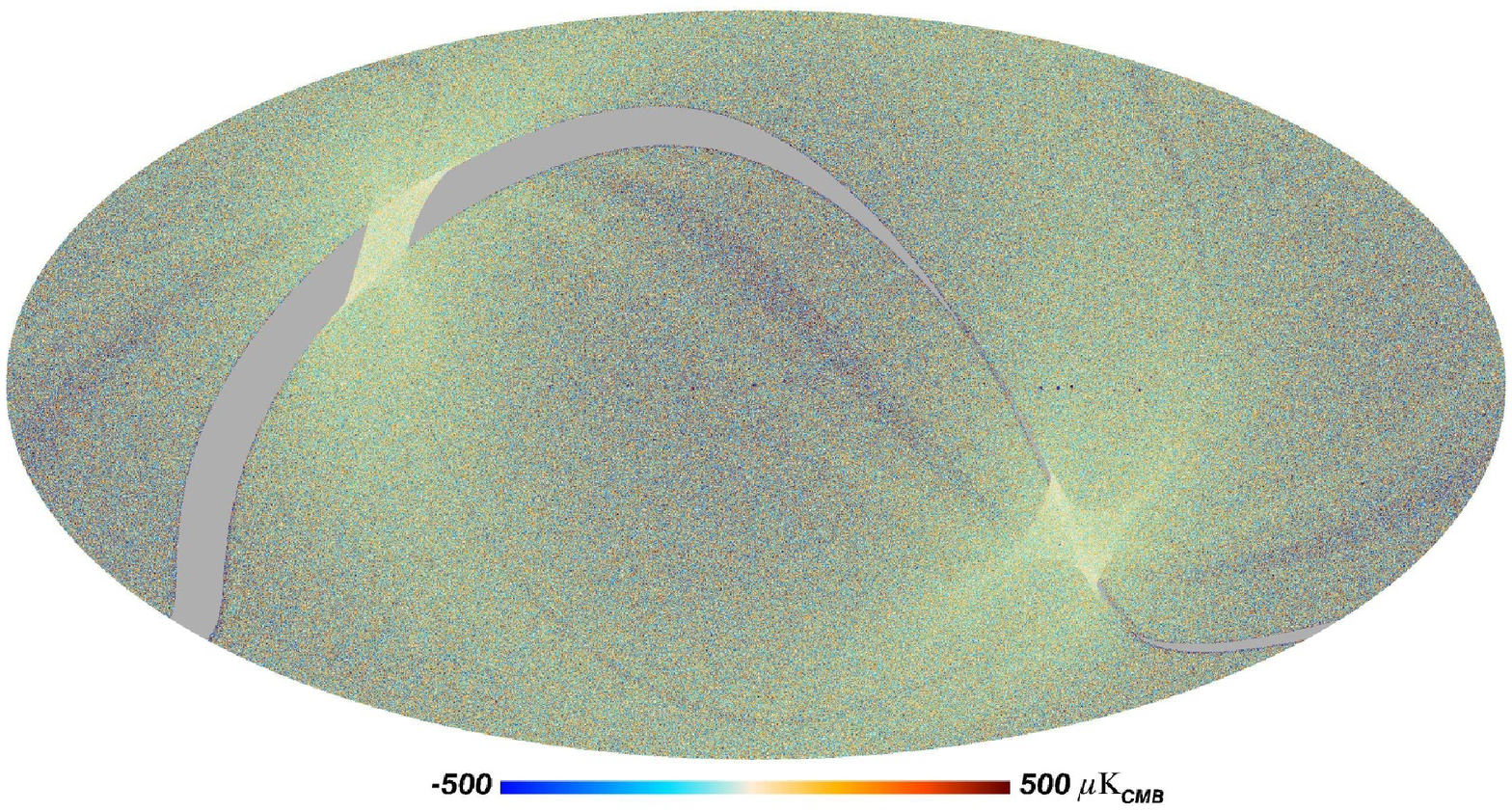}
\caption{Same as Fig.~\ref{fig:Imaps-030} for 70\,GHz.}
\label{fig:Imaps-070}
\end{centering}
\end{figure*}

Figures~\ref{fig:Imaps-030} to \ref{fig:Imaps-070} show the 30,
44, and 70\,GHz frequency maps created from LFI data. The top map
in each figure is the temperature ($I$) map based on the nominal
mission data.  The middle row is the difference between maps
($n_{\rm m}$) made of the first and second half of each stable
pointing period (half-ring maps) weighted by the hit count
calculated from Eq.~\ref{eq:hitweight}. These maps provide a
direct measure of the noise on timescales down to half the
pointing period, and are calculated as

\begin{equation}
   n_{\rm m} =  \frac{h_{\rm 1} - h_{\rm 2}}{w_{\rm hit}}\,,
\label{eq:noisemap}
\end{equation}
where the hit-count weight is
\begin{equation}
   w_{\rm hit} = \sqrt{N^{\rm hit}_{\rm full}
     \left[ \frac{1}{N^{\rm hit}_1} +
       \frac{1}{N^{\rm hit}_2} \right]}.
\label{eq:hitweight}
\end{equation}
Here $N^{\rm hit}_{\rm full} = N^{\rm hit}_1 + N^{\rm hit}_2$
is the hit count of the full map $m$,
while $N^{\rm hit}_1$ and $N^{\rm hit}_2$
are the hit counts of the half-ring maps $h_1$ and $h_2$, respectively.

The bottom row is the difference between Survey1 and Survey2, and
gives information on longer-timescale variations. A dark stripe
corresponding to two observing days is visible in the half-ring
difference maps and also (faintly) in the frequency maps.  This is
due to the fact that in the first days of observation the
instrument was affected by an occasional bit-flip change in the
gain-setting circuit of the data acquisition electronics, probably
due to cosmic ray hits.  The data acquired before a workaround for
the problem was implemented are flagged out, leading to a stripe
of reduced integration time and higher noise. One other clear
feature can be seen at the Galactic plane in the survey difference
maps, especially at 30\,GHz. There is an apparent split in
intensity (seen as a separation of red and blue) due to the beam
ellipticity: the elliptical beam had a different orientation
relative to the Galaxy in Survey1 than in Survey2.

%% file: 10-00_polarization.tex
LFI data processing has included analysis of polarization from the
beginning, but polarization results are not included in the 2013 data
release and scientific analysis because the level of systematic errors
in the maps remains above acceptable levels for cosmological work. In this section we
outline the polarization-specific steps in the data analysis, quantify
the residual systematics, and sketch how we expect to correct them for
the next data release.

To an excellent approximation (see Sect.~\ref{sec_madam}), the $Q$ and $U$
polarization maps are derived from
the difference between the calibrated signals from the two
radiometers in each RCA, the main (M) and side arm (S),
which are sensitive to orthogonal polarizations. Any differential
calibration errors between M and S cause leakage of total
intensity into the polarization maps. Such mismatch arises from
three main causes:

\begin{itemize}

\item differences between the beam profiles of M and S;

\item errors in the gain calibration;

\item differential color corrections between M and S due to
differences in their bandpasses (the ``bandpass leakage effect'').

\end{itemize}

These effects are described in \citet{leahy2010}.  Polarization imposes stringent
requirements on the accuracy of gain calibration that have driven our choice of
calibration scheme \citep{planck2013-p02b}.  The control of systematics for
polarization also requires precise cancellation of the M and S signals.
This underlies our use of identical pointings for the data from the two radiometers
in each horn, despite a small amount of beam squint between the polarizations,
and also the decision to use horn-uniform weighting.  Neither of these choices
results in significant degradation to the total intensity maps, although they
are slightly sub-optimal from the standpoint of noise.

Bandpass leakage requires a special step in the calibration.
The principle instrumental factor controlling bandpass leakage is the effective
frequency mismatch between M and S detectors, $a = (\nu_{\mathrm S} - \nu_{\mathrm M})/2\nu_0$,
which must be combined with estimates of the ``leakage amplitude'' of the
foreground emission, $L=(\beta_\mathrm{fg}-\beta_\mathrm{CMB})T_\mathrm{fg}$.
Because the leakage amplitude relies on products from component separation, maps of
$L$ are available at no higher resolution than that of the 30\,GHz channel
(33\parcm16), and are most reliable at lower resolution (1\deg),
where our analysis can incorporate the {\tt WMAP\/} 22\,GHz maps.

In principle, the $a$-factors can be estimated from the bandpass profiles
measured in the ground calibration campaign \citep{zonca2009}, but as
anticipated by \citet{leahy2010}, more accurate values can be found from
the flight data; a detailed description of our approach to this will be
discussed in a future paper. We estimate that our $a$-factors
are currently accurate to about 0.05\,\%, based on the scatter in values
derived from multiple calibrators.  A bigger problem at present is
accurate evaluation of the leakage amplitude, which requires not only
excellent separation of CMB and foreground emission, but also accurate
estimates of the foreground spectral index within the band.  Currently, the
combined uncertainty in the bandpass leakage correction is about 0.3\,\% of
the local foreground intensity. This is comparable to the mean polarization
fraction along the Galactic plane.

One other parameter must be calibrated for polarization: the
precise orientation of the polarization response for each feed horn.  This could not
be calibrated on the ground; however, as noted by \citet{leahy2010},
knowledge of horn orientation was expected to be better than 1\deg\ from
construction tolerances alone. In flight, we check these values by
observations of the Crab nebula, and the results confirm the orientations
to within a few degrees, which is sufficient for analysis of the $E$-mode
spectra.  At a higher level of precision our estimates of the response
orientations are still affected by the uncertainty in $a$-factors, and so
as yet we have no evidence to reject the nominal orientation angles.

\begin{figure*} [th!]
\centering
\includegraphics[width=6cm]{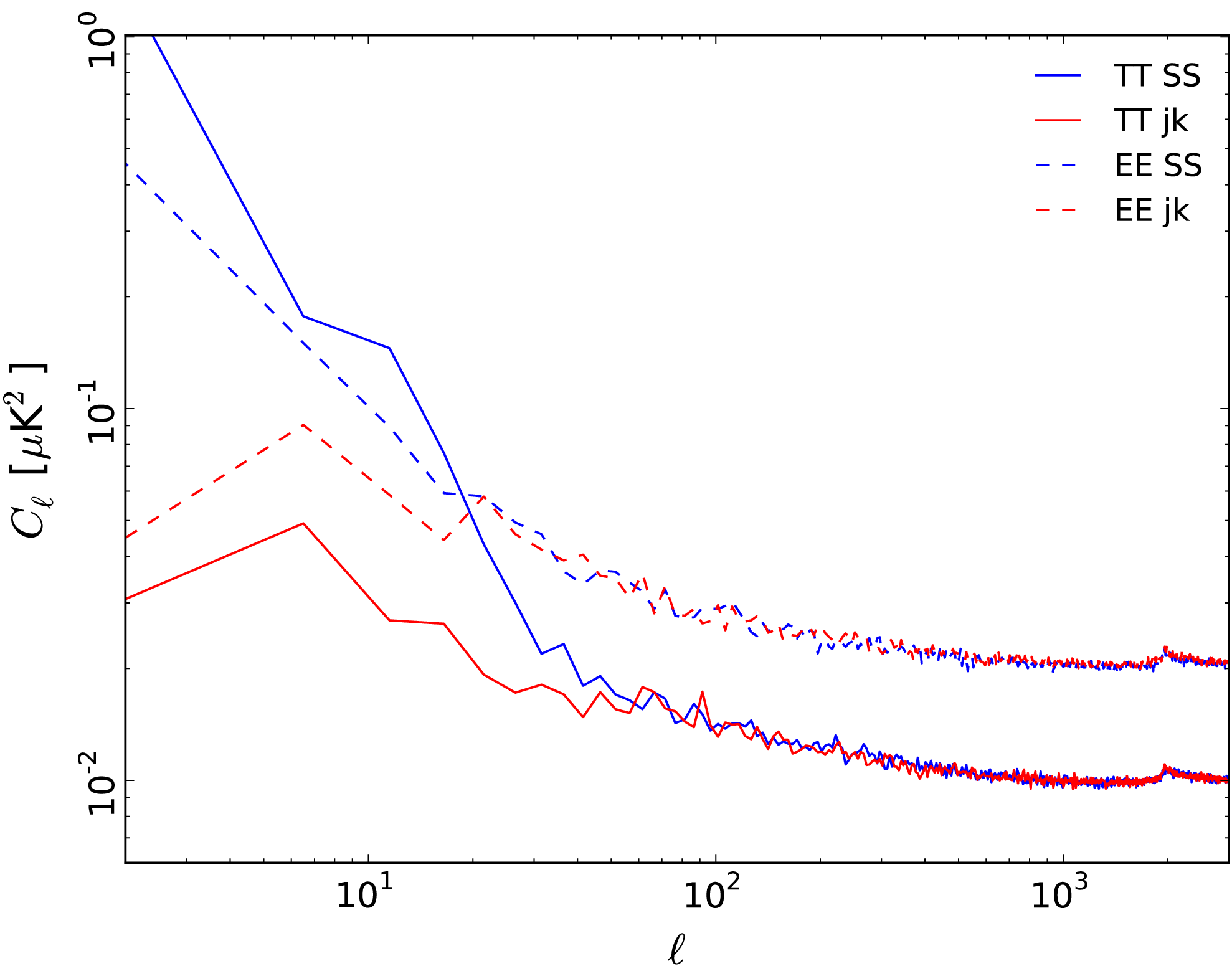}
\includegraphics[width=6cm]{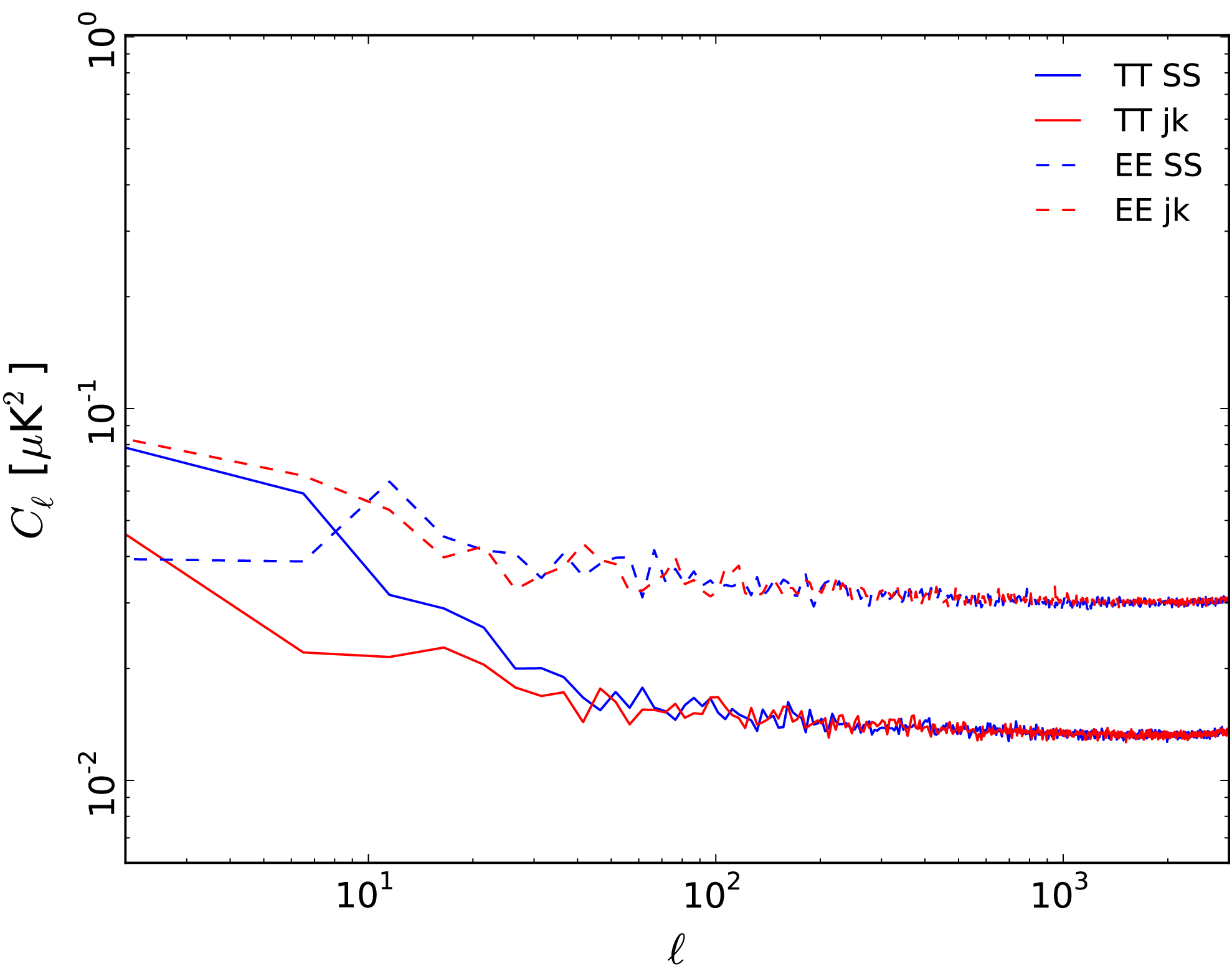}
\includegraphics[width=6cm]{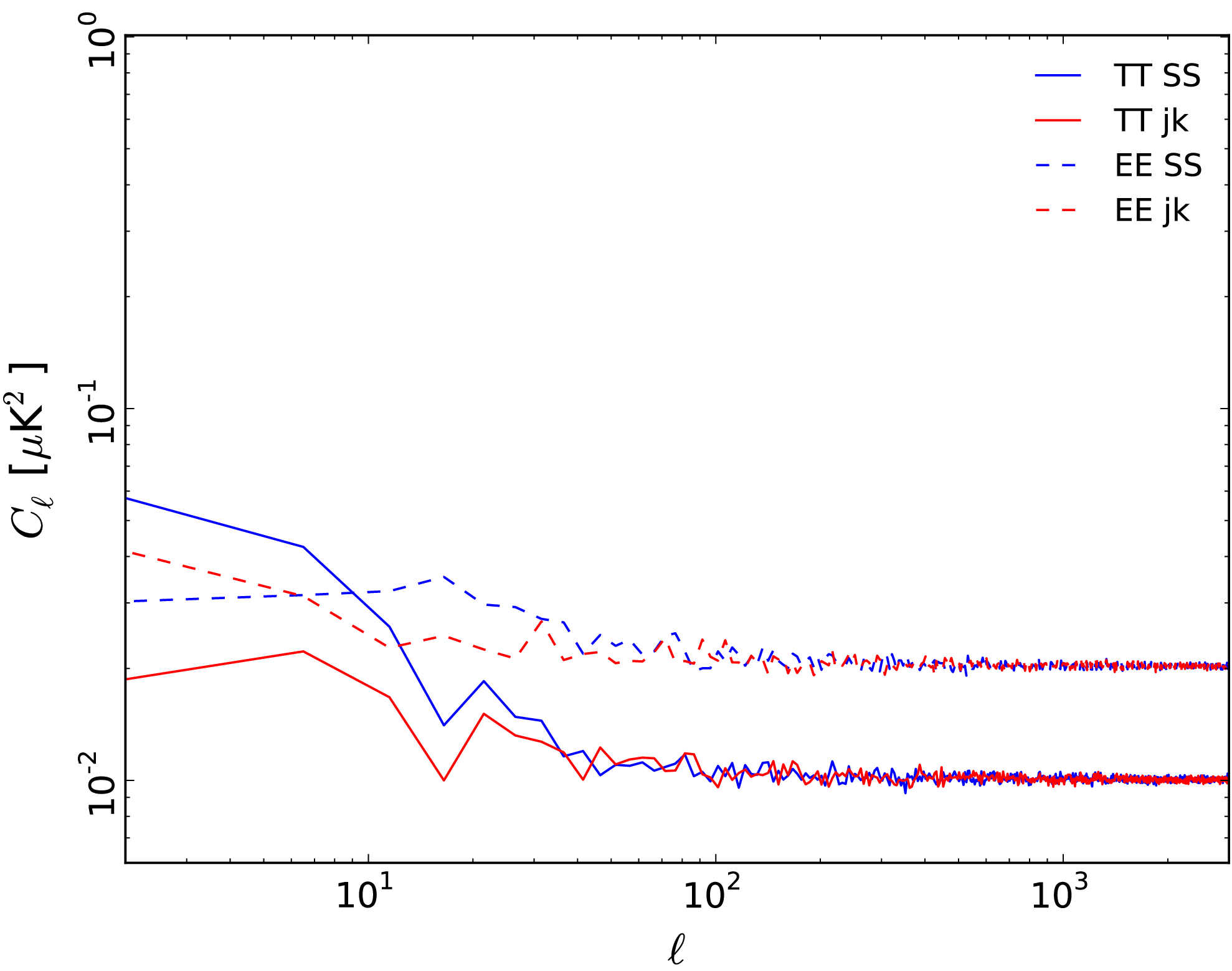}
\caption{Null-test results comparing power spectra from survey
difference maps (SS) to those from half-ring difference maps (hr).
Some excess at low multipoles is clearly visible at 30\,GHz
(\emph{left}), where the main source has been identified as
sidelobe pickup. At 44 (\emph{center}) and 70\,GHz (\emph{right}),
there is less low-$\ell$ contribution both in $TT$ and in $EE$,
although residuals are still present. For multipoles larger than
few tens, null-test $EE$ spectra follow the expected level of
noise as traced by half-ring differences. }
\label{nulltestspectra}
\end{figure*}

The most interesting cosmological signal visible in LFI
polarization is the large-scale ($\ell < 10$) $E$-mode peak due to
reionization, at a typical brightness level of 0.3\,$\mu$K.
Crucial tests of the reliability of this signal are that the
$B$-mode and $EB$ cross-correlation spectra should contain
negligible signal, since a cosmological $B$-mode signal at this
level would correspond to a tensor-to-scalar ratio significantly
larger than current upper limits, while the cosmological $EB$ mode
is precisely zero in most models. The primary LFI channel for
cosmology is 70\,GHz, which has the least foreground contamination
of all \Planck\ channels.  Our likelihood pipeline estimates these
spectra using a conservative Galactic mask, and corrects for
residual foregrounds based on the \Planck\ 30\,GHz map or {\tt
WMAP\/} maps. With our current calibration, both these spectra
contain residuals at a level comparable to the expected $E$-mode
signal from reionization. Hence, although the latter is apparently
detected, we cannot be confident that the signal is real. This
situation is better illustrated in Fig.~\ref{nulltestspectra},
where we report, at all three LFI frequencies, null-test spectra
from survey-survey differences.  At all frequencies, the null-test
$EE$ spectra are in good agreement with the noise level as
determined by half-ring difference maps at multipoles larger of
few tens; this is an indication of the data quality at these
multipoles.  However, residuals are present at very low-$\ell$,
especially at 30\,GHz and at 70\,GHz, and these preclude as yet a
proper characterization of the cosmological signal.

We have simulated the impact of numerous systematic errors to see if they
can explain the observed residuals, including foreground correction, bandpass
mismatch, Galactic stray light (i.e., leakage through the far sidelobes), and
gain errors. None of these simulations has individually generated artefacts
as large as those observed. The most likely candidate seems
to be in the combination of far sidelobes and calibration errors.
As described in \citet{planck2013-p02b}, uncertainty in the far-sidelobe
pattern is one of the dominant contributors to our calibration uncertainty,
as well as making our estimates of the additive effect of Galactic stray light
quite uncertain.

%% file: 11-00_power_spectra.tex
Temperature power spectra are computed from frequency maps using {\tt cROMAster}, an implementation of the pseudo-$C_{\ell}$ method described in \citet{master}.
We extend it to derive both auto- and cross-power spectra (see \citealt{polenta_CrossSpectra} for a comparison between the two estimators).
Noise bias and covariance matrices are computed through the FFP6 full focal plane simulations,
which include 1000 realizations of both signal and noise consistent with \Planck\ data. The angular response of the instrument is accounted for by using the beam window functions presented in \citet{planck2013-p02d}.
Coupling kernels to correct for uncompleted sky coverage are computed as described in Appendix~D of \citet{planck2013-p08}.
We mask the Galactic plane and unresolved sources using masks described in Sect.~3 of \citet{planck2013-p06}, in particular, a 70\,\% Galactic mask for 44 and 70\,GHz (70\,\% of the sky uncovered) and a 60\,\% Galactic mask for 30\,GHz.

\begin{figure}
\begin{centering}
\includegraphics[width=1\columnwidth]{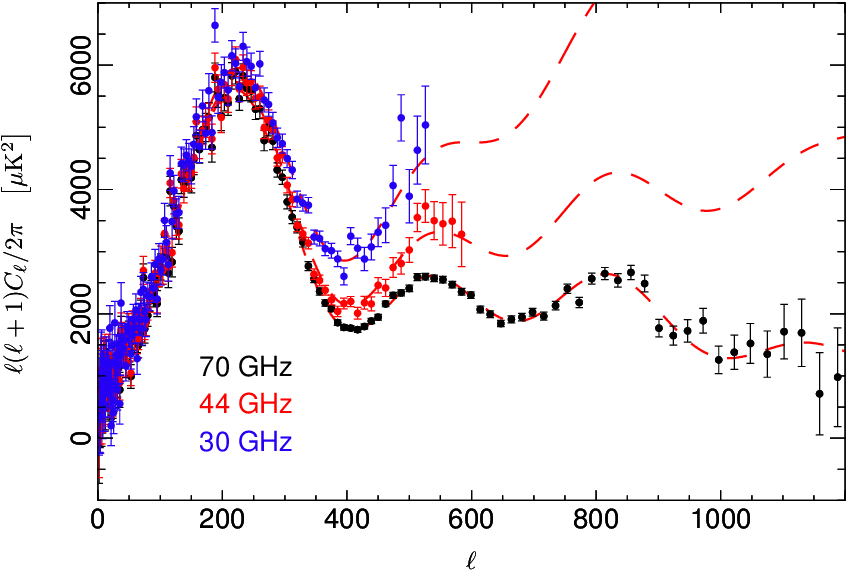}
\par\end{centering}
\caption{Temperature power spectra at 30, 44, and 70\,GHz. Diffuse
foregrounds are reduced only by masking, and a simple foreground
component corrects for residual point sources. Dashed lines
correspond to the \Planck\ Likelihood Code best fit model.
\label{fig:LFIspectra}}
\end{figure}

Figure~\ref{fig:LFIspectra} shows the 30, 44, and 70\,GHz
temperature power spectra produced from frequency maps, without
diffuse component separation but with a simple correction for
unmasked discrete source residuals. There is good agreement
between the observed spectra and the \Planck\ likelihood code
best-fit model \citep{planck2013-p08}.

%% file: 12-00_dataval_intro.tex
To assess and verify the quality of the data produced within the data-analysis pipeline, a set of null tests
is performed.  The main goal is to detect possible instrumental systematic effects.  
These include effects that are either properly corrected or accounted for
later on in the pipeline, effects related to known changes in the operational conditions of the instrument (e.g., the switchover
of the sorption cooler), or intrinsic instrument properties coupled to the sky such as stray light from sidelobes.
Such tests are also useful for a detailed analysis of the processing steps implemented in the pipeline.  Finally, such tests may discover unanticipated problems (e.g., related to different calibration approaches).

Null tests are carried out on blocks of data on different time
scales ranging from individual pointing periods to one year of observation, and 
at different instrument levels (radiometer, horn, and horn-pairs within a
given frequency, and at frequency level), for total intensity and,
when applicable, for polarization.  Such an approach is 
demanding computationally, and special tools have been built to create a 
parallel code, compute the null tests, create the output maps and
spectra, and build a report from the output from the tests.

The effects probed by null tests depend on the combination of data and time scale treated. For example, differences at
horn level between odd and even number surveys clearly reveal the impact of sidelobes, since the sky covered is exactly the same but
the orientation of the beam is not. On the other hand, differences between horns for the entire data period
may reveal calibration issues or changes in operational or instrumental conditions.

%% file: 12-01_null_test.tex
It is important to set pass-fail criteria for such null tests. In
general, test failures reveal problems in the data or in the data
analysis that must be carefully studied so that specific actions
can be taken to mitigate the problems. A simple figure of merit is
the actual level of noise in the data derived from half-ring
difference maps.  Any departure from this noise level would
indicate a problem. For example, null-test power spectra are used
in \citet{planck2013-p02a} to check the total level of systematic
effects in the data. Figure~\ref{nulltestspectra} gives results at
frequency level of survey-difference null tests for both TT and EE
spectra, compared to the noise level derived from half-ring
difference maps.

Based on such results we have analyzed our calibration pipeline,
particularly the treatment of sidelobes, which has been updated
with the inclusion of both the intermediate beam and the in-band
beam behavior based on simulations. This will be the final
approach for data calibration in the next release.

Although it is clear from these results that a proper treatment of
sidelobes is necessary for final refinements in calibration, it
is important to note that the overall amplitude of such effects is
well below the CMB signal in total intensity, leaving the analysis
of the temperature maps totally unaffected.

%% file: 12-02_half_ring_test.tex
The middle panels of Figs.~\ref{fig:Imaps-030},
\ref{fig:Imaps-044}, and \ref{fig:Imaps-070} show the noise
calculated from the hit-count-weighted half-ring difference maps
at $N_{\rm side}=1024$, as described by \citet{planck2011-1.6} and
\citet{ planck2013-p28}. As a first quality check of the maps, and
as one of the tests of the whole data processing pipeline up to
the maps, we verify both numerically and visually that the values
obtained by dividing the half-ring difference map pixel-by-pixel
by the square root of the white noise covariance map (\citet{
planck2013-p28}) are approximately Gaussian-distributed with rms
near unity.  The results are 1.0211, 1.0089, and 1.0007 for 30,
44, and 70\,GHz, respectively.

\begin{figure*}
\begin{centering}
\includegraphics[width=1.0\textwidth]{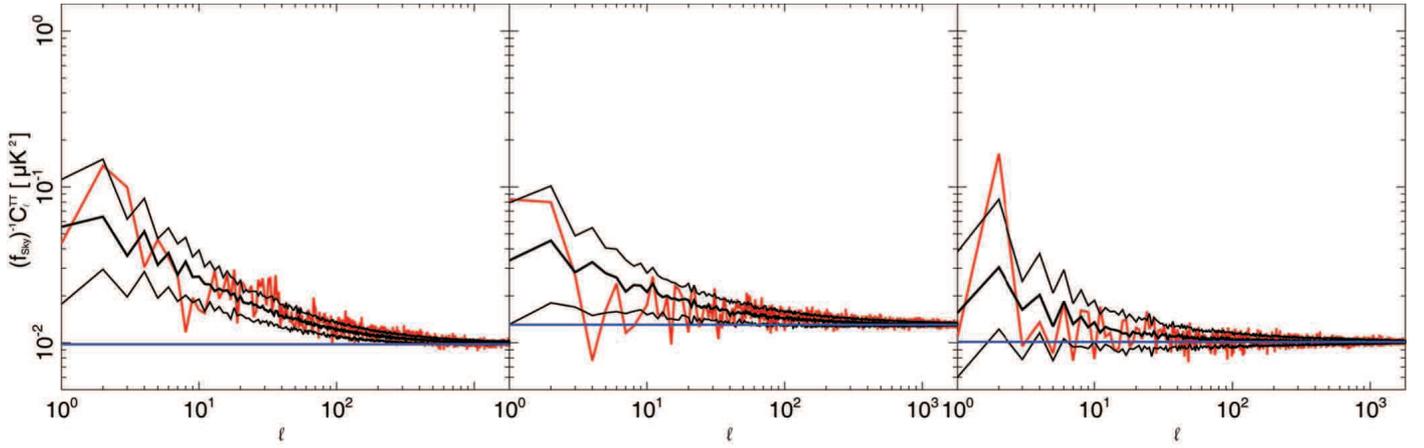}
\end{centering}
\centering{}\caption{\label{fig:HalfringClVsMC}Comparison of the
noise angular power spectra of half-ring difference maps (red),
white noise covariance maps produced by {\tt madam} (blue), and
101 full-noise MCs for each $C_{\ell}$ (black; for 16\%, 50\%
(bold), and 84\% quantiles).  In the noise MC case, no errors were
propagated from Tables~\ref{tab_white_noise_per_radiometer} and
\ref{tab_one_over_f_noise_per_radiometer}; only the median values
of the three noise parameters were used. Those plots validate our
capability to model the noise in the TOD in agreement with the
half-ring difference maps that give the most direct measure of the
noise.}
\end{figure*}

\begin{figure}
\begin{centering}
\includegraphics[width=1.0\columnwidth]{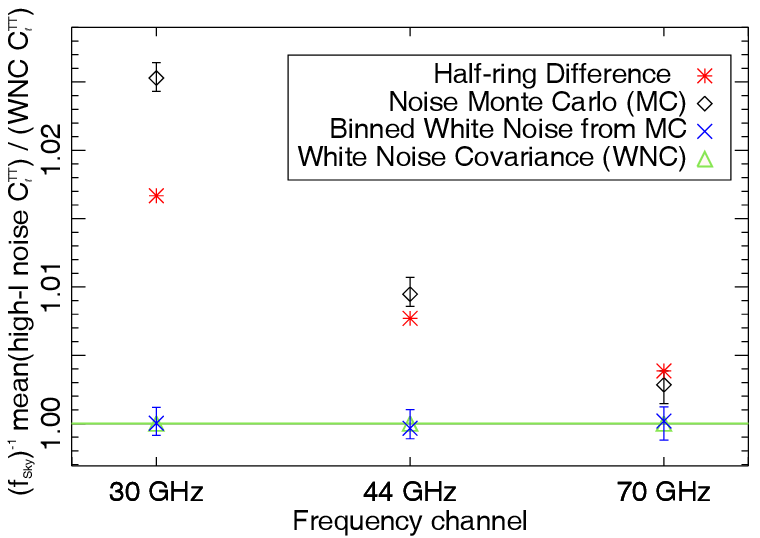}
\end{centering}
\centering{}\caption{\label{fig:highellNoiseComparison}Ratio of
the mean noise angular power at high-$\ell$ ($1150\le\ell\le1800$)
to the white noise estimate from white noise covariance matrices.
The half-ring differences are hit-count-weighted. One hundred and
one full noise MC maps were made. The binned white noise is from
the noise MC. The ``reference'' white noise levels (green) are
$9.8\times10^{-15}\,$K$^2$, $13.1\times10^{-15}\,$K$^2$,  and
$10.1\times10^{-15}\,$K$^2$, for 30, 44, and 70\,GHz,
respectively.  In the noise MCs, no errors were propagated from
Tables~\ref{tab_white_noise_per_radiometer} and
\ref{tab_one_over_f_noise_per_radiometer}; only the median values
of the three noise parameters were used.  Therefore the error bars
in the noise MCs represent only the statistical variance in the
101~realizations. If the uncertainty of the estimation of the
three noise parameters were propagated to the noise MC, the error
bars would be much larger.}
\end{figure}

The half-ring difference maps $n_{\rm m}$ are a direct measure of
the noise in the actual maps. Models of the noise, required for
NCVMs and MC noise realizations, must be validated by comparison
to the half-ring difference maps. For this purpose, we calculate
the temperature and polarization ($E$- and $B$-mode) auto- and
cross-spectra of the half-ring difference maps with {\tt anafast},
and compare these to the results from the white noise covariance
matrices (WNCM) calculated by both {\tt Madam} and MC noise
simulations (Sect.~\ref{sec_noise_simulations}).
Figure~\ref{fig:HalfringClVsMC} gives an example. Further, we
calculate the mean $C_{\ell}$ for the high-$\ell$ tails
($1150\le\ell\le1800$) of the noise angular power spectra and take
the ratio to the WNCM estimate
(Figure~\ref{fig:highellNoiseComparison}). As expected, there is
some residual $1/f$ noise even in the high-$\ell$ region, i.e.,
the full noise MCs lead to slightly higher noise predictions than
the WNCM or binned white noise from the noise MCs.  The residual
$1/f$ noise is of the order of 2.5\,\% at 30\,GHz, 1.0\,\% at
44\,GHz, and 0.1\,\% at 70\,GHz.  We find good consistency between
the noise MCs and the direct noise calculation from the half-ring
difference maps: the high-$\ell$ noise from MCs is only 0.8\,\%
higher at 30\,GHz,  0.2\,\% higher at 44\,GHz,  and  0.1\,\% lower
at 70\,GHz than the result from the half-ring differences. The
error bars of the noise MC do not include at this stage the
uncertainty of noise parameters indicated in
Tables~\ref{tab_white_noise_per_radiometer} and
\ref{tab_one_over_f_noise_per_radiometer}.  More such comparisons
are reported in \citet{ planck2013-p28}.

%% file: 12-03_consistency_scatter.tex
We have tested the consistency between the 30, 44, and 70\,GHz maps by
comparing the power spectra in the multipole range around the
first acoustic peak.  To do so, we remove the
estimated contribution from residual, unresolved point sources from the
spectra presented in Sec.~\ref{sec_power_spectra}. We then
build the scatter plots for the three frequency pairs, i.e., 
70 vs.~30\,GHz, 70 vs.~44\,GHz, and 44 vs.~30\,GHz, and perform a linear
fit accounting for errors on both axis.

\begin{figure*} [th]
\centering
\includegraphics[width=6cm]{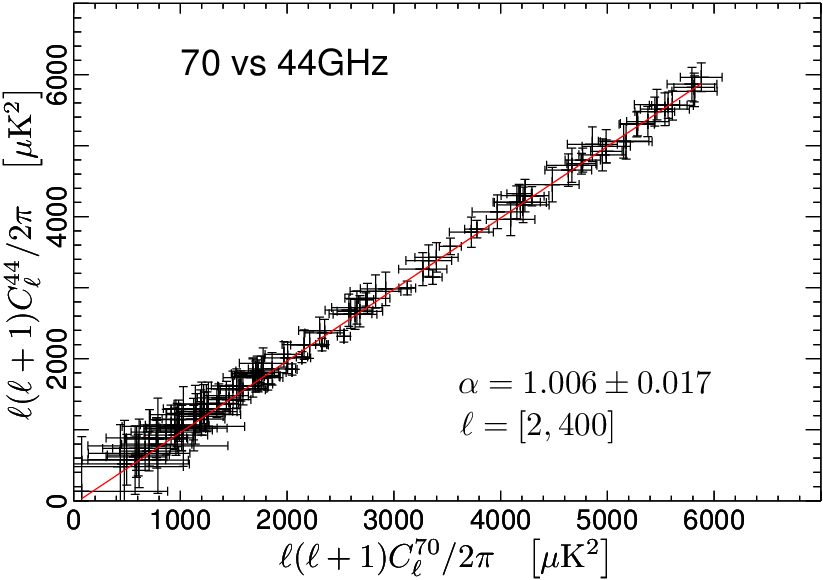}
\includegraphics[width=6cm]{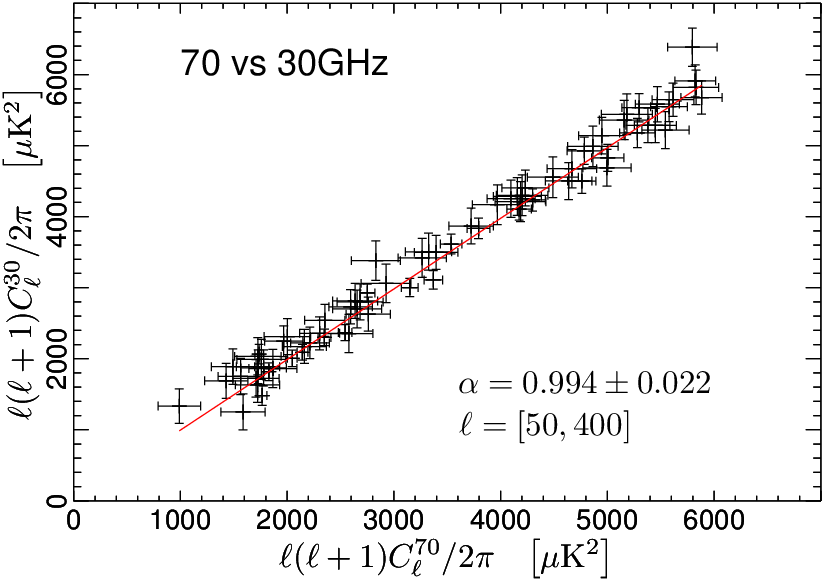}
\includegraphics[width=6cm]{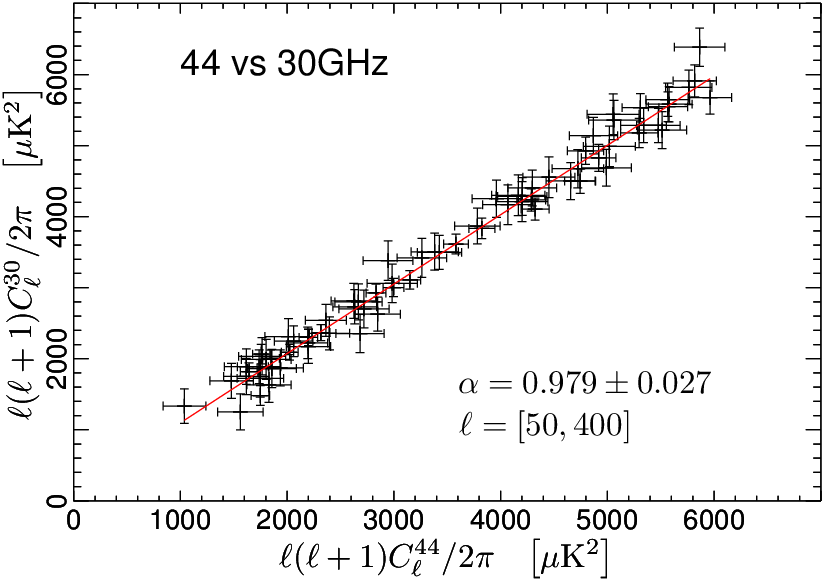}
\caption{Consistency between estimates of the angular power spectra at frequencies.  \emph{Left} to \emph{Right}: 70 vs.~44\,GHz; 70 vs.~30\,GHz; 44 vs.~30\,GHz. Solid red lines are the best fit of the linear regressions, whose slopes $\alpha$ are consistent with unity within the errors.
\label{fig:spectrascatterplots}}
\end{figure*}

The results in Fig.~\ref{fig:spectrascatterplots} show
that the three power spectra are consistent within the errors.
Moreover, the current error budget does not account
for foreground removal, calibration, and window function
uncertainties. Hence the resulting agreement between spectra at
different frequencies can reasonably be considered even more
significant. We also compared the flux densities of compact
sources at the three LFI frequencies, derived from the PCCS \citep{planck2013-p05}, and
find these in acceptable agreement.

%% file: 12-04_int_consis.tex
\begin{figure*} [th]
\centering
\includegraphics[width=6cm]{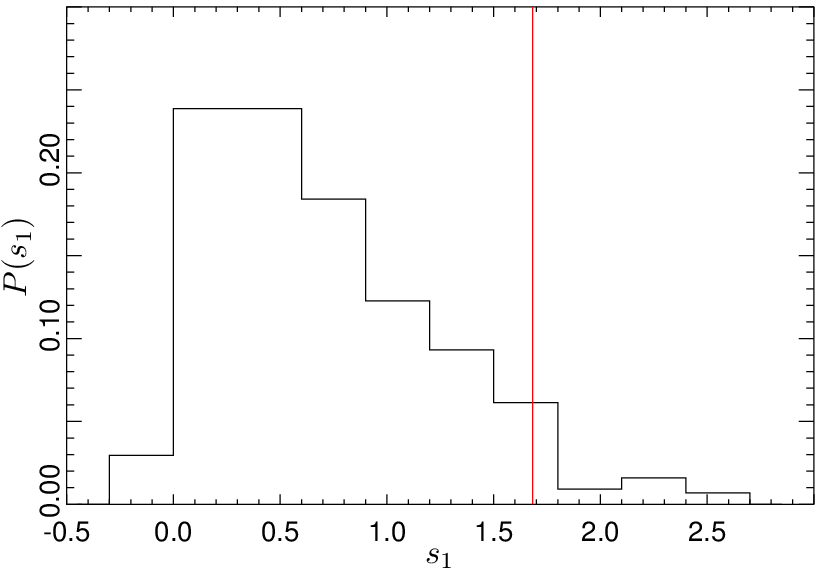}
\includegraphics[width=6cm]{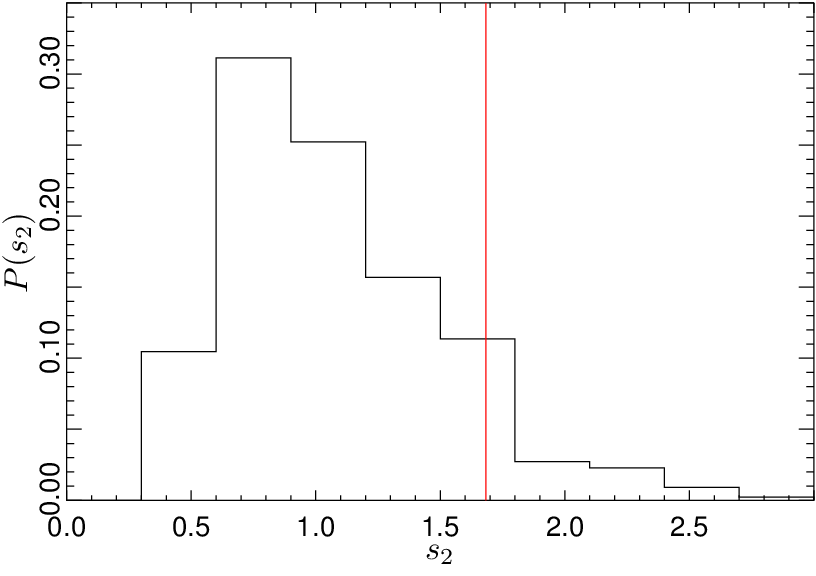}
\includegraphics[width=6cm]{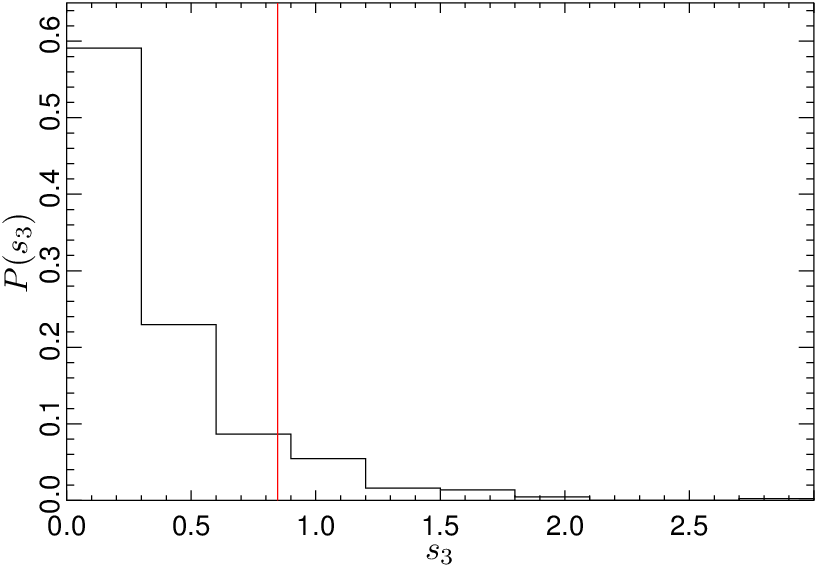}

\caption{Empirical distributions estimated via FFP6 of the $s_{1}$ (left), $s_{2}$ (middle), and $s_{3}$ (right)
statistics (see text). The vertical line represents 70\,GHz data.
\label{fig:hausman}}
\end{figure*}

We use the Hausman test \citep{polenta_CrossSpectra}
to assess the consistency of auto-
and cross-spectrum estimates at 70\,GHz.  Define the statistic
\begin{equation}
H_{\ell}=\left(\hat{C_{\ell}}-\tilde{C_{\ell}}\right)/\sqrt{{\rm Var}\left\{ \hat{C_{\ell}}-\tilde{C_{\ell}}\right\} }\,,
\end{equation}
where $\hat{C_{\ell}}$ and $\tilde{C_{\ell}}$ represent auto- and
cross-spectra, respectively.  To combine information
from different multipoles into a single quantity, we define
\begin{equation}
B_{L}(r) \equiv \frac{1}{\sqrt{L}}\sum_{\ell=2}^{[Lr]}H_{\ell},r\in\left[0,1\right]\,,
\end{equation}
where $[.]$ denotes integer part. The distribution of $B_{L}(r)$
converges (in a functional sense) to a Brownian motion process,
which can be studied through the statistics
$s_{1}=\textrm{sup}_{r}B_{L}(r)$,
$s_{2}=\textrm{sup}_{r}|B_{L}(r)|$, and
$s_{3}=\int_{0}^{1}B_{L}^{2}(r)dr$. Using the FFP6 simulations,
we derive the empirical distribution for all three test
statistics, and then compare them with the results obtained from the \Planck\ data themselves
(see Fig.~\ref{fig:hausman}). The Hausman test shows no
statistically significant inconsistencies between the two spectral
estimates.

\begin{figure}
\begin{centering}
\includegraphics[width=1\columnwidth]{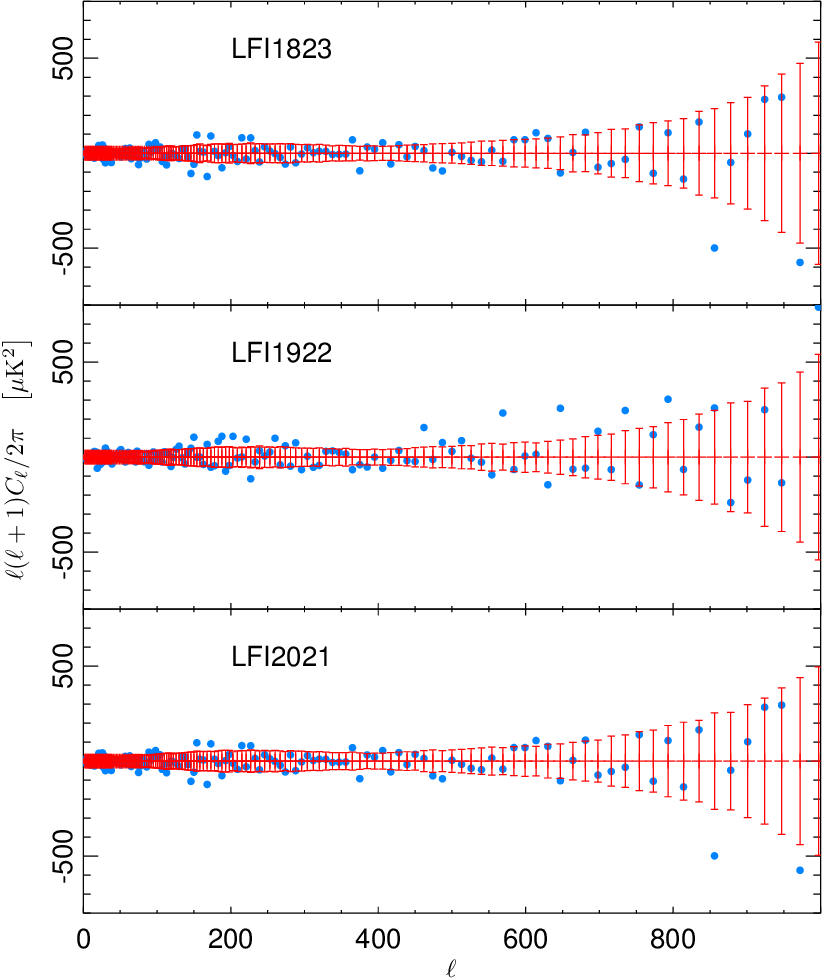}
\par\end{centering}

\caption{Residuals between the auto-spectra of the horn pair
maps and the power spectrum of the full 70\,GHz frequency map.
Error bars are derived from the FFP6 simulations.
\label{fig:70GHz-residuals}}
\end{figure}

As a further test, we estimate the temperature power
spectrum for each of the three horn-pair maps, and compare the
results with the spectrum obtained from all 12 radiometers
shown above.  Figure~\ref{fig:70GHz-residuals} shows the
difference between the horn-pair and the combined spectra.
Again, the error bars have been estimated from the FFP6
simulations.  A $\chi^{2}$ analysis of the residual shows
that they are compatible with the null hypothesis, confirming the
strong consistency of the estimates.

%% file: 12-05_systematics.tex
Known systematic effects in LFI are reported in detail in ~\citet{planck2013-p02a} and summarized in Table~\ref{tab_summary_systematic_effects_maps}, which lists both the rms and the difference between the 99\,\% and the 1\,\% quantiles in the pixel value distributions.  We refer to it as the peak-to-peak (p-p) difference even though it neglects outliers,  as it effectively approximates the peak-to-peak variation of the effect on the map.

\begin{table}[tmb]                 
\begingroup
\newdimen\tblskip \tblskip=5pt
\caption{Effect of known systematic uncertainties on maps$^a$}                          
\label{tab_summary_systematic_effects_maps}                            
\nointerlineskip
\vskip -3mm
\footnotesize
\setbox\tablebox=\vbox{
   \newdimen\digitwidth
   \setbox0=\hbox{\rm 0}
   \digitwidth=\wd0
   \catcode`*=\active
   \def*{\kern\digitwidth}
   \newdimen\signwidth
   \setbox0=\hbox{+}
   \signwidth=\wd0
   \catcode`!=\active
   \def!{\kern\signwidth}
{\tabskip=0pt
\halign{
\hbox to 1.3in{#\leaderfil}\tabskip=1.5em&
\hfil#\hfil\tabskip=0.8em&
\hfil#\hfil\tabskip=1.5em&
\hfil#\hfil\tabskip=0.8em&
\hfil#\hfil\tabskip=1.5em&
\hfil#\hfil\tabskip=0.8em&
\hfil#\hfil\tabskip=0pt\cr
\noalign{\doubleline}
\omit&\multispan6\hfil E{\sc ffect in maps} [$\mu$K$_{\rm cmb}$]\hfil\cr
\noalign{\vskip -3pt}
\omit&\multispan6\hrulefill\cr
\noalign{\vskip 2pt}
\omit&\multispan2\hfil 30\,GHz\hfil&\multispan2\hfil 44\,GHz\hfil&\multispan2\hfil 70\,GHz\hfil\cr
\noalign{\vskip -3pt}
\omit&\multispan2\hrulefill&\multispan2\hrulefill&\multispan2\hrulefill\cr
\noalign{\vskip 2pt}
\omit\hfil S{\sc ystematic}\hfil&p-p&rms&p-p&rms&p-p&rms\cr
\noalign{\vskip 3pt\hrule\vskip 5pt}
      Bias fluctuations&   *0.08&0.01&0.10&0.02&0.23&0.06\cr
\noalign{\vskip 4pt}
      Thermal fluctuations&*0.61&0.11&0.40&0.08&1.17&0.20\cr
\noalign{\vskip 4pt}
      1-Hz spikes&         *0.87&0.17&0.14&0.03&0.60&0.12\cr
\noalign{\vskip 4pt}
      Sidelobe pickup&    18.95&4.53&1.92&0.57&6.39&1.91\cr
\noalign{\vskip 4pt}
      ADC non-linearity&   *3.87&1.01&0.89&0.19&0.92&0.19\cr
\noalign{\vskip 4pt}
      Gain residuals&      *4.33&1.16&4.74&0.97&6.51&1.10\cr
\noalign{\vskip 10pt}
      Total&               21.02&4.83&5.61&1.13&7.87&2.00\cr
\noalign{\vskip 5pt\hrule\vskip 3pt}
}
}}
\endPlancktable                    
\tablenote a Calculated on a pixel of size equal to the average beam FWHM.\par
\endgroup
\end{table}                        

Our analysis ~\citep{planck2013-p02a} shows that systematic uncertainties are at least two orders of magnitude below the
CMB temperature anisotropy power spectrum, and are dominated by stray light pick-up from far sidelobes and imperfect
photometric calibration.

%% file: 13_infrast.tex
The computer cluster used for the maps productions has ten 64-bit
nodes with two single-core CPUs and 16\,GB of RAM, and ten 64-bit
nodes with two motherboards, each with two six-core CPUs and
72\,GB of RAM. The total RAM available exceeds 1.5\,TB, sufficient
to allow for the creation of all maps until the end of the
mission. The dual-motherboard nodes are connected through an
InfiniBand 40\,Gbit network interface, while a 1\,Gbit interface
is provided for the other connections.

The hardware infrastructure includes a front-end machine (two quad-core CPUs,
8\,GB of RAM), which is the access point for users and hosts
the pbs server, and a control machine running the LDAP authentication
server and the DNS and DHCP services.

The software used for system management and synchronization
includes {\tt kickstart} and {\tt puppet}, while
parallelization of the computations is guaranteed by the {\tt torque}
resource manager and the {\tt maui} scheduler.

Data products are stored and organized into three different
servers that host the Level~1, Level~2, and test databases
(Fig.~\ref{dpcpipeline}). For each database, there is
associated RAID~6 storage, with up to 40\,TB formatted with the JFS
filesystem.

%% file: 14_conclusion.tex
We have described the pipeline used to process the first 15.5
months of \Planck/LFI data from Level\,1 through temperature
frequency maps, and the assessment of the quality of the products,
which is largely based on null tests.  Companion papers provide
full descriptions of three critical aspects of the data analysis
and products delivered: \citet{planck2013-p02a} analyzes
systematic effects and assesses their impact;
\citet{planck2013-p02b} describes photometric calibration; and
\citet{planck2013-p02d} describes beam patterns and window
functions. The \Planck\ Explanatory Supplement
\citep{planck2013-p28} provides a detailed description of all the
products delivered in this release.

The Level\,1 pipeline has not changed since the start of the
mission in 2009, and has been running flawlessly and
continuously, demonstrating the robustness of the design and
development approach. In contrast, the Level\,2 pipeline has been largely
restructured (see~\citealt{planck2011-1.6} for a description of the initial pipeline).
The major improvements involved new procedures for pointing reconstruction,
detailed estimation of systematic effects, and photometric calibration.
These improvements allow us to obtain, as reported in
Table~\ref{tab_summary_performance}, a final calibration uncertainty of
the order of 0.6\%, and also to propagate, using simulations, known systematic
effects into the final product maps. The impact of the combination of
all known systematic effects is at least two orders of
magnitude below the CMB temperature anisotropy power spectrum.

Particular emphasis is given to null tests, which are routinely
applied to various subsets of the data in order to assess the
scientific quality of the LFI products. The null test procedure,
described in Sec.~\ref{sec_dataval_intro}, allowed us to detect
and solve a number of problems in the Level\,2 pipeline that
emerged during the processing period. In fact, the pipeline is
still being optimized and more improvements are planned for the
next data release. Future improvements will be aimed at obtaining
high-quality polarization results, which require control of
spurious effects at sub-microkelvin level, as well as better
characterization of the beams, taking into account second order
effects such as the bandpass response of each diode and Galactic
stray light (i.e., leakage through the far sidelobes).

%% file: 15_acknow.tex
\begin{acknowledgements}

  \Planck\ is too large a project to allow full acknowledgement of all
  contributions by individuals, institutions, industries, and funding
  agencies. The main entities involved in the mission operations are
  as follows. The European Space Agency (ESA) operates the satellite via its
  Mission Operations Centre located at ESOC (Darmstadt, Germany) and
  coordinates scientific operations via the Planck Science Office
  located at ESAC (Madrid, Spain). Two Consortia, comprising around 50
  scientific institutes within Europe, the USA, and Canada, and funded
  by agencies from the participating countries, developed the
  scientific instruments LFI and HFI, and continue to operate them via
  Instrument Operations Teams located in Trieste (Italy) and Orsay
  (France). The Consortia are also responsible for scientific
  processing of the acquired data. The Consortia are led by the
  Principal Investigators: J.L. Puget in France for HFI (funded
  principally by CNES and CNRS/INSU-IN2P3-INP) and N. Mandolesi in Italy
  for LFI(funded principally via ASI). NASA US Planck Project,
  based at JPL and involving scientists at many US institutions,
  contributes significantly to the efforts of these two Consortia. The
  author list for this paper has been selected by the Planck Science
  Team, and is composed of individuals from all of the above entities
  who have made multi-year contributions to the development of the
  mission. It does not pretend to be inclusive of all contributions.
  The \Planck -LFI project is developed by an International Consortium
  lead by Italy and involving Canada, Finland, Germany, Norway, Spain,
  Switzerland, UK, USA. The Italian contribution to \Planck\ is
  supported by the Italian Space Agency (ASI) and INAF.  This work was supported by the Academy of Finland grants 253204, 256265, and 257989. This work was granted access to the HPC resources of CSC made available within the Distributed European Computing Initiative by the PRACE-2IP, receiving funding from the European Community's Seventh Framework Programme (FP7/2007-2013) under grant agreement RI-283493.  We thank CSC -- IT Center for Science Ltd (Finland) for computational resources.  We acknowledge financial support provided by the Spanish Ministerio
  de Ciencia e Innovaci{\~o}n through the Plan Nacional del Espacio y
  Plan Nacional de Astronomia y Astrofisica.  We acknowledge the
  Max Planck Institute for Astrophysics Planck Analysis Centre (MPAC)
  funded by the Space Agency of the German Aerospace Center (DLR)
  under grant 50OP0901 with resources of the German Federal Ministry
  of Economics and Technology, and by the Max Planck Society. This
  work has made use of the Planck satellite simulation package
  (Level-S), which is assembled by the Max Planck Institute for
  Astrophysics Planck Analysis Centre (MPAC) \cite{reinecke2006}. We
  acknowledge financial support provided by the National Energy
  Research Scientific Computing Center, which is supported by the
  Office of Science of the U.S. Department of Energy under Contract
  No. DE-AC02-05CH11231. Some of the results in this paper have been
  derived using the HEALPix package \cite{gorski2005}.
The development of Planck has been supported by: ESA; CNES and CNRS/INSU-IN2P3-INP (France); ASI, CNR, and INAF (Italy); NASA and DoE (USA); STFC and UKSA (UK); CSIC, MICINN, JA and RES (Spain); Tekes, AoF and CSC (Finland); DLR and MPG (Germany); CSA (Canada); DTU Space (Denmark); SER/SSO (Switzerland); RCN (Norway); SFI (Ireland); FCT/MCTES (Portugal); and PRACE (EU). A description of the Planck Collaboration and a list of its members,  including the technical or scientific activities in which they have been involved, can be found at \burl{http://www.sciops.esa.int/index.php?project=planck&page=Planck_Collaboration}.
\end{acknowledgements}